\documentclass[aps,pra,reprint,longbibliography]{revtex4-1}
\usepackage{graphicx,color}
\usepackage{amsmath, amssymb}
\usepackage{longtable}
\usepackage{natbib}
\usepackage{bm}

\begin{document}
\title{
Dynamical Jahn-Teller effect of fullerene anions
}
\thanks{This article is dedicated to Professor Isaac Bersuker on the occasion of his 90th birthday}

\author{Dan Liu} 
\author{Naoya Iwahara}
\email{naoya.iwahara@kuleuven.be}
\author{Liviu F. Chibotaru}
\email{liviu.chibotaru@kuleuven.be}
\affiliation{Theory of Nanomaterials Group, University of Leuven, Celestijnenlaan 200F, B-3001 Leuven, Belgium}
\date{\today}

\begin{abstract}
The dynamical Jahn-Teller effect of C$_{60}^{n-}$ anions ($n = $ 1-5) is studied using the numerical diagonalization of the linear $p^n \otimes 8d$ Jahn-Teller Hamiltonian with the currently established coupling parameters.
It is found that in all anions the Jahn-Teller effect stabilizes the low-spin states, resulting in the violation of Hund's rule. 
The energy gain due to the Jahn-Teller dynamics is found to be comparable to the static Jahn-Teller stabilization. 
The Jahn-Teller dynamics influences the thermodynamic properties via strong variation of the density of vibronic states with energy. 
Thus, the large vibronic entropy in the low-spin states enhances the effective spin gap of C$_{60}^{3-}$ quenching the spin crossover. 
From the calculations of the effective spin gap in function of the Hund's rule coupling,
we found that the latter should amount 40 $\pm$ 5 meV in order to cope with the violation of Hund's rule and to reproduce the large spin gap. 
With the obtained numerical solutions the matrix elements of electronic operators for the low-lying vibronic levels and the vibronic reduction factors are calculated for all anions. 
\end{abstract}

\maketitle

\section{Introduction}
\label{Introduction}
Fullerene based compounds show diverse phenomena such as superconductivity and metal-insulator transition in alkali-doped fullerides 
\cite{Gunnarsson1997, Gunnarsson2004, Haddon1991, Hebard1991, Tanigaki1991, Winter1992, Kerkoud1996, Knupfer1997, Brouet2001, Ganin2008, Ihara2010, Klupp2012, Iwahara2013, Potocnik2014, Iwahara2015, Zadik2015, Iwahara2016, Nomura2016, Mitrano2016, Kasahara2017, Nava2017},
ferro- and antiferromagnetisms in intercalated fullerides \cite{Margadonna2001, Durand2003, Chibotaru2005} 
and various organic-fullerene compounds \cite{Allemand1991, Kawamoto1997, Sato1997, Kambe2007, Amsharov2011, Francis2012, Konarev2013}.
One of the peculiarities of the fullerene materials is that the molecular properties of C$_{60}$ ion persist in crystals, for example, the Jahn-Teller (JT) dynamics of C$_{60}$ ions \cite{Auerbach1994, Manini1994, OBrien1996, Chancey1997} is not quenched.
Nonetheless, the dynamical JT effect in crystalline materials has not been thoroughly understood in the past due to the lack of precise knowledge of vibronic coupling parameters characterizing the JT effect, the complexity of the JT dynamics itself, and the interplay of the vibronic coupling and the other interactions in crystals such as bielectronic and electron transfer interactions.

The orbital vibronic coupling constants of C$_{60}$ have been intensively studied via their extraction from spectroscopy \cite{Gunnarsson1995, Winter1996, Hands2008} and with various theoretical methods \cite{Varma1991, Schluter1992, Faulhaber1993, Antropov1993, Breda1998, Manini2001, Saito2002, Frederiksen2008, LaflammeJanssen2010}.
Theoretically derived parameters depended on the applied method and gave for the static JT stabilization energy of monoanion $E_\text{JT}^{(1)}$ values ranging range from 30 to 90 meV.
In particular, it has been a long standing problem that all the theoretical calculations predict at most a half of $E_\text{JT}^{(1)}$ derived from photoelectron spectrum available at the time \cite{Gunnarsson1995}.
The latter was recorded at high temperature (ca 200 K) and with low resolution.
A decade later, a new photoelectron spectrum of C$_{60}^-$ became available \cite{Wang2005},
recorded at low temperature (70-90 K) with sufficiently high resolution to show clear vibronic structure.
From this spectrum, the vibronic coupling parameters were derived \cite{Iwahara2010} via the simulations involving a large spectrum of vibronic states of the linear $t_{1u} \otimes (2a_g \oplus 8h_g)$ Jahn-Teller Hamiltonian. 
In this derivation, effect of thermal excitations, ignored in the treatment of Ref. \cite{Gunnarsson1995}, was also taken into account. 
The derived coupling parameters were found to be in good agreement \cite{Iwahara2010} with those extracted from the density functional theory (DFT) calculations with the hybrid functional B3LYP \cite{Saito2002, LaflammeJanssen2010, Iwahara2010}.
The accuracy of the coupling parameters with B3LYP functional was supported by the {\it GW} approximation \cite{Faber2011}.
With this advancement, it is now possible to address the actual situation of the JT dynamics of C$_{60}$ anions.

In this work, we study the low-energy vibronic states of isolated C$_{60}^{n-}$ ions ($n = $ 1-5) with the established coupling parameters. 
To this end, the low-lying vibronic states of C$_{60}^{n-}$ were obtained by numerical diagonalization of the JT Hamiltonian including all the JT active modes and the bielectronic interaction. 
With the obtained vibronic states, the matrix elements of the electronic irreducible tensor operators and spin gaps were calculated. 
Present results on the low-energy vibronic structure of C$_{60}$ anions give us solid ground to access the real situation of fullerene based materials.

\section{Vibronic and electronic interactions in C$_{60}^{n-}$}
\label{Vibronic}
The three-fold degenerate $t_{1u}$ lowest unoccupied molecular orbital (LUMO) level of C$_{60}$ ($I_h$ symmetry) is highly electronegative and upon electron doping (in fullerides) the LUMOs become partially filled.
The $(t_{1u})^n$ electron configurations split into electronic terms due to the bielectronic interaction.  
For $n =$ 1-5, the $t_{1u}$ orbitals couple to the molecular vibrations of the C$_{60}$ cage (vibronic coupling).
According to the selection rule,
\begin{eqnarray}
 [t_{1u} \otimes t_{1u}] &=& a_g \oplus h_g,
\label{Eq:selection}
\end{eqnarray}
where the square bracket in Eq. (\ref{Eq:selection}) stands for the symmetrized product, the $t_{1u}$ orbital linearly couples to totally symmetric $a_g$ vibrations and five-fold degenerate $h_g$ vibrational modes \cite{Jahn1937, Bersuker1989, Chancey1997}.
Since the $a_g$ modes are irrelevant to the JT effect, we will not consider them in this work. 
The model Hamiltonian describing the low-energy states of C$_{60}^{n-}$ is given by 
\begin{eqnarray}
 \hat{H} &=& \hat{H}_\text{bi} + \hat{H}_0 + \hat{H}_\text{JT},
\label{Eq:H}
\end{eqnarray}
where $\hat{H}_\text{bi}$ is the bielectronic part, $\hat{H}_0$ is the Hamiltonian of the harmonic oscillators of all JT active $h_g$ modes, and $\hat{H}_\text{JT}$ is the linear vibronic coupling term \cite{Auerbach1994, Manini1994, OBrien1996, Chancey1997}.
The analysis including the quadratic vibronic coupling \cite{Dunn1995, Alqannas2013} with the coupling parameters for C$_{60}^{n-}$ anions will be presented elsewhere.
$\hat{H}_0$ is written as 
\begin{eqnarray}
 \hat{H}_0 &=& \sum_{\mu = 1}^8 \sum_{m = -2}^2 \hslash \omega_\mu \left( \hat{n}_{\mu m} + \frac{1}{2} \right), 
\label{Eq:H0}
\end{eqnarray}
where $\mu$ distinguish the vibrational frequency $\omega_\mu$ of the $h_g$ modes, $m$ ($= -2, -1, 0, +1, +2$) is the $z$ component of the $h_g$ mode in the spherical form ($m$ stands for vibrational angular momentum \cite{OBrien1971, Auerbach1994}), and $\hat{n}_{\mu m}$ is the vibrational quantum number operator. 
The eigenstate of $\hat{H}_0$ is described by the set of vibrational quantum numbers $\bm{n} = \{..., n_{\mu m}, ...\}$.
The forms of $\hat{H}_\text{bi}$ and $\hat{H}_\text{JT}$ depend on the number of electrons $n$ and, therefore, we will discuss them separately.
The JT Hamiltonian matrices for $n = 1,5$ and for $n = 2,3,4$ are, respectively, the same as those 
in Refs. \cite{OBrien1969, Auerbach1994, Manini1994, OBrien1996, Chancey1997} and Refs. \cite{OBrien1996, Chancey1997}.
Their derivation is given in Appendix \ref{A:HJT}.

Further we make use of the following notations.
Within $I_h$ symmetry, the $a_g$, $t_{1u}$, and $h_g$ irreducible representations transform as the ones of SO(3) group with angular momenta $l = 0,1,2$, respectively \cite{Altmann1994}.
Thus, the electronic states are specified by atomic $LS$ terms \cite{Condon1953}. 
The orbital part of the $LS$ term is written as $|LM_L\rangle$ ($M_L = -L, -L+1, ..., L$), and 
the projection operator into the term is $\hat{I}_L = \sum_{M_L=-L}^L |LM_L\rangle\langle LM_L|$.
The only bielectronic parameter, the Hund's rule coupling parameter, is denoted $J_\text{H}$.
The dimensionless vibronic coupling constant to the $\mu$ mode is $g_\mu$
and the dimensionless normal coordinate is $\hat{q}_{\mu m} = [\hat{b}^\dagger_{\mu m} + (-1)^m \hat{b}_{\mu, -m}]/\sqrt{2}$, 
where $\hat{b}^\dagger_{\mu m}$ ($\hat{b}_{\mu, -m}$) is the creation (annihilation) operator corresponding to the vibrational $\mu m$ mode \cite{Auerbach1994}.

\subsection{$n=1,5$}
\label{Sec:C60-}
Since there is only one electron (hole), $\hat{H}_\text{bi} = 0$.
We use the zero point energy of $\hat{H}_0$ as the origin of energy. 
The vibronic coupling for the $p^1$ system is given by 
\cite{OBrien1969, Auerbach1994, Manini1994, OBrien1996, Chancey1997}: 
\begin{eqnarray}
 \hat{H}_\text{JT} &=& 
 \sum_{\mu=1}^8 \hslash \omega_\mu g_\mu 
 \left(|P,-1\rangle, |P,0\rangle, |P,+1\rangle \right)
\nonumber\\
 &\times&
 \begin{pmatrix}
  \frac{1}{2} \hat{q}_{\mu, 0}         & \frac{\sqrt{3}}{2} \hat{q}_{\mu, +1} &  \sqrt{\frac{3}{2}} \hat{q}_{\mu, +2}\\
 -\frac{\sqrt{3}}{2} \hat{q}_{\mu, -1} & -\hat{q}_{\mu, 0}                    & -\frac{\sqrt{3}}{2} \hat{q}_{\mu, +1} \\
  \sqrt{\frac{3}{2}} \hat{q}_{\mu, -2} & \frac{\sqrt{3}}{2} \hat{q}_{\mu, -1} &  \frac{1}{2} \hat{q}_{\mu, 0}
 \end{pmatrix}
 \begin{pmatrix}
  \langle P,-1|\\
  \langle P,0| \\ 
  \langle P,+1| \\
 \end{pmatrix}.
\nonumber\\
\label{Eq:HJT1}
\end{eqnarray}
The JT Hamiltonian for the $p^5$ system is of the same form as Eq. (\ref{Eq:HJT1}) except for the opposite sign of entering $g_\mu$, a usual situation for the single-electron operator under electron-hole transformation. 
In the presence of the vibronic coupling, neither vibrational nor electronic angular momenta for the JT active $d$ modes and the $p$ orbitals, respectively, commute with the Hamiltonian. 
However, the square of the projections of the total angular momentum, $\hat{J}_q$ ($q = -1, 0, +1$), which is the sum of the vibrational and the electronic angular momenta (Appendix \ref{A:J}), and any projection $\hat{J}_q$ commute \cite{OBrien1971, Romestain1971}.
Thus, the eigenstate of Eq. (\ref{Eq:H}) is characterized by the total angular momentum $J$ ($ = 0, 1, 2, ...$), the $z$ component $M_J$ ($= -J, -J+1, ..., J$), and the other quantum numbers $\alpha$.
The general form of the vibronic state is 
\begin{eqnarray}
 |\Psi_{\alpha JM_J}\rangle &=& \sum_{M_L}|PM_L\rangle |\chi_{PM_L; \alpha J M_J}\rangle,
\label{Eq:Psi1}
\end{eqnarray}
where, $|PM_L\rangle$ indicates the orbital part of the $^2P$ term and $|\chi_{PM_L; \alpha J M_J}\rangle$ is the nuclear part 
\footnote{
Note that the nuclear part \unexpanded{$|\chi\rangle$} is not normalized, and thus the weights of LS terms in the vibronic state are not equal 
(see also Eq. (\ref{Eq:chi}).
}.
Eq. (\ref{Eq:Psi1}) expresses the entangled state of orbital and nuclear degrees of freedom. 
According to the general rule for the ground vibronic states of linear dynamical JT systems, the irreducible representations of the ground vibronic state is the same with the electronic state \cite{Ham1968, Bersuker1989}. Thus, $J = 1$ is expected for any vibronic coupling parameters, and indeed various analyses and numerical calculations support the conclusion \cite{OBrien1969, OBrien1971, Auerbach1994, Manini1994, OBrien1996, Iwahara2010}.

\subsection{$n=2,4$}
\label{Sec:C602-}
The $p^2$ ($p^4$) configurations split into one spin triplet term and two spin singlet terms, $^3P \oplus {}^1S \oplus {}^1D$, which is described by \cite{Condon1953}
\begin{eqnarray}
 \hat{H}_\text{bi} &=& -2J_\text{H} \hat{I}_{P} + 3J_\text{H} \hat{I}_{S}. 
\label{Eq:HH2}
\end{eqnarray}
The sum of the $^1D$ term energy and the zero point energy is used as the origin of energy.

The JT coupling for the triplet term (${}^3P$) with $n =$ 2 (4) is of the same form as Eq. (\ref{Eq:HJT1}) for $n =$ 5 (1). 
As in $p^1$ and $p^5$ systems, the vibronic states are specified by $\alpha, J, M_J$, Eq. (\ref{Eq:Psi1}), and the spin projection $M_S$ ($S = 1, M_S = -1, 0, 1$).
The vibronic level corresponds to that for $p^5$ ($p^1$) with the Hund's shift ($-2J_\text{H}$). 
Despite the even number of electrons, the form of the Hamiltonian indicates that the lowest vibronic states possess odd vibronic angular momenta, which looks contradictory to the selection rule on angular momentum established earlier (Eq. (36) in Ref. \cite{Auerbach1994}).
This issue will be resolved elsewhere \cite{Iwahara2018}. 

In the case of singlet states ($n=2$), the $^1D$ term linearly couples to the JT modes, and the $^1S$ and $^1D$ couples in the manner of pseudo JT effect \cite{OBrien1996, Chancey1997}:
\begin{widetext}
\begin{eqnarray}
 \hat{H}_\text{JT} &=& 
 \sum_{\mu=1}^8 \hslash \omega_\mu g_\mu
 \left(|S\rangle, |D,-2\rangle, |D,-1\rangle, |D,0\rangle, |D,+1\rangle, |D,+2\rangle\right)
\nonumber\\
 &\times&
 \begin{pmatrix}
  0 & \sqrt{2} \hat{q}_{\mu, -2} & \sqrt{2} \hat{q}_{\mu, -1} & \sqrt{2} \hat{q}_{\mu, 0} & \sqrt{2} \hat{q}_{\mu, 1} & \sqrt{2} \hat{q}_{\mu, 2} \\
  \sqrt{2} \hat{q}_{\mu, 2}      & \hat{q}_{\mu, 0} & \sqrt{\frac{3}{2}} \hat{q}_{\mu, 1} & \hat{q}_{\mu, 2} & 0 & 0 \\
 -\sqrt{2} \hat{q}_{\mu, 1}      & -\sqrt{\frac{3}{2}} \hat{q}_{\mu, -1} & -\frac{1}{2} \hat{q}_{\mu, 0} & \frac{1}{2} \hat{q}_{\mu, 1} & \sqrt{\frac{3}{2}} \hat{q}_{\mu, 2} & 0 \\
  \sqrt{2} \hat{q}_{\mu, 0}      & \hat{q}_{\mu, -2} & -\frac{1}{2} \hat{q}_{\mu, -1} & -\hat{q}_{\mu, 0} & -\frac{1}{2} \hat{q}_{\mu, 1} & \hat{q}_{\mu, 2} \\
 -\sqrt{2} \hat{q}_{\mu, -1}     & 0 & \sqrt{\frac{3}{2}} \hat{q}_{\mu, -2} & \frac{1}{2} \hat{q}_{\mu, -1} & -\frac{1}{2} \hat{q}_{\mu, 0} & -\sqrt{\frac{3}{2}} \hat{q}_{\mu, 1} \\
  \sqrt{2} \hat{q}_{\mu, -2}     & 0 & 0 & \hat{q}_{\mu, -2} & \sqrt{\frac{3}{2}} \hat{q}_{\mu, -1} & \hat{q}_{\mu, 0}
 \end{pmatrix}
 \begin{pmatrix}
 \langle S|\\
 \langle D,-2|\\
 \langle D,-1|\\
 \langle D,0|\\
 \langle D,+1|\\
 \langle D,+2|\\
 \end{pmatrix}.
\label{Eq:HJT2}
\end{eqnarray}
\end{widetext}
The Hamiltonian for $n = 4$ electrons is of the same form except for the sign change of the entire right hand side of Eq. (\ref{Eq:HJT2}). 
Since the total angular momenta (Eq. (\ref{Eq:J})) commute with the bielectronic part, the vibronic state is characterized by $\alpha, J, M_J$. 
Therefore, the vibronic states have the form:
\begin{eqnarray}
 |\Psi^\text{LS}_{\alpha J M_J}\rangle &=& |S\rangle |\chi_{S;\alpha J M_J}\rangle + \sum_{M_L}|D M_L\rangle |\chi_{D M_L;\alpha J M_J}\rangle,
\nonumber\\
\label{Eq:Psi2}
\end{eqnarray}
where, the superscript of $\Psi$ stands for low-spin state, $|S\rangle$ and $|DM_L\rangle$ are the $^1S$ and $^1D$ term states, and $|\chi_{S;\alpha J M_J}\rangle$ and $|\chi_{DM_L;\alpha J M_J}\rangle$ are the corresponding nuclear parts \cite{Note1}. 
Because of the existence of two $LS$ terms, the general rule on the ground states discussed above does not apply.
The ground state can be either $J =$ 0 or 2 as shown by the numerical simulation of $p^2 \otimes d$ JT system with single effective JT mode in Ref. \cite{OBrien1996} (see Fig. 1 in the reference).

\subsection{$n=3$}
\label{Sec:C603-}
The $p^3$ configurations split into one spin quartet term and two doublet terms: ${}^{4}S\oplus {}^{2}P\oplus {}^{2}D$ \cite{Condon1953}.
Thus, the bielectronic interaction is 
\begin{eqnarray}
 \hat{H}_\text{bi} &=& -3J_\text{H} \hat{I}_S + 2J_\text{H} \hat{I}_P.
\label{Eq:HH3}
\end{eqnarray}
The sum of the $^2D$ term energy and the zero point energy is used as the origin of energy. 

Since the quartet term is orbitally non-degenerate, it does not couple to the JT active $h_g$ modes.
The eigenstates are specified by the set of vibrational quantum numbers $\bm{n}$ and the spin quantum numbers, and the corresponding energy levels are
the sum of the term energy ($-3J_\text{H}$) and the vibrational energy.

The spin doublet terms couple to the vibrational modes in the manner of pseudo JT coupling \cite{OBrien1996, Chancey1997}:
\begin{widetext}
\begin{eqnarray}
\hat{H}_\text{JT} &=& \sum_{\mu = 1}^8 \hslash\omega_\mu g_\mu 
 \left(
  |P,-1\rangle, |P,0\rangle, |P,+1\rangle
 \right)
\nonumber\\
 &\times&
 \begin{pmatrix}
  -\sqrt{\frac{3}{2}} \hat{q}_{\mu, -1} & -\frac{3}{2} \hat{q}_{\mu, 0} & -\frac{3}{2} \hat{q}_{\mu, 1} & -\sqrt{\frac{3}{2}} \hat{q}_{\mu, 2} & 0 \\
  \sqrt{3} \hat{q}_{\mu, -2} & \frac{\sqrt{3}}{2} \hat{q}_{\mu, -1} & 0 & -\frac{\sqrt{3}}{2} \hat{q}_{\mu, 1} & -\sqrt{3} \hat{q}_{\mu, 2} \\
  0 & \sqrt{\frac{3}{2}} \hat{q}_{\mu, -2} & \frac{3}{2} \hat{q}_{\mu, -1} & \frac{3}{2} \hat{q}_{\mu, 0} & \sqrt{\frac{3}{2}} \hat{q}_{\mu, 1}\\
 \end{pmatrix}
 \begin{pmatrix}
  \langle D,-2|\\
  \langle D,-1|\\
  \langle D, 0|\\
  \langle D,+1|\\
  \langle D,+2|\\
 \end{pmatrix}
 + \text{H.c.}
\label{Eq:HJT3}
\end{eqnarray}
\end{widetext}
Despite the fact that both spin doublet terms are orbitally degenerate, the vibronic coupling within these terms does not exist, which is explained by the seniority selection rule for the matrix elements of half-filled system \cite{Racah1942, Racah1943} (see also Appendix \ref{A:HJT}).
The simultaneously commuting operators with the Hamiltonian are the total angular momentum $\hat{\bm{J}}^2$, one of the components, for example, $\hat{J}_0$ 
(\ref{Eq:J}) and the ``inversion operator'' \cite{Iwahara2013},
\begin{eqnarray}
\hat{P} = (\hat{I}_P - \hat{I}_D)\exp(i\pi \hat{N}),
\label{Eq:P}
\end{eqnarray}
where $\hat{N} = \sum_{\mu m} \hat{n}_{\mu m}$.
The eigenvalues of $\hat{P}$ are $\pm 1$, and the parity is inherited from the seniority of the electronic terms.
The vibronic state is characterized by the quantum numbers of angular momentum $J$, its $z$ component $M_J$, parity $P$, spin quantum numbers $S = 1/2$ and $M_S$, and other quantum number $\alpha$ 
The vibronic states are represented as
\begin{eqnarray}
 |\Psi^\text{LS}_{\alpha J M_J P}\rangle 
&=& \sum_{M_L} |PM_L\rangle |\chi_{P M_L;\alpha J M_J P}\rangle 
\nonumber\\
&+& \sum_{M_L}|D M_L\rangle |\chi_{D M_L;\alpha J M_J P}\rangle,
\label{Eq:Psi_vibro}
\end{eqnarray}
where $|PM_L\rangle$ and $|DM_L\rangle$ are the $^2P$ and $^2D$ term states, and $|\chi_{PM_L;\alpha J M_JP}\rangle$ and $|\chi_{DM_L;\alpha J M_JP}\rangle$ are the corresponding nuclear parts \cite{Note1}. 
As in the case of C$_{60}^{2/4-}$, the irreducible representation of the ground state can be $J =$ 1 or 2 depending on the balance of the strengths of the vibronic coupling and bielectronic interactions (see Fig. 2 in Ref. \cite{OBrien1996} for the numerical simulation of $p^3 \otimes d$ JT model).

\section{Computational method}
The vibronic coupling parameters and the Hund's rule coupling parameter obtained by DFT calculations with hybrid (B3LYP) functional were used: $g_\mu$'s were taken from 
Ref. \cite{Iwahara2010} (Table I. (6)) and $J_\text{H} =$ 44 meV \cite{Iwahara2013}.
The frequencies $\omega_\mu$ were taken from the Raman scattering data of pristine C$_{60}$ crystal \cite{Bethune1991}.
The validity of $g_\mu$ and $J_\text{H}$ is discussed in Sec. \ref{Sec:coupling}.

The theoretical description of single mode $p^n \otimes d$ JT model has been developed within the weak or the strong limit of the vibronic couplings (e.g., Refs. \cite{Auerbach1994, Manini1994, OBrien1996, Chancey1997, Sookhun2003, Dunn2005}).
However, the static JT stabilization energies of C$_{60}^{n-}$ are comparable to the vibrational frequencies of JT active modes and far from these limits, requiring accurate numerical treatment to access the actual situations of low-energy states of C$_{60}^{n-}$.
The vibronic states of the single mode model have also been numerically investigated (e.g., Refs. \cite{Auerbach1994, OBrien1996}), whereas multimode effect is essential for the correct distribution of the low-energy vibronic levels \cite{Iwahara2013} (For further discussion, see Appendix \ref{A:eff}).
Thus, to derive precise low-energy vibronic states of C$_{60}^{n-}$ anions, numerical diagonalization of the $p^n \otimes 8d$ JT Hamiltonian with multiplet splitting is carried out.

The JT Hamiltonian matrix is calculated using the product of the electronic term and the eigenstates of harmonic oscillator $\hat{H}_0$ as the basis:
\begin{eqnarray}
 \left\{|LM_L\rangle |\bm{n}\rangle \left| 0 \le \sum_{\mu = 1}^8 \sum_{m = -2}^2 n_{\mu m} \le 7 \right. \right\}.
\label{Eq:basis}
\end{eqnarray}
We stress that all eight $h_g$ modes (40 vibrational coordinates) are included in the basis (\ref{Eq:basis}).
With this basis, the nuclear part of the vibronic state is expanded as 
\begin{eqnarray}
 |\chi_{LM_L; \alpha J M_J (P)}\rangle &=& \sum_{\bm{n}} |\bm{n}\rangle \chi_{L M_L \bm{n}; \alpha J M_J (P)},
\label{Eq:chi}
\end{eqnarray}
where $\chi_{L M_L \bm{n}; \alpha J M_J (P)} = \langle \bm{n}|\chi_{L M_L; \alpha J M_J (P)}\rangle$. 
The Hamiltonian matrix was numerically diagonalized using Lanczos algorithm. 
The Lanczos iteration was continued until the changes in energy of all the target states become less than 10$^{-4}$ in units of the lowest $h_g$ frequency.

\section{Results and Discussions}

\begin{table}[tb]
\begin{ruledtabular}
\caption{
Contributions to the ground vibronic energy (Total) of C$_{60}^{n-}$ (meV).
$\langle \hat{H}_\text{bi}\rangle$, Static, and Dynamic indicate the 
bielectronic energy, static JT, and dynamical JT stabilization energies, respectively,
as contributions to the total energy.
}
\label{Table:E}
\begin{tabular}{ccccc}
$n$ & Total & $\langle\hat{H}_\text{bi}\rangle$ & Static & Dynamic  \\
\hline
1,5 &  $-96.5$ &    -  & $-50.3$  & $-46.2$  \\  
2,4 & $-244.4$ & 39.3  & $-195.6$ & $-88.1$  \\
3   & $-196.2$ & 41.0  & $-147.6$ & $-89.6$  \\
\end{tabular}
\end{ruledtabular}
\end{table}

\begin{figure}
\begin{center}
\includegraphics[width=8cm]{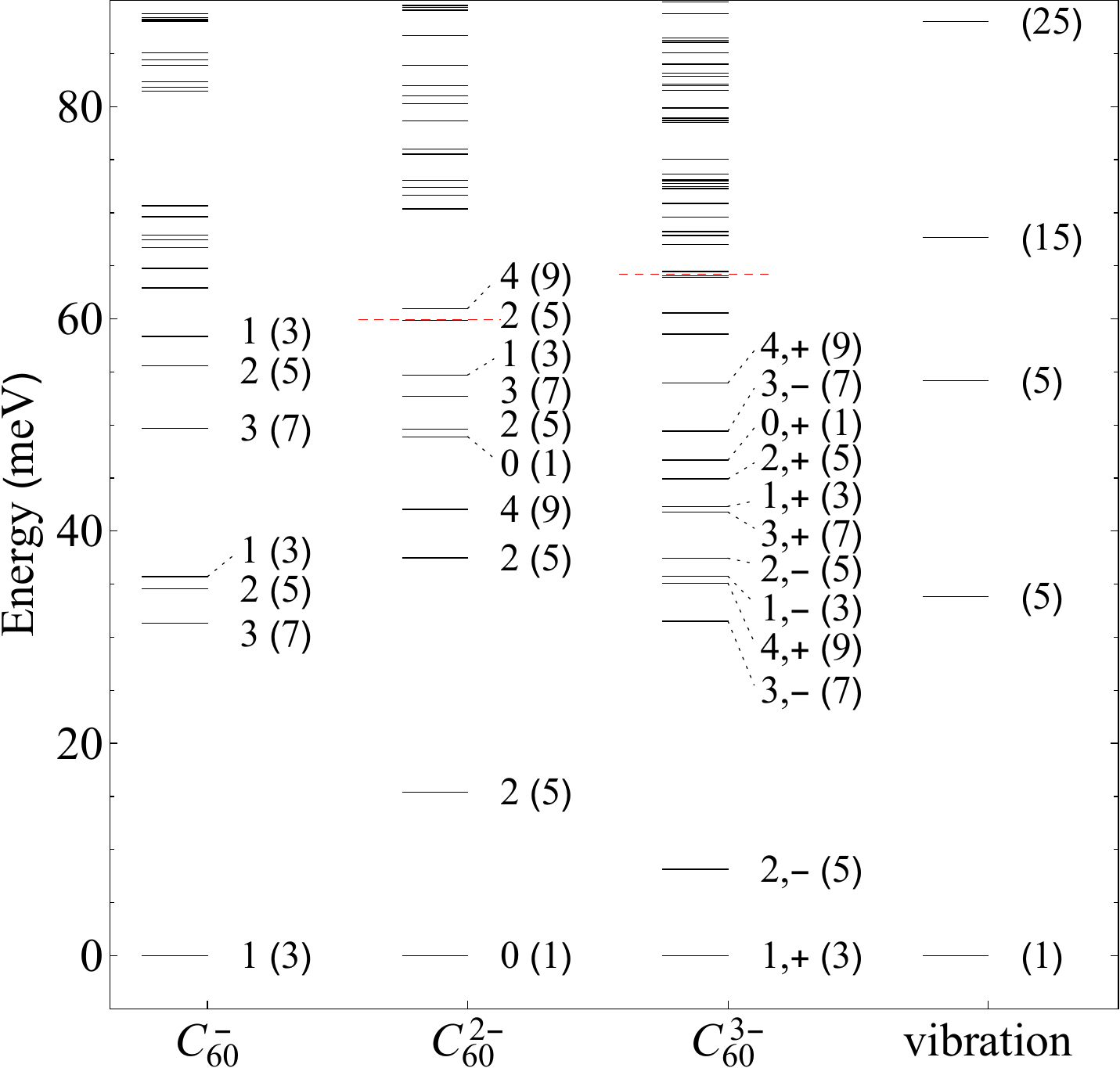}
\end{center}
\caption{
(Color online)
Low-lying vibronic levels with respect to the ground vibronic level of each C$_{60}^{n-}$ anion ($n=1,2,3$) and the zero-vibrational level of neutral C$_{60}$ (meV).
The numbers next to the energy levels are $J$ for C$_{60}^-$ and C$_{60}^{2-}$ and $(J,P)$ for C$_{60}^{3-}$ and the numbers in the parenthesis are the degeneracy.
The horizontal red dashed lines indicate the ground high-spin levels. 
}
\label{Fig:E}
\end{figure}

\subsection{Low-energy vibronic states}
\label{Sec:vibronic}
The obtained low-energy vibronic states of C$_{60}^{n-}$ ($n=$ 1-5) are presented below.
The ground-state energy is decomposed into the bielectronic, static JT, and dynamic JT contributions.
The first one is defined by the expectation value of $\hat{H}_\text{bi}$ in the ground vibronic state, $\langle \hat{H}_\text{bi} \rangle$.
The static JT energy is calculated by subtracting the bielectronic energy in the adiabatic state from the energy at the minima of the adiabatic potential energy surface (APES). The remaining part is the dynamical JT contribution
\footnote{
Note that due to bielectronic interaction the static JT energy for $n = 2, 4$ and 3 is slightly smaller than the expected respective values 
$4E_\text{JT}^{(1)}$ and $3E_\text{JT}^{(1)}$, where $E_\text{JT}^{(1)}$ is the static JT energy for $n = 1$ (Table \ref{Table:E}).
}. 

The vibronic states are further analyzed in terms of the weight of the vibronic basis with $N$ vibrational excitations, 
\begin{eqnarray}
 w_N(\alpha J M_J (P)) &=& \sum_{LM_L} \sideset{}{'}\sum_{\bm{n}} \left|\chi_{LM_L \bm{n}; \alpha J M_J (P)}\right|^2,
\label{Eq:wN}
\end{eqnarray}
where the sum over $\bm{n}$ is taken under the constraint $\sum_{\mu m} n_{\mu m} = N$.

\subsubsection{$n = 1, 5$}
\label{Sec:vibronic1}
The ground vibronic state is characterized by $J = 1$ and the energy level is $-96.5$ meV. 
The contributions from the static ($E^{(1)}_\text{JT} = -\sum_\mu \hslash \omega_\mu g_\mu^2/2$) and the dynamic JT effect to the ground energy is almost the same
(see Table \ref{Table:E}).
The static JT energy is of the order of vibrational frequencies of the JT active modes, thus the vibronic coupling is classified as intermediate.
This particularly implies that the coupling is not weak enough to allow the description of the total stabilization within second order perturbation theory:
the ground state energy within the perturbation theory, $\frac{5}{2} E^{(1)}_\text{JT}$ \cite{Manini1994}, is larger by a half of $E_\text{JT}^{(1)}$ than the present one. 
The deviation is also seen in the contributions to the vibronic state. 
Within the second order of perturbation theory, the ratio of the weights (\ref{Eq:wN}) for $N = 0$ and 1, $w_0/w_1$, is about 1.
On the other hand, the weights (\ref{Eq:wN}) for the vibronic bases with $N=0,1,2,3,4$ vibrational excitations are 0.524, 0.364, 0.094, 0.016, 0.002, respectively.
The weight for $N=1$, $w_1$, is reduced and those for $N \ge 2$ become finite in the numerical ground state. 

The low-energy vibronic levels are shown in Fig. \ref{Fig:E} (see also Table S1 \cite{SM}). 
The low-lying excited levels characterized by $J = 3, 2, 1$ appear at around 30 meV. 
The energy gap between the ground and the $T_{2u}$ level has been estimated to be about 30 meV from the energy difference between the zero-phonon and side bands of near infrared absorption spectra \cite{Tomita2005}
\footnote{
The vibronic level with $J=3$ splits into $T_{2u}$ and $G_u$ levels \cite{Altmann1994} due to weak higher order vibronic coupling. 
Although the side band is attributed to the ground $T_{1u}$ to the $T_{2u}$ excitations, 
all the quasi degenerate levels $(J = 3,2,1)$ including the $T_{2u}$ vibronic level are populated and contribute to the side band. 
}.
The experimental and the present excitation energies agree well with each other.

\subsubsection{$n = 2, 4$}
\label{Sec:vibronic2}
Although the spin triplet term is lower than the singlet terms, Eq. (\ref{Eq:HH2}), the order is inverted by the vibronic coupling.
The ground vibronic state is spin-singlet characterized by $J = 0$ and the corresponding energy level is $-244.4$ meV. 
The energy contains contributions from the bielectronic coupling, static and dynamic JT effects (Table \ref{Table:E}). 
The bielectronic energy $\langle \hat{H}_\text{bi} \rangle$ amounts to 30 \% of the energy gap between $^1S$ and $^1D$ terms because of their mixing by the pseudo JT coupling.
To derive the static JT contribution, the potential energy in $\hat{H}$, Eq. (\ref{Eq:H}), is minimized with respect to all the $q_{\mu \theta}$ ($ = q_{\mu, 0}$) coordinates (the other JT active coordinates are kept to zero). 
The lowest energy of the APES is $-163.1$ meV at $q_{\mu \theta} = 0.986 \times 2g_\mu$ (for $n = 2$) and the expectation value of the bielectronic interaction in the ground adiabatic electronic state is 32.5 meV
\footnote{
The bielectronic energy for the ground adiabatic state is smaller than for vibronic ground state (Table \ref{Table:E}) because the JT dynamics contribute to a stronger mixing of the electronic terms of a given spin multiplicity. 
}.
Subtracting the latter from the minima of the APES, we obtain the static JT contribution of $-195.6$ meV.
The effect of the bielectronic energy on the APES is small, and the magnitude of the JT distortion and the static JT energy are close to the case of the absence of the bielectronic interaction ($q_\theta = 2g$, $E_\text{JT} = 4E_\text{JT}^{(1)}$) \cite{Auerbach1994, OBrien1996}.
The remaining part of the ground energy corresponds to the dynamical contribution, which is about 40 \% of the static one. 
The vibronic coupling becomes ca two times larger in C$_{60}^{2-}$ than in C$_{60}^-$, and thus, many vibronic basis functions (\ref{Eq:basis}) with higher vibronic excitations contribute to the ground vibronic states. 
The weights of the vibronic basis with $N=$ 0-4 vibrational excitations to the ground state (\ref{Eq:wN}) are 0.150, 0.368, 0.255, 0.145, 0.058, 
respectively. 

Figure \ref{Fig:E} shows the low-energy vibronic levels of C$_{60}^{2-}$ (see also Table S2 \cite{SM}). 
The first exited vibronic level ($J=2$) appears at about 15 meV above the ground one (Fig. \ref{Fig:E}). 
The lowest excited level is estimated 22 meV from near infrared absorption spectra of C$_{60}^{2-}$ \cite{Tomita2006}, which agrees well with the present energy gap.
The gap is about a half of the first vibrational excitation due to the moderate JT effect of C$_{60}^{2-}$.
The red dotted line in Fig. \ref{Fig:E} indicates the lowest vibronic level with $S = 1$.
The breakdown of the Hund's rule also occurs within static JT effect: 
The lowest high-spin and low-spin energy levels are $-138.2$ meV and $-163.1$ meV, respectively.
This breakdown is caused by the weak $J_\text{H}$ due to the delocalization of the molecular orbitals over the relatively large C$_{60}$ cage and the enhanced vibronic coupling in C$_{60}^{2-}$.
The situations here is similar to that of alkali metal clusters (static JT system) \cite{Rao1985} and impurity in semiconductor (dynamical JT system) \cite{Ham1993}, where the violation of the Hund rule via JT effect is observed as well.

\subsubsection{$n = 3$}
\label{Sec:vibronic3}
As in the previous case 
with $n = 2,4$, the low-spin states ($S=1/2$) are more stabilized than the high-spin ones ($S=3/2$) by JT effect in C$_{60}^{3-}$. 
The ground state of C$_{60}^{3-}$ is characterized by $J = 1$ and $P = +1$ and the energy level is $-196.2$ meV \cite{Iwahara2013}. 
The expectation value of the bielectronic energy in the ground vibronic state is 41.0 meV, which is about a half of the splitting of $^2P$ and $^2D$ terms.
The static JT contribution is calculated from the minima of the APES. 
Minimizing the sum of the Hund and the potential terms with respect to $q_{\mu \epsilon}$ [$= (q_{\mu, -2} + q_{\mu, +2})/\sqrt{2}$] distortions, 
the minima of the potential was obtained $-110.1$ meV at $q_{\mu \epsilon} = 0.989 \times \sqrt{3}g_\mu$.
Subtracting the expectation value of  coupling in the ground adiabatic state (37.6 meV \cite{Note4}) from the energy of the potential minima, we obtain the static JT contribution of $-147.6$ meV. 
Both the static JT energy and the JT distortion are close to those without the bielectronic interaction ($q_\epsilon = \sqrt{3}g$, $E_\text{JT} = 3E_\text{JT}^{(1)}$) \cite{Auerbach1994, OBrien1996}.
The stabilization by the JT dynamics is $-89.6$ meV, which amounts to as much as about 60 \% of the static JT energy. 
Compared to C$_{60}^{-}$, the vibronic coupling is enhanced by a factor of $\sqrt{3}$ times larger in C$_{60}^{3-}$, the vibronic state becomes more involved.
Thus, the weights of the vibronic basis with $N = $ 0-4 vibrational excitations (\ref{Eq:wN}) in the ground state are 0.204, 0.390, 0.221, 0.132, 0.038, respectively. 

The low-lying vibronic levels are shown in Fig. \ref{Fig:E} (see also Table S3 \cite{SM}). 
The lowest excitation lies at only about 8 meV, and the higher excited states appear above 35 meV. 
The distribution of the vibronic energy levels is significantly different from that of vibrational levels.
Contrary to C$_{60}^{2-}$, the Hund's rule still holds for static JT stabilization. 
Compared to the energy of $^4S$ term (the red dotted line in Fig. \ref{Fig:E}), the minimum energy of APES is higher by 20 meV.
Therefore, the Hund's rule is violated in C$_{60}^{3-}$ due to the existence of the JT dynamics, pretty similar to the case of double acceptor in semiconductor \cite{Ham1993}.
The different behaviour of C$_{60}^{2-}$ is explained by a stronger vibronic coupling.

\subsection{Matrix elements of electronic operators and the vibronic reduction factors}
\label{Sec:reduction}

\begin{table}
\begin{ruledtabular}
\caption{
Matrix elements of irreducible tensor electronic operator, $\langle \hat{O}_l \rangle = \langle \Psi_{JM_J^0(P)}| \hat{O}_{l0}^{LL'} | \Psi_{J'M_J^0(P')} \rangle$.
In the case with the matrix element corresponds to the vibronic reduction factor $K$, it is marked by \checkmark in the last column. 
}
\label{Table:O}
\begin{tabular}{cccccccccc}
$n$  & $L$ & $L'$ & $M_L^0$ & $l$ & $J$ $(P)$ & $J'$ $(P')$ & $M_J^0$ & $\langle \hat{O}_l \rangle$ & $K$\\
\hline
1    & 1 & 1 & 1 & 1 & 1 & 1 & 1 & 0.353 & \checkmark \\
     &   &   &   & 2 & 1 & 1 & 1 & 0.602 & \checkmark \\
2    & 0 & 0 & 0 & 0 & 0 & 0 & 0 & 0.298 & \checkmark \\ 
     &   &   &   &   & 2 & 2 & 2 & 0.257 \\
     & 0 & 2 & 0 & 2 & 0 & 2 & 0 & 0.251 & \\ 
     &   &   &   &   & 2 & 2 & 2 & 0.123 \\ 
     & 2 & 2 & 2 & 0 & 0 & 0 & 0 & 0.702 \\
     &   &   &   &   & 2 & 2 & 2 & 0.743 & \\ 
     &   &   &   & 1 & 2 & 2 & 2 & 0.170 & \\ 
     &   &   &   & 2 & 0 & 2 & 0 & 0.322 \\
     &   &   &   &   & 2 & 2 & 2 & 0.246 & \\ 
     &   &   &   & 3 & 2 & 2 & 2 & 0.075 & \\ 
     &   &   &   & 4 & 2 & 2 & 2 & 0.265 & \\ 
3    & 1 & 1 & 1 & 0 & 1, $+$ & 1, $+$ & 1 & 0.465 & \checkmark \\
     &   &   &   &   & 2, $-$ & 2, $-$ & 2 & 0.438 \\
     &   &   &   & 1 & 1, $+$ & 1, $+$ & 1 & 0.228 & \checkmark \\
     &   &   &   &   & 2, $-$ & 2, $-$ & 2 & 0.097 \\
     &   &   &   & 2 & 1, $+$ & 1, $+$ & 1 & 0.319 & \checkmark \\ 
     &   &   &   &   & 2, $-$ & 2, $-$ & 2 & $-0.367$\\
     & 1 & 2 & 1 & 1 & 1, $+$ & 2, $-$ & 1 & 0.215 & \\ 
     &   &   &   &   & 2, $-$ & 1, $+$ & 1 & 0.129 \\
     &   &   &   & 2 & 1, $+$ & 2, $-$ & 1 & $-0.347$ & \\ 
     &   &   &   &   & 2, $-$ & 1, $+$ & 1 & $-0.283$ \\
     &   &   &   & 3 & 1, $+$ & 2, $-$ & 1 & 0.254 & \\ 
     &   &   &   &   & 2, $-$ & 1, $+$ & 1 & $-0.108$ \\
     & 2 & 2 & 2 & 0 & 1, $+$ & 1, $+$ & 1 & 0.535 \\
     &   &   &   &   & 2, $-$ & 2, $-$ & 2 & 0.562 & \\ 
     &   &   &   & 1 & 1, $+$ & 1, $+$ & 1 & 0.098 \\
     &   &   &   &   & 2, $-$ & 2, $-$ & 2 & 0.177 & \\ 
     &   &   &   & 2 & 1, $+$ & 1, $+$ & 1 & $-0.163$ \\
     &   &   &   &   & 2, $-$ & 2, $-$ & 2 & 0.300 & \\ 
     &   &   &   & 3 & 2, $-$ & 2, $-$ & 2 & 0.192 & \\ 
     &   &   &   & 4 & 2, $-$ & 2, $-$ & 2 & 0.267 & \\ 
\end{tabular}
\end{ruledtabular}
\end{table}

Any electronic operator acting on the orbital part of $LS$ terms can be expressed by the linear combinations of irreducible tensor operators \cite{Abragam1970, Varshalovich1988}:
\begin{eqnarray}
\hat{O} &=& \sum_{LL'} \sum_{lm} a_{lm}^{LL'} \hat{O}_{lm}^{LL'}, 
\\
\hat{O}_{lm}^{LL'} &=& \sum_{M_LM_L'} 
 \frac{\langle LM_L| L' M_L', lm \rangle}{\langle LM_L^0| L' M_L^0, l0 \rangle}
 |LM_L\rangle \langle L'M_L'|,
\label{Eq:O}
\end{eqnarray}
where, 
$l$ is the rank ($|L-L'| \le l \le L + L'$), $m$ $(m = -l, -l+1, ..., l)$ is the component, $a^{LL'}_{lm}$ is a coefficient for the expansion, 
$M_L^0 = \min(L,L')$ and $\langle jm| j_1 m_1, j_2 m_2\rangle$ is a Clebsch-Gordan coefficient with Condon-Shortley phase convention \cite{Condon1953, Varshalovich1988}.
Therefore, it is sufficient to consider the matrix elements of irreducible tensor operators (\ref{Eq:O}). 
Moreover, 
it is sufficient to calculate several matrix elements of $\hat{O}_{lm}^{LL'}$ with specific $m$, e.g., $m = 0$, whereas the rest of them can be calculated by using the relation:
\begin{eqnarray}
\frac{
 \langle \Psi_{JM_J(P)}| \hat{O}_{lm}^{LL'} | \Psi_{J'M_J'(P')} \rangle 
}{\langle \Psi_{JM_J^0(P)}| \hat{O}_{l0}^{LL'} | \Psi_{J'M_J^0(P')} \rangle}
 &=& 
 \frac{\langle JM_J|J'M_J', lm\rangle}{\langle JM_J^0|J'M_J^0, l0\rangle},
\end{eqnarray}
where, $M_J^0 = \min(J,J')$, and the denominators of both sides are non-zero.

For the ground vibronic term, the irreducible representations of the vibronic states $\Psi$ ($\Psi'$) are coinciding with the irreducible representations of the electronic states $\psi$ ($\psi'$), 
and the ratio of the corresponding matrix elements is called vibronic reduction factor \cite{Child1961, Ham1968, Bersuker1989}:
\begin{eqnarray}
 K &=& \frac{\langle \Psi| \hat{O} |\Psi' \rangle}{\langle \psi| \hat{O} |\psi' \rangle},
\label{Eq:K}
\end{eqnarray}
where, $\langle \psi| \hat{O} |\psi' \rangle \ne 0$ is assumed. 
The denominator of Eq. (\ref{Eq:O}) is introduced for the normalization of the tensor operator, $\langle LM_L^0|\hat{O}^{LL'}_{l0}|L'M_L^0\rangle = 1$. 
Therefore, when $J = L, J' = L', M_J^0 = M_L^0 = M_0$, $\langle \Psi_{JM_0(P)}| \hat{O}_{l0}^{LL'} | \Psi_{J'M_0(P')} \rangle$ reduces to vibronic reduction factor $K$ (\ref{Eq:K}).
In the calculations below, the phase factors of the vibronic states are fixed so that the coefficient for the vibronic basis with no vibrational excitation become positive.

\subsubsection{$n=1,5$}
Any electronic operators acting on the $^2P$ term can be expressed by the tensor operators (\ref{Eq:O}) of ranks $l = 0, 1, 2$.
The tensor operator of rank 0 is simply the identity operator, and the non-zero matrix element is 1.
The matrix elements of the other tensor operators are listed in Table \ref{Table:O}.
The matrix elements for the operators of rank 1 and 2 are 0.353 and 0.602, respectively; they correspond to the vibronic reduction factors. 
The reduction factor for the first rank operator was recently calculated to be ca 0.3 \cite{Ponzellini} with the sets of vibronic coupling parameters extracted from the photoelectron spectra \cite{Iwahara2010}.
Since these vibronic coupling parameters are slightly larger than the DFT values used in this work
\footnote{The vibronic coupling parameters derived from the photoelectron spectra \cite{Iwahara2010} could be slightly overestimated because the dependence of intensities on the absorbed photon energy ($\hslash \omega_\text{ph}$) was neglected since $\omega_\mu/\omega_\text{ph} \ll 1$.
Within the second order perturbation theory, the intensity is proportional to the product of $g_\mu^2$ and $\omega_\text{ph}$. 
Using this relation, the vibronic coupling parameters for high frequency modes are estimated to be reduced by about 3-4 \%.
},
the reduction factor is slightly smaller than the present one.

\subsubsection{$n=2,4$}
Because the low-spin electronic terms are $^1S$ and $^1D$, the irreducible tensor operators (\ref{Eq:O}) of ranks $l =$ 0-4 are considered. 
The matrix elements of the tensor operators for the two lowest vibronic states of C$_{60}^{2-}$ are calculated in Table \ref{Table:O}.
Using the selection rule on the angular momenta, only the non-zero matrix elements are shown.

\subsubsection{$n=3$}
The electronic operators acting on the low-spin electronic terms ($^2P$ and $^2D$) are expressed by the irreducible tensor operators (\ref{Eq:O}) of ranks $l = $ 0-4.
The matrix elements were calculated within the ground and the first excited vibronic states (Table \ref{Table:O}).
The matrix elements which become zero due to the selection rule are not shown. 

Since the parity (\ref{Eq:P}) characterizes the vibronic states in C$_{60}^{3-}$, there is a selection rule related to $P$:
Suppose the parity of the $LS$ term with orbital angular momentum $L$ ($L'$) is $\tilde{P}$ ($\tilde{P}'$), 
$\langle \Psi_{JM_JP}^\text{LS}|\hat{O}_{lm}^{LL'}|\Psi_{J'M_J'P'}^\text{LS}\rangle \ne 0$, then $PP' = \tilde{P} \tilde{P}'$. 
This is proved by using $\hat{P}\hat{O}^{LL'}_{lm}\hat{P} = \tilde{P}\tilde{P}'\hat{O}^{LL'}_{lm}$, where the latter can be checked by substituting Eq. (\ref{Eq:O}) into both sides. 
Calculating the matrix elements of both sides between $|\Psi_{JM_JP}^\text{LS}\rangle$ and $|\Psi_{J'M_J'P'}^\text{LS}\rangle$,
and then simplifying the expression, we obtain
\begin{eqnarray}
\left(PP'-\tilde{P}\tilde{P}'\right)\langle \Psi_{JM_JP}^\text{LS}|\hat{O}_{lm}^{LL'}|\Psi_{J'M_J'P'}^\text{LS}\rangle = 0.
\label{Eq:selection_P}
\end{eqnarray}
Thus, the matrix element $\langle \Psi_{JM_JP}^\text{LS}|\hat{O}_{lm}^{LL'}|\Psi_{J'M_J'P'}^\text{LS}\rangle$ is only non-zero when $PP' = \tilde{P}\tilde{P}'$.

\subsection{Thermodynamic properties}
\label{Sec:thremo}

\begin{figure}
\begin{center}
\begin{tabular}{ll}
(a) & (b) \\
\includegraphics[width=4.2cm]{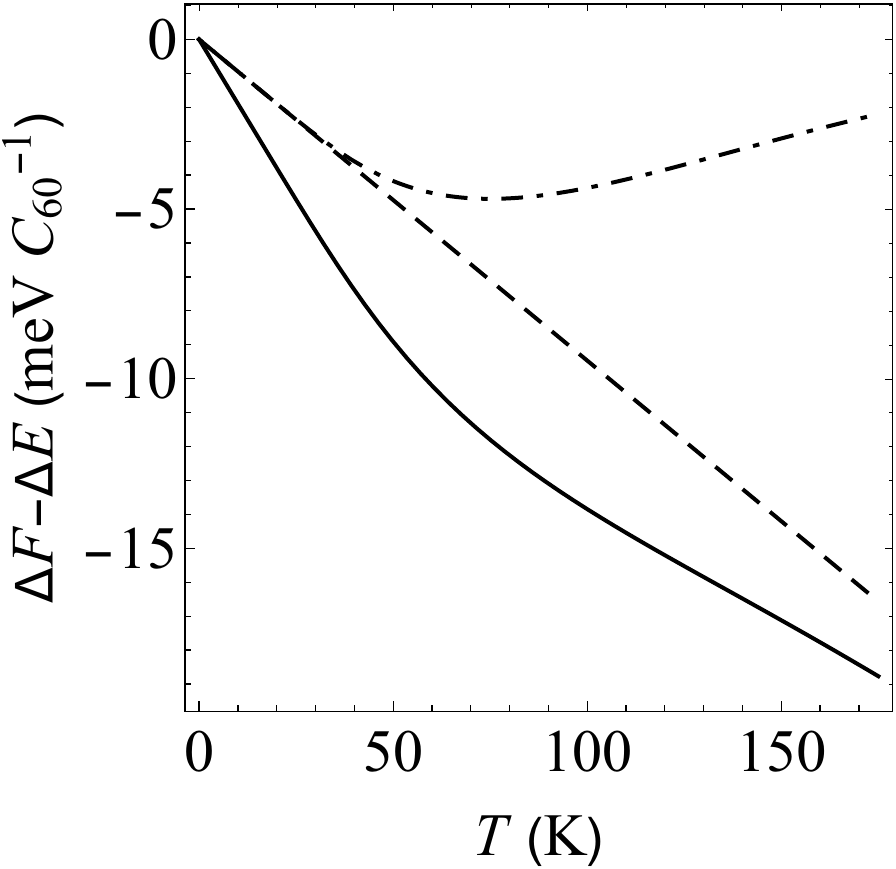}
&
\includegraphics[width=4.2cm]{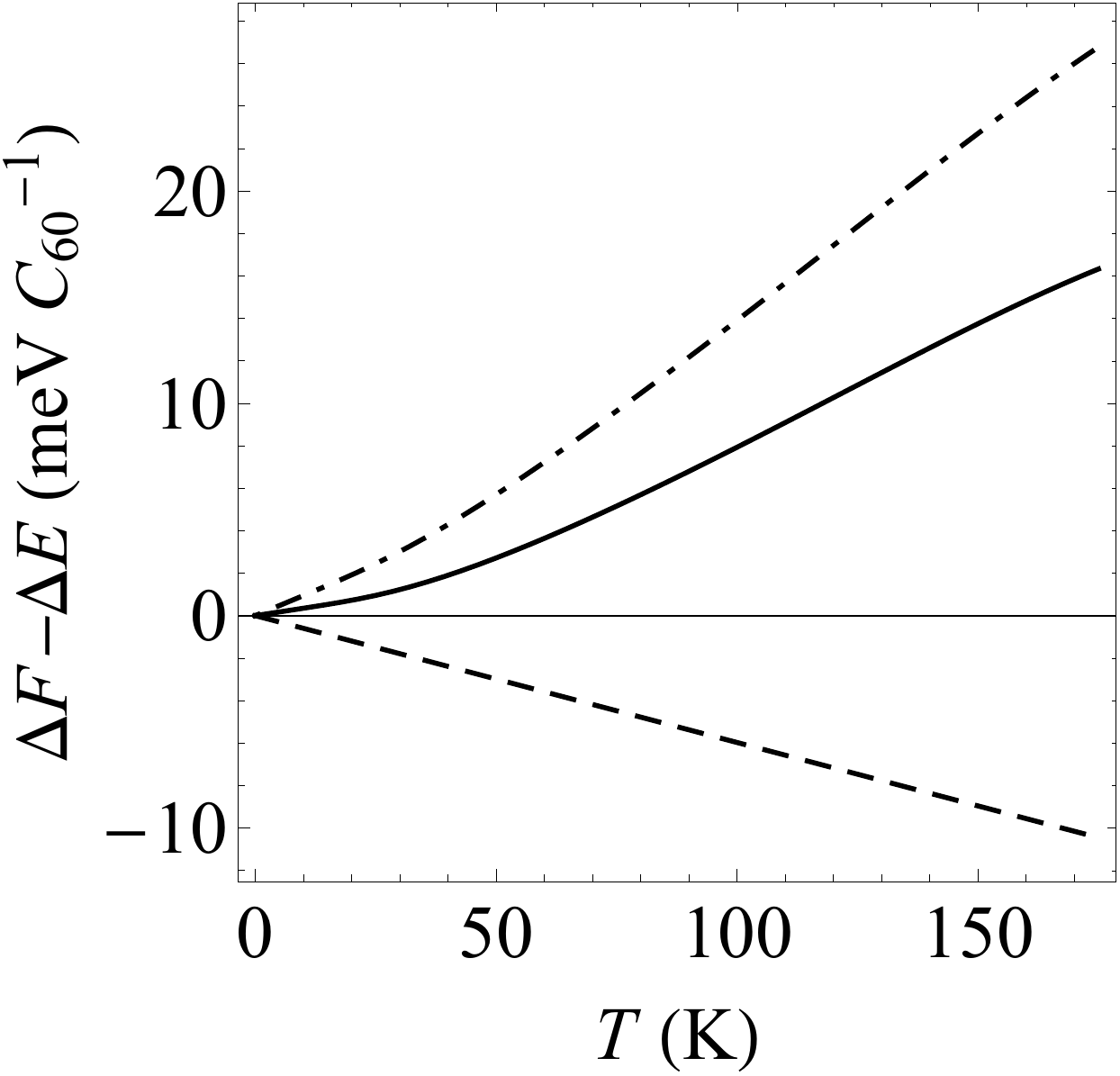}
\end{tabular}
\end{center}
\caption{
The spin gaps $\Delta F - \Delta E$ of (a) C$_{60}^{2-}$ and (b) C$_{60}^{3-}$ (meV).
The solid, dashed, and dot-dashed lines indicate $\Delta F - \Delta E$, and the contributions from spin multiplicity and vibronic states, respectively. 
}
\label{Fig:SF}
\end{figure}

\begin{table*}[tb]
\caption{
Experimental and theoretical spin gaps of C$_{60}^{n-}$ (in meV). 
}
\label{Table:spingap}
\begin{ruledtabular}
\begin{tabular}{lllc}
                       & $\Delta F$ & Method & Ref. \\
\hline
\multicolumn{4}{c}{$n=2,4$} \\
Isolated C$_{60}^{2-}$ and C$_{60}^{4-}$ & 59.9-41.1 & Theory (dynamic JT) at $T =$ 0 and 175 K & present \\
C$_{60}^{2-}$ in DMSO  & 74 $\pm$ 12       & EPR signal under the melting point of DMSO  & \cite{Trulove1995} \\ 
Na$_2$C$_{60}$         & 140 $\pm$ 20      & $^{13}$C NMR $1/T_1$                & \cite{Brouet2001} \\ 
                       & 125               & Data of Ref. \cite{Brouet2001} with different fitting function & \cite{Brouet2002c} \\ 
                       & 100               & $^{23}$Na NMR line shift            & \cite{Brouet2002a}\\ 
K$_4$C$_{60}$          & 50                & $^{13}$C NMR $1/T_1$                & \cite{Zimmer1994} \\ 
                       & 70                & $^{13}$C NMR $1/T_1$                & \cite{Brouet2002a} \\ 
Rb$_4$C$_{60}$         & 57, 51            & $^{13}$C NMR line shift and $1/T_1$ & \cite{Zimmer1995} \\ 
                       & 65                & $^{13}$C NMR $1/T_1$                & \cite{Kerkoud1996} \\
                       & 90                & Data of Ref. \cite{Kerkoud1996} with different fitting function & \cite{Brouet2002c} \\
                       & 52 $\pm$ 4        & SQUID                               & \cite{Lukyanchuk1995} \\ 
(NH$_3$)$_2$NaK$_2$C$_{60}$ & 65 $\pm$ 3, 76 $\pm$ 3\footnotemark[1] & $^{13}$C NMR $1/T_1$      & \cite{Ricco2003} \\ 
(Cp$^*_2$Co$^+$)$_2$C$_{60}^{2-}$(C$_6$H$_4$Cl$_2$, C$_6$H$_5$CN)$_2$ 
                       & 91 $\pm$ 1        & EPR signal                          & \cite{Konarev2003} \\ 
(TMP$^+$)$_2\cdot$(C$_{60}^{2-}$)$\cdot$(C$_6$H$_4$Cl$_2$)$_2$ 
                       & 63 $\pm$ 1        & EPR signal                          & \cite{Konarev2013} \\ 
\multicolumn{4}{l}{\{DB-18-crown-6$\cdot$[Na$^+$]$\cdot$(C$_6$H$_5$CN)$_2$\}$_2\cdot$(C$_{60}^{2-}$)$\cdot$C$_6$H$_5$CN$\cdot$C$_6$H$_4$Cl$_2$} \\
                       & 60 $\pm$ 1        & EPR signal                          & \cite{Konarev2013} \\ 
\{Cryptand[2,2,2](Na$^+$)\}$_2\cdot$(C$_{60}^{2-}$) 
                       & 67 $\pm$ 2        & EPR signal                          & \cite{Konarev2013} \\ 
(PPN$^+$)$_2\cdot$(C$_{60}^{2-}$)$\cdot$(C$_6$H$_4$Cl$_2$)$_2$
                       & 66                & EPR signal                          & \cite{Konarev2013} \\ 
(Me$_4$N$^+$)$_2$(C$_{60}^{2-}$)$\cdot$(TPC)$_2\cdot$2C$_6$H$_4$Cl$_2$ &    
                         58 $\pm$ 1        & EPR signal                          & \cite{Konarev2017} \\ 
\multicolumn{4}{c}{$n=3$} \\
Isolated C$_{60}^{3-}$ & 64.2-80.5         & Theory (dynamic JT) at $T=$ 0 and 175 K & present \\
Na$_2$CsC$_{60}$       & 110               & $^{13}$C NMR $1/T_1$                         & \cite{Brouet2001} \\ 
                       & 85                & Data of Ref. \cite{Brouet2001} with different fitting function & \cite{Brouet2002c} \\
Rb$_3$C$_{60}$         & 75                & $^{13}$C NMR $1/T_1$                         & \cite{Brouet2002c} \\ 
A15 Cs$_3$C$_{60}$     & $\approx$ 100     & $^{13}$C NMR $1/T_1$                         & \cite{Jeglic2009} \\ 
\end{tabular}
\end{ruledtabular}
\footnotetext[1]{This material is a charge density wave insulator, and the two spin gaps 
presumably correspond to differently charged fullerene sites, C$_{60}^{(3-)+\delta}$ and C$_{60}^{(3-)-\delta}$.}
\end{table*}

The dense vibronic spectrum influences thermodynamic quantities such as the effective spin gap which has often been addressed with magnetic resonance techniques. 
The spin gap $\Delta F$ defines the overall thermal population of high-spin states $p^\text{HS}$:
\begin{eqnarray}
 p^\text{HS} &=& \frac{Z^\text{HS}}{Z^\text{LS} + Z^\text{HS}}
\nonumber\\
 &=& \frac{e^{-\Delta F/k_\text{B} T}}{1 + e^{-\Delta F/k_\text{B} T}},
\end{eqnarray}
where, $Z^\text{LS}$ and $Z^\text{HS}$ are the partition functions for the low- and high-spin states, respectively, $k_\text{B}$ is Boltzmann's constant, $T$ is temperature. 
$\Delta F$ is defined as a difference of Helmholtz free energies of high- and low-spin states: 
\begin{eqnarray}
 \Delta F &=& -k_\text{B}T \left( \ln Z^\text{HS} - \ln Z^\text{LS} \right)
\nonumber\\
          &=& \Delta E - T \left(S^\text{HS} - S^\text{LS}\right),
\label{Eq:spingap}
\end{eqnarray}
where, $\Delta E$ is the energy gap between the ground high-spin and low-spin vibronic levels and $S^\text{HS}$, $S^\text{LS}$ are the entropies corresponding to high-spin and low-spin states. 
In the simulations, the highest temperature of $T = 175$ K is determined so that the highest calculated vibronic levels are not populated more than a few \% \cite{SM}.

\subsubsection{$n=2,4$}
\label{Sec:SF2}
The energy gap $\Delta E$ for C$_{60}^{2-}$ is 59.9 meV. 
Fig. \ref{Fig:SF}(a) shows the entropy part of the spin gap, $-T (S^\text{HS} - S^\text{LS})$, in function of temperature.
With the rise of temperature, the spin gap decreases, which is explained by the large difference between the degeneracies of the low- and high-spin states.
The low-spin ground state is nondegenerate, whereas the lowest high-spin state is nine-fold degenerate due to the triple spin degeneracy and vibronic states. 

The spin gap of C$_{60}^{2-}$ anion has been investigated in solutions and crystals with various experimental methods (see Table \ref{Table:spingap})
\footnote{
The activation energy of C$_{60}^{2-}$ in gas phase has been estimated to be 120 $\pm$ 20 meV by analyzing the decay rate from C$_{60}^{2-}$ to C$_{60}^-+e^-$, where $e^-$ is an electron \cite{Tomita2006}. 
However, the singlet-triplet excitation is relatively small value in their analysis and many approximations are employed for the treatment of the complicated process, and hence, the error bar of the gap would be large. 
}.
The population of the triplet state of C$_{60}^{2-}$ can be detected as sharp peak in electron paramagnetic resonance (EPR) spectra \cite{Yoshizawa1993}, 
and the spin gap in frozen dimethyl sulfoxide (DMSO) has been derived ca 74 meV from the temperature dependence of the intensity \cite{Trulove1995}.
The same technique has been used for the studies of various C$_{60}$ based organic salts, and their singlet-triplet gaps were estimated to be 58-90 meV \cite{Konarev2003, Konarev2013, Konarev2017}.
The spin gaps of various alkali-doped fullerides have also been evaluated from the spin-lattice relaxation time $1/T_1$ and line shift of nuclear magnetic resonance (NMR) measurements \cite{Zimmer1994, Zimmer1995, Kerkoud1996, Brouet2001, Brouet2002a, Brouet2002c, Ricco2003} and the bulk magnetic susceptibility \cite{Lukyanchuk1995}.
In Na$_2$C$_{60}$, the gaps were estimated 100-140 meV for Na$_2$C$_{60}$ \cite{Brouet2001, Brouet2002a, Brouet2002c}, 50-70 meV for K$_4$C$_{60}$ \cite{Zimmer1994, Brouet2002a}, 50-90 meV for Rb$_4$C$_{60}$ \cite{Zimmer1995, Kerkoud1996, Brouet2002c, Lukyanchuk1995}, and 70 meV for nonmagnetic insulator (NH$_3$)$_2$NaK$_2$C$_{60}$ \cite{Ricco2003}.

Overall, the experimental spin gaps tend to be larger than calculated here (Fig. \ref{Fig:SF}(a)). 
The discrepancy could be explained by the effect of the low-symmetric environment, which could (partially) quench the JT dynamics and also induce symmetry lowering of C$_{60}$.
Understanding of the detailed interplay between the JT effect and the low-lying crystal field in the di- and tetravalent fullerene materials requires further analysis.

\subsubsection{$n=3$}
\label{Sec:SF3}
The energy gap $\Delta E$ of C$_{60}^{3-}$ is calculated as 64.2 meV. 
The entropy part of the spin gap $\Delta F(T)$ is shown in Fig. \ref{Fig:SF} (b).
Contrary to the case of C$_{60}^{2-}$ (Fig. \ref{Fig:SF}(a)), the spin gap continuously increases as temperature rises.
At $T = 175$ K, the gap is enhanced by 25 \% of $\Delta E$, and keeps rising for higher temperatures.
This different behaviour comes from the partial cancellation of different contributions to the entropy.
The reduction of the effective spin gap due to the spin multiplicity (dashed line) is cancelled by the contribution from the dense vibronic spectrum of the low-spin states (dot-dashed line). 

The value of spin gap of about 80 meV at $T = 175$ K is in line with the experimental estimates: 85-110 meV for Na$_2$CsC$_{60}$ \cite{Brouet2001, Brouet2002c}, 75 meV for Rb$_3$C$_{60}$ \cite{Brouet2002c},
and 0.1 eV for A15 Cs$_3$C$_{60}$ \cite{Jeglic2009}.
The derivation of $\Delta F$ of A15 Cs$_3$C$_{60}$ was also attempted by other group from the NMR measurements, however, clear features were not observed owing to the large spin gap \cite{Ihara2011}.
The existence of the excited spin quartet were detected by EPR measurement at $T \agt 130$ K in a series of organic fullerene compounds, nonetheless the spin gaps were not estimated \cite{Boeddinghaus2014}.
The good agreement between the theoretical and the experimental data could be explained by the presence of the JT dynamics in cubic alkali-doped fullerides \cite{Iwahara2013, Iwahara2015}, which makes the present treatment more adequate to the experimental situation in the fullerides. 

The increase of effective spin gap with temperature makes the system difficult to exhibit spin crossover.
In the study of spin crossover, the role of vibrational degrees of freedom is often discussed via the enhanced entropic effect in excited high-spin terms, resulting from the softening of vibrations \cite{Gutlich1994}. 
The present study shows that in JT systems the situation can be opposite, i.e., the spin crossover can be suppressed by vibronic entropy contribution.

\subsection{Coupling parameters}
\label{Sec:coupling}

\begin{figure*}[tb]
\begin{tabular}{ll}
(a) & (b) \\
\includegraphics[width=8cm]{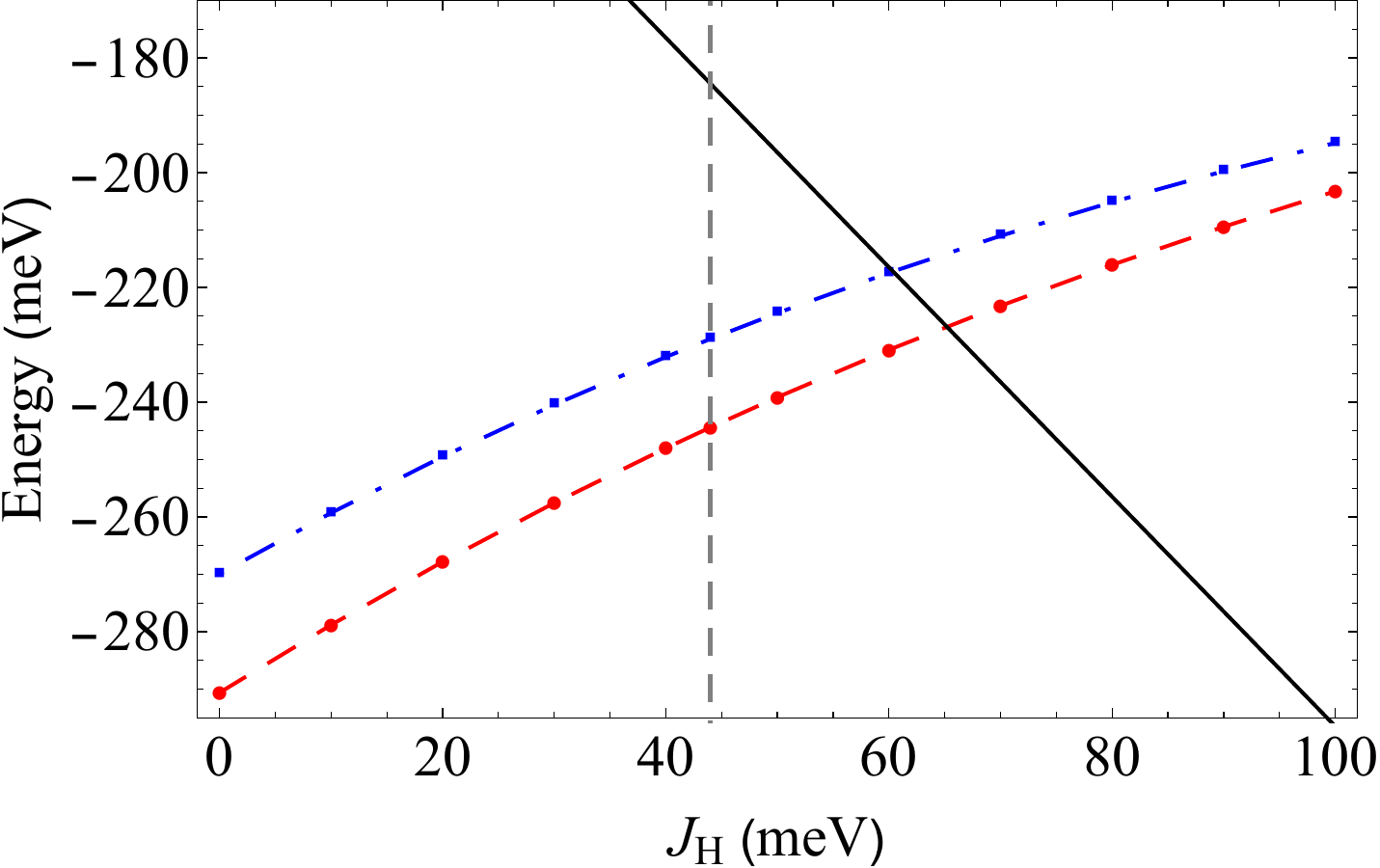}
&
\includegraphics[width=8cm]{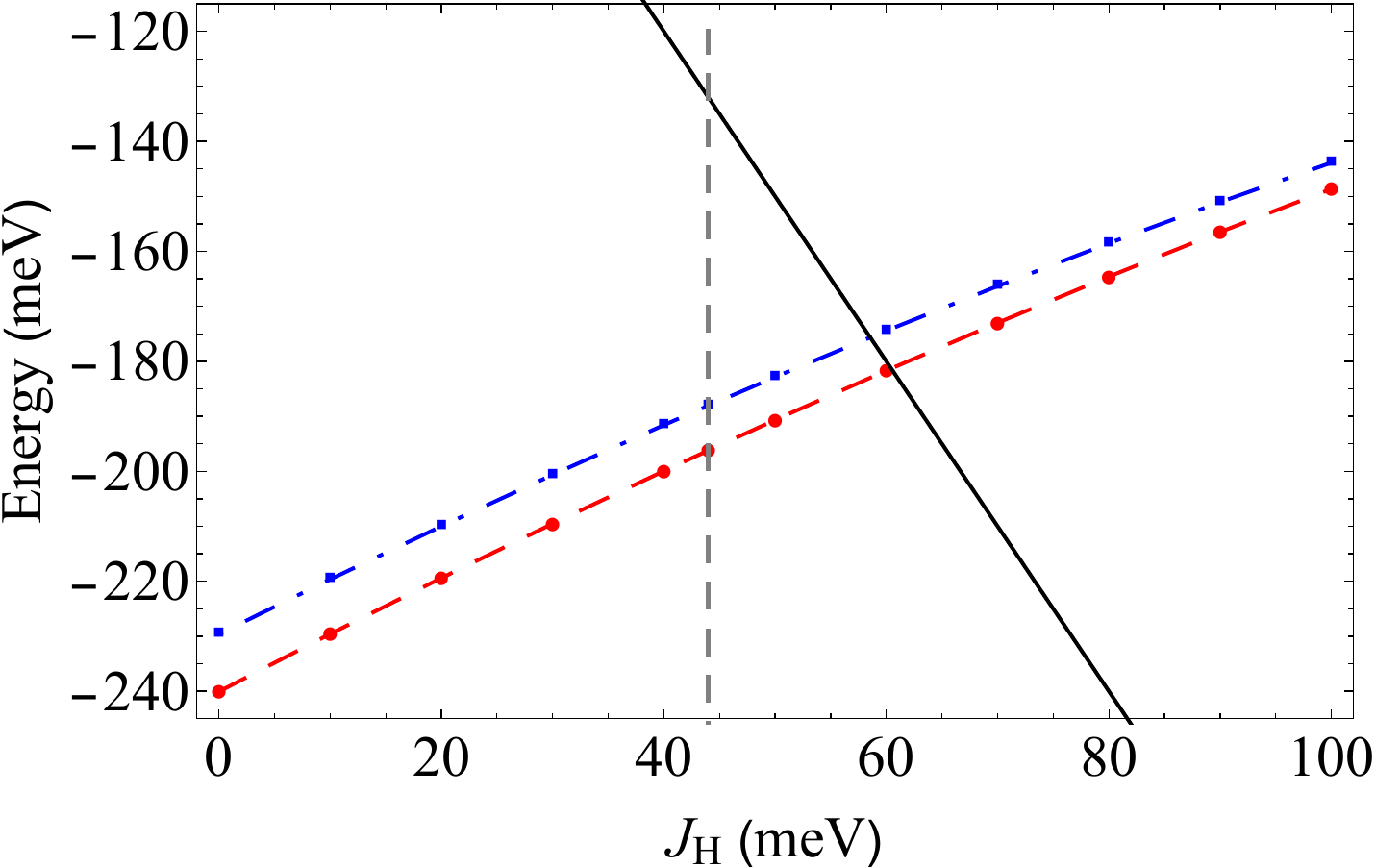}
\end{tabular}
\caption{(Color online) Lowest energy levels for the low- and high-spin states of (a) C$_{60}^{2-}$ and (b) C$_{60}^{3-}$ in function of Hund's rule coupling parameter $J_\text{H}$.
The black solid lines show the high-spin energy level, 
the red dashed and the blue dot-dashed lines
are the ground and the first excited vibronic levels of low-spin multiplicity, and the grey vertical dashed line indicates the Hund's rule coupling parameter used in this work ($J_\text{H} = 44$ meV).
}
\label{Fig:E_JH}
\end{figure*}

So far, we have studied the vibronic spectrum by using the coupling parameters derived from DFT calculations with hybrid functional \cite{Iwahara2010, Iwahara2013}.
The DFT values of $g_\mu$ \cite{Saito2002, LaflammeJanssen2010, Iwahara2010} are close to the coupling constants extracted from the experimental data \cite{Iwahara2010} and also to the parameters derived from $GW$ calculations \cite{Faber2011}.
As discussed in Sec. \ref{Sec:vibronic1}, the energy gap between the ground and the first excited states (Fig. \ref{Fig:E}) agrees well with the experimental estimate \cite{Tomita2006}, which supports the reliability of $g_\mu$ used in this work.

On the other hand, the value of $J_\text{H}$ is still under debate.
By employing the same DFT approach to a Cs$_3$C$_{60}$ cluster, we derived the Hund's rule coupling ($J_\text{H} = $ 44 meV) \cite{Iwahara2013}, which is in line with the expected value of about 50 meV for fullerides \cite{Martin1993}.
The present Hund's coupling parameter is slightly larger than that calculated for C$_{60}$ anions within local density approximation (32 meV) \cite{Luder2002} and those for $A_3$C$_{60}$ ($A =$ K, Rb, Cs) within generalized gradient approximation (30-37 meV) \cite{Nomura2012}. 
The slightly larger value is obtained because of the presence of a fraction of exact exchange in the hybrid functional. 
As discussed in the previous sections, with the use of the $J_\text{H}$ and $g_\mu$, we found that the excitation energy of C$_{60}^{2-}$ agrees well with the experimental one (Sec. \ref{Sec:vibronic2}) and the spin gap of C$_{60}^{3-}$ is close to experimental values (Sec. \ref{Sec:SF3}).
On the other hand, large $J_\text{H} \approx$ 0.1 eV has been proposed based on post Hartree-Fock calculations \cite{Nikolaev2002}.
The range of $J_\text{H}$ can be now narrowed down by comparing the present theoretical and experimental data. 

With the increase of $J_\text{H}$, the high-spin state is stabilized and the effect of pseudo JT coupling between the low-spin terms becomes weaker. 
As a result, for $J_\text{H} \agt$ 60 meV, the high-spin state becomes more stable than the low-spin state (Fig. \ref{Fig:E_JH}).
Therefore, $J_\text{H} \approx$ 0.1 eV should be ruled out. 
Furthermore, in order to reproduce the spin gap of about 50-100 meV, $J_\text{H}$ should be about 40 $\pm$ 5 meV.

\section{Conclusions}
\label{Conclusions}
In this work, we studied the low-energy vibronic states and spin gaps of fullerene anions.
The vibronic states have been derived by the numerical diagonalization of the linear $p^n \otimes 8d$ Jahn-Teller Hamiltonian with the realistic vibronic and Hund coupling parameters.
Analyzing the ground vibronic states, the contribution of the JT dynamics to the total stabilization was found to be comparable to the static one, which enables the JT dynamics to be unquenched in fullerene materials. 
In the case of $n = 2, 3, 4$, it was confirmed that the ground state turns out to be low-spin one violating the Hund's rule due to the strong JT effect. 
Particularly, in the case of $n = 3$, the violation occurs due to the dynamical JT stabilization.
The density of vibronic states becomes higher at lower energy in comparison with that of harmonic oscillator (Fig. \ref{Fig:E}), leading to the large entropic effect. 
We demonstrate that the latter makes the spin gap of C$_{60}^{3-}$ larger as the temperature rises, which is a new mechanism controlling the spin crossover.
Finally, in order to narrow down the range of the Hund's rule coupling $J_\text{H}$, the low- and high-spin states in function of $J_\text{H}$ were simulated.
It was shown that $J_\text{H}$ has to be about 40 meV to reproduce the low-spin ground state and the large spin gap. 
The current research gives the fundamental information on the dynamical Jahn-Teller effect in C$_{60}^{n-}$, which is indispensable to understand the spectroscopic and electronic properties of C$_{60}^{n-}$ molecules and the extended systems containing C$_{60}$ anions.

\section*{Acknowledgment}
D.L. gratefully acknowledges funding by the China Scholarship Council.
N.I. is supported by Japan Society for the Promotion of Science (JSPS) Overseas Research Fellowship.

%

\appendix
\section{Derivation of the vibronic Hamiltonian}
\label{A:HJT}
The matrix form of the vibronic Hamiltonian of C$_{60}^{n-}$ has been given in Refs. \cite{OBrien1996, Chancey1997}, whereas the derivation was not shown. 
Therefore, for completeness, we give a derivation of the vibronic Hamiltonian. 

In the second quantization form, the linear vibronic interaction is written as \cite{Auerbach1994, Manini1994}
\begin{eqnarray}
 \hat{H}_\text{JT} &=& \sum_{nn'\sigma} \sum_{\mu m} 
 \hslash \omega_\mu g_\mu 
 \sqrt{\frac{5}{2}}
\nonumber\\
&\times&
 (-1)^m \langle l_pn|l_pn', l_d-m\rangle \hat{c}_{n\sigma}^\dagger \hat{c}_{n'\sigma} \hat{q}_{\mu m},
\label{Eq:H_second_quantization}
\end{eqnarray}
where, $\hat{c}_{n\sigma}^\dagger$ ($\hat{c}_{n\sigma}$) is the electron creation (annihilation) operator,
$\hat{q}_{\mu m} = [\hat{b}^\dagger_{\mu m} + (-1)^m \hat{b}_{\mu,-m}]/\sqrt{2}$ is the normal coordinate, 
$\langle jm|j_1m_1, j_2m_2\rangle$ is the Clebsch-Gordan coefficient of SO(3) group \cite{Condon1953, Varshalovich1988},
$l_p=1$ and $l_d=2$ are the orbital angular momenta, 
$n, m$ are the projections, and $\sigma = \pm 1/2$ is the projection of the electron spin $s = 1/2$.
The coefficient $\sqrt{5/2}$ is introduced to reproduce the Hamiltonian given in Ref. \cite{OBrien1969}.
In order to obtain the matrix form in the basis of electronic terms, we derive tensor form of the Hamiltonian. 
Then, applying Wigner-Eckart theorem, we obtain the JT Hamiltonian matrices.

Since the electron annihilation operator is not an irreducible tensor, we transformed it into \cite{Judd1967}
\begin{eqnarray}
 \tilde{c}_{n'\sigma} &=& (-1)^{l_p+s-n'-\sigma} \hat{c}_{-n'-\sigma}.
\end{eqnarray}
The product of tensor operators $\hat{c}^\dagger \tilde{c}$ is reduced as:
\begin{eqnarray}
 \sum_{\sigma} \hat{c}_{n\sigma}^\dagger \hat{c}_{n'\sigma}
&=&
 \sum_{\sigma}
 (-1)^{l_p+s+n'+\sigma} \hat{c}_{n\sigma}^\dagger \tilde{c}_{-n'-\sigma}
\nonumber\\
&=&
 \sum_{\sigma}
 (-1)^{l_p+s+n'+\sigma} 
 \sum_{kq} \sum_{\kappa \pi} 
 \left(\hat{c}^\dagger \tilde{c}\right)^{(k\kappa)}_{q \pi} 
\nonumber\\
&\times&
 \langle kq|l_pn, l_p-n'\rangle \langle \kappa \pi|s\sigma,s-\sigma\rangle.
\end{eqnarray}
Here, $\left(\hat{c}^\dagger \tilde{c}\right)^{(k\kappa)}_{q \pi}$ is irreducible double tensor operator of ranks $k$ and $\kappa$ and 
components $q$ and $\pi$ for orbital and spin parts, respectively. 
Since $\langle 00|s\sigma,s-\sigma\rangle = (-1)^{s-\sigma}/\sqrt{[s]}$, 
\begin{eqnarray}
\sum_{\sigma} (-1)^{s-\sigma} \langle \kappa \pi|s\sigma,s-\sigma\rangle = \sqrt{[s]} \delta_{\kappa 0} \delta_{\pi 0},
\end{eqnarray}
and thus, 
\begin{eqnarray}
\sum_{\sigma} \hat{c}_{n\sigma}^\dagger \hat{c}_{n'\sigma}
&=& \sum_{kq} (-1)^{l_p+n'} \sqrt{[k]} \hat{U}^{(k)}_q \langle kq|l_pn, l_p-n'\rangle,
\nonumber\\
\end{eqnarray}
where, $\hat{U}^{(k)}_q = -\sqrt{[s]/[k]} \left(\hat{c}^\dagger \tilde{c}\right)^{(k0)}_{q0}$ is Racah's $U^{(k)}$ operator \cite{Racah1943, Judd1967}, and $[k] = 2k+1$.
Substituting this equation in the Hamiltonian (\ref{Eq:H_second_quantization}), and using 
$\langle j m| j_1 m_1, j_2 m_2\rangle = (-1)^{j_1 - m_1} \sqrt{[j]/[j_2]} \langle j_2 m_2|j m, j_1 -m_1\rangle$ \cite{Varshalovich1988},
we obtain the tensor form of $\hat{H}_\text{JT}$:
\begin{eqnarray}
 \hat{H}_\text{JT} &=& \sum_{nn'\sigma} \sum_{\mu m} 
 \hslash \omega_\mu g_\mu \sqrt{\frac{5[l_p]}{2}} 
 (-1)^m 
  \hat{U}^{(2)}_{-m} \hat{q}_{\mu m}.
\label{Eq:HJT_tensor}
\end{eqnarray}

To derive the matrix form of $\hat{H}_\text{JT}$ using the electronic terms $\{|{}^{2S+1}_{\phantom{2s+}v}LM_LM_S\rangle\}$ 
as the basis, 
we use Wigner-Eckart theorem \cite{Varshalovich1988} for the calculation of the matrix elements of $\hat{U}^{(2)}_{-m}$:
\begin{eqnarray}
  \langle {}^{2S+1}_{\phantom{2s+}v}L M_LM_S|\hat{U}^{(2)}_{-m}|{}^{2S'+1}_{\phantom{2s+}v'}L'M'_LM'_S\rangle 
 = 
 \delta_{SS'} \delta_{M_SM_S'}
\nonumber\\
 \times 
 \frac{\langle {}^{2S+1}_{\phantom{2s+}v}L\Vert \hat{U}^{(2)}\Vert {}^{2S+1}_{\phantom{2;~~}v'}L'\rangle}{\sqrt{[L]}} 
 \langle LM_L|L'M_L' ,2-m\rangle.
\nonumber\\
\label{Eq:WE}
\end{eqnarray}
Here, $v$ is the seniority of $LS$ term. 
The reduced matrix element for $p^1$ is 1 and those for $p^n$ $(n=2,3)$ are shown in Tables V and VI in Ref. \cite{Racah1943}.
The reduced matrix elements of more than half-filled systems are obtained by multiplying $(-1)^{\frac{1}{2}(v-v')+1}$ with the element of the corresponding less than half-filled system (see Eq. (46) in Ref. \cite{Judd1967}).
As a result, the reduced matrix elements for $N=5$ and $N=4$ are obtained by changing the signs of those for $N=1$ and $N=2$, respectively.
Combining the tensor form (\ref{Eq:HJT_tensor}), Wigner-Eckart theorem (\ref{Eq:WE}), and the reduced matrix elements, we obtain Eqs. (\ref{Eq:HJT1}), (\ref{Eq:HJT2}) and (\ref{Eq:HJT3}).

The reduced matrix elements of $\hat{U}^{(2)}$ connecting the same $LS$ term are zero in the case of the half-filled $p^3$ system due to the selection rule on seniority.
When $LS$ term with half-filled shell is characterized by seniority $v$, its conjugate state
is obtained by multiplying the phase factor $(-1)^\frac{v-1}{2}$ (see Eq. (65) in Ref. \cite{Racah1943}).
Thus, there are two classes of $LS$ terms: one is invariant and the other changes sign under conjugation.
Consider irreducible double tensor operator $\hat{T}^{(k \kappa)}$ acting on orbital (rank $k$) and spin (rank $\kappa$). 
The reduced matrix elements of $\hat{T}^{(k \kappa)}$ between the terms of the same class are zero when $k + \kappa$ is even, and those between the terms belonging to different classes are zero when $k + \kappa$ is odd (see Eq. (76) and the following description in Ref. \cite{Racah1942}).
In the present case, the $^2P$ and $^2D$ terms are characterized by seniorities $v = 1$ and $v = 3$, respectively, and thus, the former is invariant and the latter changes the sign.
Since the electronic part of the vibronic interaction (\ref{Eq:HJT_tensor}) is a rank-2 operator ($k = l_d = 2, \kappa = 0$), the matrix elements connecting the same terms become zero and those between the different terms are non-zero. 
The classification of the vibronic states of $p^3$ system by parity (\ref{Eq:P}) is also understood as a generalization of the two classes of $LS$ terms.

\section{Angular momentum}
\label{A:J}
The total angular momentum $\hat{J}_q$ ($q = -1, 0, 1$) is defined as \cite{Bersuker1989}:
\begin{eqnarray}
 \hat{J}_{q} &=& 
    \sum_\mu \sum_{mm'=-d}^d
    \langle l_d m|\hat{l}_q^\text{vib}|l_d m'\rangle 
    \hat{b}_{\mu m}^\dagger \hat{b}_{\mu m'}
\nonumber\\
&+& \sum_{\sigma}\sum_{nn' = -p}^p
    \langle l_p n|\hat{l}_q^\text{el}|l_p n'\rangle 
    \hat{c}_{n\sigma}^\dagger \hat{c}_{n'\sigma}.
\label{Eq:J}
\end{eqnarray}
Here, $\hat{l}_q^\text{vib}$ and $\hat{l}_q^\text{el}$ are the vibrational and electronic angular momentum operators.  
By the similar transformation as in Appendix \ref{A:HJT}, the electronic contribution reduces to $\hat{L}_q^\text{el}$ acting on electronic terms. 
The angular momentum (and also Eq. (\ref{Eq:P}) for $n = 3$) is conserved within the linear vibronic model.
With higher-order vibronic coupling, they no longer commute with the Hamiltonian.

\section{Vibronic states of the effective single-mode Jahn-Teller model}
\label{A:eff}
\begin{figure}[tb]
\begin{center}
\includegraphics[width=8cm]{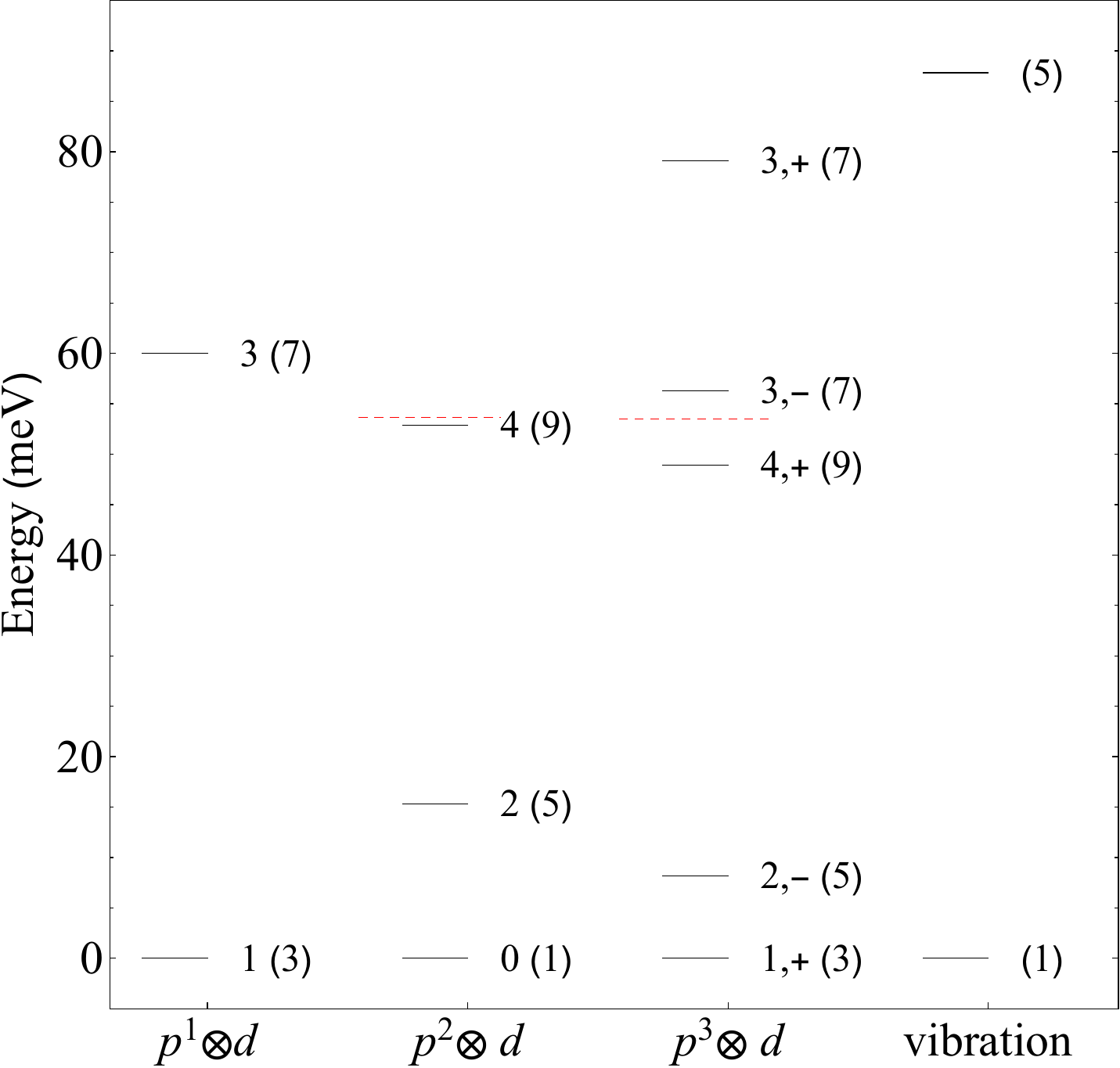}
\end{center}
\caption{
(Color online)
Low-lying vibronic levels of the effective $p^n \otimes d$ JT model with respect to the ground vibronic level of each $n$ and the zero-vibrational level (meV).
The numbers next to the vibronic levels are $J$ for $n = 1, 2$ and $(J,P)$ for $n = 3$ and the numbers in the parenthesis are the degeneracy.
The horizontal red dashed lines indicate the ground high-spin levels. 
}
\label{Fig:Eeff}
\end{figure}

In order to reveal the difference between the effective single-mode and multimode JT models, the low-energy vibronic levels of the single mode $p^n \otimes d$ JT models were calculated (Fig. \ref{Fig:Eeff}). 
The effective mode is defined so that the static JT energy $E^{(1)}_\text{JT}$ and the lowest vibronic excitation energy of C$_{60}^{3-}$ are reproduced. 
The vibronic coupling parameter and frequency for the effective mode are $g = 1.07$ and $\omega = 87.8$ meV, respectively \cite{Iwahara2013}.
The vibronic basis includes up to 20 vibrational excitations in total. 

The obtained ground vibronic energies are $-91.2$, $-232.9$, $-185.5$ meV for $n = 1, 2, 3$, respectively, which are in good agreement with the energies of C$_{60}^{n-}$ (Table \ref{Table:E}).
The energy gap between the ground and first excited levels of $p^2 \otimes d$ model is also close to the gap for C$_{60}^{2-}$ (Fig. \ref{Fig:E}). 
On the other hand, the description of the excited states becomes worse in the single mode model.
Apparently, the number of the vibronic levels of the single mode model is significantly reduced compared with that of the multimode model because of the much smaller vibrational degrees of freedom. 
Besides, the order of the excited levels can be interchanged. 
For instance, the second and the third excited vibronic levels are inverted compared with $p^3 \otimes 8d$ JT system as discussed before \cite{Iwahara2013}. 

In the case of strong coupling, the vibronic levels of the effective model is described as the sum of the energies of fast radial harmonic oscillation and slow pseudorotation in the trough.
However, in reality, the effective modes, particularly, the pseudorotational modes, accompany the cloud of the non-effective vibrations \cite{Polinger1979, Manini1998}. 
The reconstructed vibrational and pseudorotational energies will be superimposed on the vibronic levels of the single mode $p^n \otimes d$ JT system, which would to some extent reproduce the dense vibronic levels of C$_{60}^{n-}$ with correct order.


\begin{thebibliography}{102}%
\makeatletter
\providecommand \@ifxundefined [1]{%
 \@ifx{#1\undefined}
}%
\providecommand \@ifnum [1]{%
 \ifnum #1\expandafter \@firstoftwo
 \else \expandafter \@secondoftwo
 \fi
}%
\providecommand \@ifx [1]{%
 \ifx #1\expandafter \@firstoftwo
 \else \expandafter \@secondoftwo
 \fi
}%
\providecommand \natexlab [1]{#1}%
\providecommand \enquote  [1]{``#1''}%
\providecommand \bibnamefont  [1]{#1}%
\providecommand \bibfnamefont [1]{#1}%
\providecommand \citenamefont [1]{#1}%
\providecommand \href@noop [0]{\@secondoftwo}%
\providecommand \href [0]{\begingroup \@sanitize@url \@href}%
\providecommand \@href[1]{\@@startlink{#1}\@@href}%
\providecommand \@@href[1]{\endgroup#1\@@endlink}%
\providecommand \@sanitize@url [0]{\catcode `\\12\catcode `\$12\catcode
  `\&12\catcode `\#12\catcode `\^12\catcode `\_12\catcode `\%12\relax}%
\providecommand \@@startlink[1]{}%
\providecommand \@@endlink[0]{}%
\providecommand \url  [0]{\begingroup\@sanitize@url \@url }%
\providecommand \@url [1]{\endgroup\@href {#1}{\urlprefix }}%
\providecommand \urlprefix  [0]{URL }%
\providecommand \Eprint [0]{\href }%
\providecommand \doibase [0]{http://dx.doi.org/}%
\providecommand \selectlanguage [0]{\@gobble}%
\providecommand \bibinfo  [0]{\@secondoftwo}%
\providecommand \bibfield  [0]{\@secondoftwo}%
\providecommand \translation [1]{[#1]}%
\providecommand \BibitemOpen [0]{}%
\providecommand \bibitemStop [0]{}%
\providecommand \bibitemNoStop [0]{.\EOS\space}%
\providecommand \EOS [0]{\spacefactor3000\relax}%
\providecommand \BibitemShut  [1]{\csname bibitem#1\endcsname}%
\let\auto@bib@innerbib\@empty
\bibitem [{\citenamefont {Gunnarsson}(1997)}]{Gunnarsson1997}%
  \BibitemOpen
  \bibfield  {author} {\bibinfo {author} {\bibfnamefont {O.}~\bibnamefont
  {Gunnarsson}},\ }\bibfield  {title} {\enquote {\bibinfo {title}
  {Superconductivity in fullerides},}\ }\href {\doibase
  10.1103/RevModPhys.69.575} {\bibfield  {journal} {\bibinfo  {journal} {Rev.
  Mod. Phys.}\ }\textbf {\bibinfo {volume} {69}},\ \bibinfo {pages} {575}
  (\bibinfo {year} {1997})}\BibitemShut {NoStop}%
\bibitem [{\citenamefont {Gunnarsson}(2004)}]{Gunnarsson2004}%
  \BibitemOpen
  \bibfield  {author} {\bibinfo {author} {\bibfnamefont {O.}~\bibnamefont
  {Gunnarsson}},\ }\href@noop {} {\emph {\bibinfo {title} {{Alkali-Doped
  Fullerides: Narrow-Band Solids with Unusual Properties}}}}\ (\bibinfo
  {publisher} {World Scientific},\ \bibinfo {address} {Singapore},\ \bibinfo
  {year} {2004})\BibitemShut {NoStop}%
\bibitem [{\citenamefont {Haddon}\ \emph {et~al.}(1991)\citenamefont {Haddon},
  \citenamefont {Hebard}, \citenamefont {Rosseinsky}, \citenamefont {Murphy},
  \citenamefont {Duclos}, \citenamefont {Lyons}, \citenamefont {Miller},
  \citenamefont {Rosamilia}, \citenamefont {Fleming}, \citenamefont {Kortan},
  \citenamefont {Glarum}, \citenamefont {Makhija}, \citenamefont {Muller},
  \citenamefont {Eick}, \citenamefont {Zahurak}, \citenamefont {Tycko},
  \citenamefont {Dabbagh},\ and\ \citenamefont {Thiel}}]{Haddon1991}%
  \BibitemOpen
  \bibfield  {author} {\bibinfo {author} {\bibfnamefont {R.~C.}\ \bibnamefont
  {Haddon}}, \bibinfo {author} {\bibfnamefont {A.~F.}\ \bibnamefont {Hebard}},
  \bibinfo {author} {\bibfnamefont {M.~J.}\ \bibnamefont {Rosseinsky}},
  \bibinfo {author} {\bibfnamefont {D.~W.}\ \bibnamefont {Murphy}}, \bibinfo
  {author} {\bibfnamefont {S.~J.}\ \bibnamefont {Duclos}}, \bibinfo {author}
  {\bibfnamefont {K.~B.}\ \bibnamefont {Lyons}}, \bibinfo {author}
  {\bibfnamefont {B.}~\bibnamefont {Miller}}, \bibinfo {author} {\bibfnamefont
  {J.~M.}\ \bibnamefont {Rosamilia}}, \bibinfo {author} {\bibfnamefont {R.~M.}\
  \bibnamefont {Fleming}}, \bibinfo {author} {\bibfnamefont {A.~R.}\
  \bibnamefont {Kortan}}, \bibinfo {author} {\bibfnamefont {S.~H.}\
  \bibnamefont {Glarum}}, \bibinfo {author} {\bibfnamefont {A.~V.}\
  \bibnamefont {Makhija}}, \bibinfo {author} {\bibfnamefont {A.~J.}\
  \bibnamefont {Muller}}, \bibinfo {author} {\bibfnamefont {R.~E.}\
  \bibnamefont {Eick}}, \bibinfo {author} {\bibfnamefont {S.~M.}\ \bibnamefont
  {Zahurak}}, \bibinfo {author} {\bibfnamefont {R.}~\bibnamefont {Tycko}},
  \bibinfo {author} {\bibfnamefont {G.}~\bibnamefont {Dabbagh}}, \ and\
  \bibinfo {author} {\bibfnamefont {F.~A.}\ \bibnamefont {Thiel}},\ }\bibfield
  {title} {\enquote {\bibinfo {title} {{Conducting film of C$_{60}$ and
  C$_{70}$ by alkali-metal doping}},}\ }\href@noop {} {\bibfield  {journal}
  {\bibinfo  {journal} {Nature}\ }\textbf {\bibinfo {volume} {350}},\ \bibinfo
  {pages} {320} (\bibinfo {year} {1991})}\BibitemShut {NoStop}%
\bibitem [{\citenamefont {Hebard}\ \emph {et~al.}(1991)\citenamefont {Hebard},
  \citenamefont {Rosseinsky}, \citenamefont {Haddon}, \citenamefont {Murphy},
  \citenamefont {Glarum}, \citenamefont {Palstra}, \citenamefont {Ramirez},\
  and\ \citenamefont {Kortan}}]{Hebard1991}%
  \BibitemOpen
  \bibfield  {author} {\bibinfo {author} {\bibfnamefont {A.~F.}\ \bibnamefont
  {Hebard}}, \bibinfo {author} {\bibfnamefont {M.~J.}\ \bibnamefont
  {Rosseinsky}}, \bibinfo {author} {\bibfnamefont {R.~C.}\ \bibnamefont
  {Haddon}}, \bibinfo {author} {\bibfnamefont {D.~W.}\ \bibnamefont {Murphy}},
  \bibinfo {author} {\bibfnamefont {S.~H.}\ \bibnamefont {Glarum}}, \bibinfo
  {author} {\bibfnamefont {T.~T.~M.}\ \bibnamefont {Palstra}}, \bibinfo
  {author} {\bibfnamefont {A.~P.}\ \bibnamefont {Ramirez}}, \ and\ \bibinfo
  {author} {\bibfnamefont {A.~R.}\ \bibnamefont {Kortan}},\ }\bibfield  {title}
  {\enquote {\bibinfo {title} {{Superconductivity at 18 K in potassium-doped
  C$_{60}$}},}\ }\href@noop {} {\bibfield  {journal} {\bibinfo  {journal}
  {Nature}\ }\textbf {\bibinfo {volume} {350}},\ \bibinfo {pages} {600}
  (\bibinfo {year} {1991})}\BibitemShut {NoStop}%
\bibitem [{\citenamefont {Tanigaki}\ \emph {et~al.}(1991)\citenamefont
  {Tanigaki}, \citenamefont {Ebbesen}, \citenamefont {Saito}, \citenamefont
  {Mizuki}, \citenamefont {Tsai}, \citenamefont {Kubo},\ and\ \citenamefont
  {Kuroshima}}]{Tanigaki1991}%
  \BibitemOpen
  \bibfield  {author} {\bibinfo {author} {\bibfnamefont {K.}~\bibnamefont
  {Tanigaki}}, \bibinfo {author} {\bibfnamefont {T.~W.}\ \bibnamefont
  {Ebbesen}}, \bibinfo {author} {\bibfnamefont {S.}~\bibnamefont {Saito}},
  \bibinfo {author} {\bibfnamefont {J.}~\bibnamefont {Mizuki}}, \bibinfo
  {author} {\bibfnamefont {J.~S.}\ \bibnamefont {Tsai}}, \bibinfo {author}
  {\bibfnamefont {Y.}~\bibnamefont {Kubo}}, \ and\ \bibinfo {author}
  {\bibfnamefont {S.}~\bibnamefont {Kuroshima}},\ }\bibfield  {title} {\enquote
  {\bibinfo {title} {{Superconductivity at 33 K in Cs$_x$Rb$_y$C$_{60}$}},}\
  }\href@noop {} {\bibfield  {journal} {\bibinfo  {journal} {Nature}\ }\textbf
  {\bibinfo {volume} {352}},\ \bibinfo {pages} {222} (\bibinfo {year}
  {1991})}\BibitemShut {NoStop}%
\bibitem [{\citenamefont {Winter}\ and\ \citenamefont
  {Kuzmany}(1992)}]{Winter1992}%
  \BibitemOpen
  \bibfield  {author} {\bibinfo {author} {\bibfnamefont {J.}~\bibnamefont
  {Winter}}\ and\ \bibinfo {author} {\bibfnamefont {H.}~\bibnamefont
  {Kuzmany}},\ }\bibfield  {title} {\enquote {\bibinfo {title}
  {{Potassium-doped fullerene K$_x$C$_{60}$ with $x =$ 0, 1, 2, 3, 4, and
  6}},}\ }\href {\doibase https://doi.org/10.1016/0038-1098(92)90796-C}
  {\bibfield  {journal} {\bibinfo  {journal} {Solid State Commun.}\ }\textbf
  {\bibinfo {volume} {84}},\ \bibinfo {pages} {935} (\bibinfo {year}
  {1992})}\BibitemShut {NoStop}%
\bibitem [{\citenamefont {Kerkoud}\ \emph {et~al.}(1996)\citenamefont
  {Kerkoud}, \citenamefont {Auban-Senzier}, \citenamefont {J\'{e}rome},
  \citenamefont {Brazovskii}, \citenamefont {Luk'yanchuk}, \citenamefont
  {Kirova}, \citenamefont {Rachdi},\ and\ \citenamefont {Goze}}]{Kerkoud1996}%
  \BibitemOpen
  \bibfield  {author} {\bibinfo {author} {\bibfnamefont {R.}~\bibnamefont
  {Kerkoud}}, \bibinfo {author} {\bibfnamefont {P.}~\bibnamefont
  {Auban-Senzier}}, \bibinfo {author} {\bibfnamefont {D.}~\bibnamefont
  {J\'{e}rome}}, \bibinfo {author} {\bibfnamefont {S.}~\bibnamefont
  {Brazovskii}}, \bibinfo {author} {\bibfnamefont {I.}~\bibnamefont
  {Luk'yanchuk}}, \bibinfo {author} {\bibfnamefont {N.}~\bibnamefont {Kirova}},
  \bibinfo {author} {\bibfnamefont {F.}~\bibnamefont {Rachdi}}, \ and\ \bibinfo
  {author} {\bibfnamefont {C.}~\bibnamefont {Goze}},\ }\bibfield  {title}
  {\enquote {\bibinfo {title} {{Insulator-metal transition in Rb$_4$C$_{60}$
  under pressure from ${}^{13}$C-NMR}},}\ }\href {\doibase
  http://dx.doi.org/10.1016/0022-3697(95)00113-1} {\bibfield  {journal}
  {\bibinfo  {journal} {J. Phys. Chem. Solids}\ }\textbf {\bibinfo {volume}
  {57}},\ \bibinfo {pages} {143} (\bibinfo {year} {1996})}\BibitemShut
  {NoStop}%
\bibitem [{\citenamefont {Knupfer}\ and\ \citenamefont
  {Fink}(1997)}]{Knupfer1997}%
  \BibitemOpen
  \bibfield  {author} {\bibinfo {author} {\bibfnamefont {M.}~\bibnamefont
  {Knupfer}}\ and\ \bibinfo {author} {\bibfnamefont {J.}~\bibnamefont {Fink}},\
  }\bibfield  {title} {\enquote {\bibinfo {title} {{Mott-Hubbard-like Behavior
  of the Energy Gap of $A_4$C$_{60}$ ($A =$ Na, K, Rb, Cs) and
  Na$_{10}${C}$_{60}$}},}\ }\href {\doibase 10.1103/PhysRevLett.79.2714}
  {\bibfield  {journal} {\bibinfo  {journal} {Phys. Rev. Lett.}\ }\textbf
  {\bibinfo {volume} {79}},\ \bibinfo {pages} {2714} (\bibinfo {year}
  {1997})}\BibitemShut {NoStop}%
\bibitem [{\citenamefont {Brouet}\ \emph {et~al.}(2001)\citenamefont {Brouet},
  \citenamefont {Alloul}, \citenamefont {Le}, \citenamefont {Garaj},\ and\
  \citenamefont {Forr\'o}}]{Brouet2001}%
  \BibitemOpen
  \bibfield  {author} {\bibinfo {author} {\bibfnamefont {V.}~\bibnamefont
  {Brouet}}, \bibinfo {author} {\bibfnamefont {H.}~\bibnamefont {Alloul}},
  \bibinfo {author} {\bibfnamefont {T.-N.}\ \bibnamefont {Le}}, \bibinfo
  {author} {\bibfnamefont {S.}~\bibnamefont {Garaj}}, \ and\ \bibinfo {author}
  {\bibfnamefont {L.}~\bibnamefont {Forr\'o}},\ }\bibfield  {title} {\enquote
  {\bibinfo {title} {{Role of Dynamic Jahn-Teller Distortions in
  ${\mathrm{Na}}_{2}{\mathrm{C}}_{60}$ and
  ${\mathrm{Na}}_{2}{\mathrm{CsC}}_{60}$ Studied by NMR}},}\ }\href {\doibase
  10.1103/PhysRevLett.86.4680} {\bibfield  {journal} {\bibinfo  {journal}
  {Phys. Rev. Lett.}\ }\textbf {\bibinfo {volume} {86}},\ \bibinfo {pages}
  {4680} (\bibinfo {year} {2001})}\BibitemShut {NoStop}%
\bibitem [{\citenamefont {Ganin}\ \emph {et~al.}(2008)\citenamefont {Ganin},
  \citenamefont {Takabayashi}, \citenamefont {Khimyak}, \citenamefont
  {Margadonna}, \citenamefont {Tamai}, \citenamefont {Rosseinsky},\ and\
  \citenamefont {Prassides}}]{Ganin2008}%
  \BibitemOpen
  \bibfield  {author} {\bibinfo {author} {\bibfnamefont {A.~Y.}\ \bibnamefont
  {Ganin}}, \bibinfo {author} {\bibfnamefont {Y.}~\bibnamefont {Takabayashi}},
  \bibinfo {author} {\bibfnamefont {Y.~Z.}\ \bibnamefont {Khimyak}}, \bibinfo
  {author} {\bibfnamefont {S.}~\bibnamefont {Margadonna}}, \bibinfo {author}
  {\bibfnamefont {A.}~\bibnamefont {Tamai}}, \bibinfo {author} {\bibfnamefont
  {M.~J.}\ \bibnamefont {Rosseinsky}}, \ and\ \bibinfo {author} {\bibfnamefont
  {K.}~\bibnamefont {Prassides}},\ }\bibfield  {title} {\enquote {\bibinfo
  {title} {{Bulk superconductivity at 38 K in a molecular system}},}\
  }\href@noop {} {\bibfield  {journal} {\bibinfo  {journal} {Nat. Mater.}\
  }\textbf {\bibinfo {volume} {7}},\ \bibinfo {pages} {367} (\bibinfo {year}
  {2008})}\BibitemShut {NoStop}%
\bibitem [{\citenamefont {Ihara}\ \emph {et~al.}(2010)\citenamefont {Ihara},
  \citenamefont {Alloul}, \citenamefont {Wzietek}, \citenamefont {Pontiroli},
  \citenamefont {Mazzani},\ and\ \citenamefont {Ricc\`o}}]{Ihara2010}%
  \BibitemOpen
  \bibfield  {author} {\bibinfo {author} {\bibfnamefont {Y.}~\bibnamefont
  {Ihara}}, \bibinfo {author} {\bibfnamefont {H.}~\bibnamefont {Alloul}},
  \bibinfo {author} {\bibfnamefont {P.}~\bibnamefont {Wzietek}}, \bibinfo
  {author} {\bibfnamefont {D.}~\bibnamefont {Pontiroli}}, \bibinfo {author}
  {\bibfnamefont {M.}~\bibnamefont {Mazzani}}, \ and\ \bibinfo {author}
  {\bibfnamefont {M.}~\bibnamefont {Ricc\`o}},\ }\bibfield  {title} {\enquote
  {\bibinfo {title} {{NMR Study of the Mott Transitions to Superconductivity in
  the Two ${\mathrm{Cs}}_{3}{\mathrm{C}}_{60}$ Phases}},}\ }\href {\doibase
  10.1103/PhysRevLett.104.256402} {\bibfield  {journal} {\bibinfo  {journal}
  {Phys. Rev. Lett.}\ }\textbf {\bibinfo {volume} {104}},\ \bibinfo {pages}
  {256402} (\bibinfo {year} {2010})}\BibitemShut {NoStop}%
\bibitem [{\citenamefont {Klupp}\ \emph {et~al.}(2012)\citenamefont {Klupp},
  \citenamefont {Matus}, \citenamefont {{Kamar\'{a}s}}, \citenamefont {Ganin},
  \citenamefont {McLennan}, \citenamefont {Rosseinsky}, \citenamefont
  {Takabayashi}, \citenamefont {McDonald},\ and\ \citenamefont
  {Prassides}}]{Klupp2012}%
  \BibitemOpen
  \bibfield  {author} {\bibinfo {author} {\bibfnamefont {G.}~\bibnamefont
  {Klupp}}, \bibinfo {author} {\bibfnamefont {P.}~\bibnamefont {Matus}},
  \bibinfo {author} {\bibfnamefont {K.}~\bibnamefont {{Kamar\'{a}s}}}, \bibinfo
  {author} {\bibfnamefont {A.~Y.}\ \bibnamefont {Ganin}}, \bibinfo {author}
  {\bibfnamefont {A.}~\bibnamefont {McLennan}}, \bibinfo {author}
  {\bibfnamefont {M.~J.}\ \bibnamefont {Rosseinsky}}, \bibinfo {author}
  {\bibfnamefont {Y.}~\bibnamefont {Takabayashi}}, \bibinfo {author}
  {\bibfnamefont {M.~T.}\ \bibnamefont {McDonald}}, \ and\ \bibinfo {author}
  {\bibfnamefont {K.}~\bibnamefont {Prassides}},\ }\bibfield  {title} {\enquote
  {\bibinfo {title} {{Dynamic Jahn-Teller effect in the parent insulating state
  of the molecular superconductor Cs$_3$C$_{60}$}},}\ }\href@noop {} {\bibfield
   {journal} {\bibinfo  {journal} {Nat. Commun.}\ }\textbf {\bibinfo {volume}
  {3}},\ \bibinfo {pages} {912} (\bibinfo {year} {2012})}\BibitemShut {NoStop}%
\bibitem [{\citenamefont {Iwahara}\ and\ \citenamefont
  {Chibotaru}(2013)}]{Iwahara2013}%
  \BibitemOpen
  \bibfield  {author} {\bibinfo {author} {\bibfnamefont {N.}~\bibnamefont
  {Iwahara}}\ and\ \bibinfo {author} {\bibfnamefont {L.~F.}\ \bibnamefont
  {Chibotaru}},\ }\bibfield  {title} {\enquote {\bibinfo {title} {{Dynamical
  Jahn-Teller Effect and Antiferromagnetism in Cs$_3$C$_{60}$}},}\ }\href
  {\doibase 10.1103/PhysRevLett.111.056401} {\bibfield  {journal} {\bibinfo
  {journal} {Phys. Rev. Lett.}\ }\textbf {\bibinfo {volume} {111}},\ \bibinfo
  {pages} {056401} (\bibinfo {year} {2013})}\BibitemShut {NoStop}%
\bibitem [{\citenamefont {Poto{\v{c}}nik}\ \emph {et~al.}(2014)\citenamefont
  {Poto{\v{c}}nik}, \citenamefont {Ganin}, \citenamefont {Takabayashi},
  \citenamefont {McDonald}, \citenamefont {Heinmaa}, \citenamefont
  {Jegli{\v{c}}}, \citenamefont {Stern}, \citenamefont {Rosseinsky},
  \citenamefont {Prassides},\ and\ \citenamefont {Ar{\v{c}}on}}]{Potocnik2014}%
  \BibitemOpen
  \bibfield  {author} {\bibinfo {author} {\bibfnamefont {A.}~\bibnamefont
  {Poto{\v{c}}nik}}, \bibinfo {author} {\bibfnamefont {A.~Y.}\ \bibnamefont
  {Ganin}}, \bibinfo {author} {\bibfnamefont {Y.}~\bibnamefont {Takabayashi}},
  \bibinfo {author} {\bibfnamefont {M.~T.}\ \bibnamefont {McDonald}}, \bibinfo
  {author} {\bibfnamefont {I.}~\bibnamefont {Heinmaa}}, \bibinfo {author}
  {\bibfnamefont {P.}~\bibnamefont {Jegli{\v{c}}}}, \bibinfo {author}
  {\bibfnamefont {R.}~\bibnamefont {Stern}}, \bibinfo {author} {\bibfnamefont
  {M.~J.}\ \bibnamefont {Rosseinsky}}, \bibinfo {author} {\bibfnamefont
  {K.}~\bibnamefont {Prassides}}, \ and\ \bibinfo {author} {\bibfnamefont
  {D.}~\bibnamefont {Ar{\v{c}}on}},\ }\bibfield  {title} {\enquote {\bibinfo
  {title} {{Jahn-Teller orbital glass state in the expanded fcc
  Cs$_{3}$C$_{60}$ fulleride}},}\ }\href@noop {} {\bibfield  {journal}
  {\bibinfo  {journal} {Chem. Sci.}\ }\textbf {\bibinfo {volume} {5}},\
  \bibinfo {pages} {3008} (\bibinfo {year} {2014})}\BibitemShut {NoStop}%
\bibitem [{\citenamefont {Iwahara}\ and\ \citenamefont
  {Chibotaru}(2015)}]{Iwahara2015}%
  \BibitemOpen
  \bibfield  {author} {\bibinfo {author} {\bibfnamefont {N.}~\bibnamefont
  {Iwahara}}\ and\ \bibinfo {author} {\bibfnamefont {L.~F.}\ \bibnamefont
  {Chibotaru}},\ }\bibfield  {title} {\enquote {\bibinfo {title} {{Dynamical
  Jahn-Teller instability in metallic fullerides}},}\ }\href {\doibase
  10.1103/PhysRevB.91.035109} {\bibfield  {journal} {\bibinfo  {journal} {Phys.
  Rev. B}\ }\textbf {\bibinfo {volume} {91}},\ \bibinfo {pages} {035109}
  (\bibinfo {year} {2015})}\BibitemShut {NoStop}%
\bibitem [{\citenamefont {Zadik}\ \emph {et~al.}(2015)\citenamefont {Zadik},
  \citenamefont {Takabayashi}, \citenamefont {Klupp}, \citenamefont {Colman},
  \citenamefont {Ganin}, \citenamefont {{Poto\v{c}nik}}, \citenamefont
  {{Jegli\v{c}}}, \citenamefont {{Ar\v{c}on}}, \citenamefont {Matus},
  \citenamefont {{Kamar\'{a}s}}, \citenamefont {Kasahara}, \citenamefont
  {Iwasa}, \citenamefont {Fitch}, \citenamefont {Ohishi}, \citenamefont
  {Garbarino}, \citenamefont {Kato}, \citenamefont {Rosseinsky},\ and\
  \citenamefont {Prassides}}]{Zadik2015}%
  \BibitemOpen
  \bibfield  {author} {\bibinfo {author} {\bibfnamefont {R.~H.}\ \bibnamefont
  {Zadik}}, \bibinfo {author} {\bibfnamefont {Y.}~\bibnamefont {Takabayashi}},
  \bibinfo {author} {\bibfnamefont {G.}~\bibnamefont {Klupp}}, \bibinfo
  {author} {\bibfnamefont {R.~H.}\ \bibnamefont {Colman}}, \bibinfo {author}
  {\bibfnamefont {A.~Y.}\ \bibnamefont {Ganin}}, \bibinfo {author}
  {\bibfnamefont {A.}~\bibnamefont {{Poto\v{c}nik}}}, \bibinfo {author}
  {\bibfnamefont {P.}~\bibnamefont {{Jegli\v{c}}}}, \bibinfo {author}
  {\bibfnamefont {D.}~\bibnamefont {{Ar\v{c}on}}}, \bibinfo {author}
  {\bibfnamefont {P.}~\bibnamefont {Matus}}, \bibinfo {author} {\bibfnamefont
  {K.}~\bibnamefont {{Kamar\'{a}s}}}, \bibinfo {author} {\bibfnamefont
  {Y.}~\bibnamefont {Kasahara}}, \bibinfo {author} {\bibfnamefont
  {Y.}~\bibnamefont {Iwasa}}, \bibinfo {author} {\bibfnamefont {A.~N.}\
  \bibnamefont {Fitch}}, \bibinfo {author} {\bibfnamefont {Y.}~\bibnamefont
  {Ohishi}}, \bibinfo {author} {\bibfnamefont {G.}~\bibnamefont {Garbarino}},
  \bibinfo {author} {\bibfnamefont {K.}~\bibnamefont {Kato}}, \bibinfo {author}
  {\bibfnamefont {M.~J.}\ \bibnamefont {Rosseinsky}}, \ and\ \bibinfo {author}
  {\bibfnamefont {K.}~\bibnamefont {Prassides}},\ }\bibfield  {title} {\enquote
  {\bibinfo {title} {{Optimized unconventional superconductivity in a molecular
  Jahn-Teller metal}},}\ }\href@noop {} {\bibfield  {journal} {\bibinfo
  {journal} {Sci. Adv.}\ }\textbf {\bibinfo {volume} {1}},\ \bibinfo {pages}
  {e1500059} (\bibinfo {year} {2015})}\BibitemShut {NoStop}%
\bibitem [{\citenamefont {Iwahara}\ and\ \citenamefont
  {Chibotaru}(2016)}]{Iwahara2016}%
  \BibitemOpen
  \bibfield  {author} {\bibinfo {author} {\bibfnamefont {N.}~\bibnamefont
  {Iwahara}}\ and\ \bibinfo {author} {\bibfnamefont {L.~F.}\ \bibnamefont
  {Chibotaru}},\ }\bibfield  {title} {\enquote {\bibinfo {title} {Orbital
  disproportionation of electronic density is a universal feature of
  alkali-doped fullerides},}\ }\href@noop {} {\bibfield  {journal} {\bibinfo
  {journal} {Nat. Commun.}\ }\textbf {\bibinfo {volume} {7}},\ \bibinfo {pages}
  {13093} (\bibinfo {year} {2016})}\BibitemShut {NoStop}%
\bibitem [{\citenamefont {Nomura}\ \emph {et~al.}(2016)\citenamefont {Nomura},
  \citenamefont {Sakai}, \citenamefont {Capone},\ and\ \citenamefont
  {Arita}}]{Nomura2016}%
  \BibitemOpen
  \bibfield  {author} {\bibinfo {author} {\bibfnamefont {Y.}~\bibnamefont
  {Nomura}}, \bibinfo {author} {\bibfnamefont {S.}~\bibnamefont {Sakai}},
  \bibinfo {author} {\bibfnamefont {M.}~\bibnamefont {Capone}}, \ and\ \bibinfo
  {author} {\bibfnamefont {R.}~\bibnamefont {Arita}},\ }\bibfield  {title}
  {\enquote {\bibinfo {title} {Exotic $s$-wave superconductivity in
  alkali-doped fullerides},}\ }\href
  {http://stacks.iop.org/0953-8984/28/i=15/a=153001} {\bibfield  {journal}
  {\bibinfo  {journal} {J. Phys.: Condens. Matter}\ }\textbf {\bibinfo {volume}
  {28}},\ \bibinfo {pages} {153001} (\bibinfo {year} {2016})}\BibitemShut
  {NoStop}%
\bibitem [{\citenamefont {Mitrano}\ \emph {et~al.}(2016)\citenamefont
  {Mitrano}, \citenamefont {Cantaluppi}, \citenamefont {Nicoletti},
  \citenamefont {Kaiser}, \citenamefont {Perucchi}, \citenamefont {Lupi},
  \citenamefont {Pietro}, \citenamefont {Pontiroli}, \citenamefont {Ricc\`{o}},
  \citenamefont {Clark}, \citenamefont {Jaksch},\ and\ \citenamefont
  {Cavalleri}}]{Mitrano2016}%
  \BibitemOpen
  \bibfield  {author} {\bibinfo {author} {\bibfnamefont {M.}~\bibnamefont
  {Mitrano}}, \bibinfo {author} {\bibfnamefont {A.}~\bibnamefont {Cantaluppi}},
  \bibinfo {author} {\bibfnamefont {D.}~\bibnamefont {Nicoletti}}, \bibinfo
  {author} {\bibfnamefont {S.}~\bibnamefont {Kaiser}}, \bibinfo {author}
  {\bibfnamefont {A.}~\bibnamefont {Perucchi}}, \bibinfo {author}
  {\bibfnamefont {S.}~\bibnamefont {Lupi}}, \bibinfo {author} {\bibfnamefont
  {P.~Di}\ \bibnamefont {Pietro}}, \bibinfo {author} {\bibfnamefont
  {D.}~\bibnamefont {Pontiroli}}, \bibinfo {author} {\bibfnamefont
  {M.}~\bibnamefont {Ricc\`{o}}}, \bibinfo {author} {\bibfnamefont {S.~R.}\
  \bibnamefont {Clark}}, \bibinfo {author} {\bibfnamefont {D.}~\bibnamefont
  {Jaksch}}, \ and\ \bibinfo {author} {\bibfnamefont {A.}~\bibnamefont
  {Cavalleri}},\ }\bibfield  {title} {\enquote {\bibinfo {title} {{Possible
  light-induced superconductivity in K$_3$C$_{60}$ at high temperature}},}\
  }\href@noop {} {\bibfield  {journal} {\bibinfo  {journal} {Nature}\ }\textbf
  {\bibinfo {volume} {530}},\ \bibinfo {pages} {461} (\bibinfo {year}
  {2016})}\BibitemShut {NoStop}%
\bibitem [{\citenamefont {Kasahara}\ \emph {et~al.}(2017)\citenamefont
  {Kasahara}, \citenamefont {Takeuchi}, \citenamefont {Zadik}, \citenamefont
  {Takabayashi}, \citenamefont {Colman}, \citenamefont {McDonald},
  \citenamefont {Rosseinsky}, \citenamefont {Prassides},\ and\ \citenamefont
  {Iwasa}}]{Kasahara2017}%
  \BibitemOpen
  \bibfield  {author} {\bibinfo {author} {\bibfnamefont {Y.}~\bibnamefont
  {Kasahara}}, \bibinfo {author} {\bibfnamefont {Y.}~\bibnamefont {Takeuchi}},
  \bibinfo {author} {\bibfnamefont {R.~H.}\ \bibnamefont {Zadik}}, \bibinfo
  {author} {\bibfnamefont {Y.}~\bibnamefont {Takabayashi}}, \bibinfo {author}
  {\bibfnamefont {R.~H.}\ \bibnamefont {Colman}}, \bibinfo {author}
  {\bibfnamefont {R.~D.}\ \bibnamefont {McDonald}}, \bibinfo {author}
  {\bibfnamefont {M.~J.}\ \bibnamefont {Rosseinsky}}, \bibinfo {author}
  {\bibfnamefont {K.}~\bibnamefont {Prassides}}, \ and\ \bibinfo {author}
  {\bibfnamefont {Y.}~\bibnamefont {Iwasa}},\ }\bibfield  {title} {\enquote
  {\bibinfo {title} {{Upper critical field reaches 90 tesla near the Mott
  transition in fulleride superconductors}},}\ }\href@noop {} {\bibfield
  {journal} {\bibinfo  {journal} {Nat. Commun.}\ }\textbf {\bibinfo {volume}
  {8}},\ \bibinfo {pages} {14467} (\bibinfo {year} {2017})}\BibitemShut
  {NoStop}%
\bibitem [{\citenamefont {Nava}\ \emph {et~al.}(2018)\citenamefont {Nava},
  \citenamefont {Giannetti}, \citenamefont {Georges}, \citenamefont {Tosatti},\
  and\ \citenamefont {Fabrizio}}]{Nava2017}%
  \BibitemOpen
  \bibfield  {author} {\bibinfo {author} {\bibfnamefont {A.}~\bibnamefont
  {Nava}}, \bibinfo {author} {\bibfnamefont {C.}~\bibnamefont {Giannetti}},
  \bibinfo {author} {\bibfnamefont {A.}~\bibnamefont {Georges}}, \bibinfo
  {author} {\bibfnamefont {E.}~\bibnamefont {Tosatti}}, \ and\ \bibinfo
  {author} {\bibfnamefont {M.}~\bibnamefont {Fabrizio}},\ }\bibfield  {title}
  {\enquote {\bibinfo {title} {{Cooling quasiparticles in A$_3$C$_{60}$
  fullerides by excitonic mid-infrared absorption}},}\ }\href {\doibase
  doi:10.1038/nphys4288} {\bibfield  {journal} {\bibinfo  {journal} {Nat.
  Phys.}\ }\textbf {\bibinfo {volume} {14}},\ \bibinfo {pages} {154} (\bibinfo
  {year} {2018})}\BibitemShut {NoStop}%
\bibitem [{\citenamefont {Margadonna}\ \emph {et~al.}(2001)\citenamefont
  {Margadonna}, \citenamefont {Prassides}, \citenamefont {Shimoda},
  \citenamefont {Takenobu},\ and\ \citenamefont {Iwasa}}]{Margadonna2001}%
  \BibitemOpen
  \bibfield  {author} {\bibinfo {author} {\bibfnamefont {S.}~\bibnamefont
  {Margadonna}}, \bibinfo {author} {\bibfnamefont {K.}~\bibnamefont
  {Prassides}}, \bibinfo {author} {\bibfnamefont {H.}~\bibnamefont {Shimoda}},
  \bibinfo {author} {\bibfnamefont {T.}~\bibnamefont {Takenobu}}, \ and\
  \bibinfo {author} {\bibfnamefont {Y.}~\bibnamefont {Iwasa}},\ }\bibfield
  {title} {\enquote {\bibinfo {title} {{Orientational ordering of
  ${\mathrm{C}}_{60}$ in the antiferromagnetic
  $({\mathrm{NH}}_{3}){\mathrm{K}}_{3}{\mathrm{C}}_{60}$ phase}},}\ }\href
  {\doibase 10.1103/PhysRevB.64.132414} {\bibfield  {journal} {\bibinfo
  {journal} {Phys. Rev. B}\ }\textbf {\bibinfo {volume} {64}},\ \bibinfo
  {pages} {132414} (\bibinfo {year} {2001})}\BibitemShut {NoStop}%
\bibitem [{\citenamefont {Durand}\ \emph {et~al.}(2003)\citenamefont {Durand},
  \citenamefont {Darling}, \citenamefont {Dubitsky}, \citenamefont {Zaopo},\
  and\ \citenamefont {Rosseinsky}}]{Durand2003}%
  \BibitemOpen
  \bibfield  {author} {\bibinfo {author} {\bibfnamefont {P.}~\bibnamefont
  {Durand}}, \bibinfo {author} {\bibfnamefont {G.~R.}\ \bibnamefont {Darling}},
  \bibinfo {author} {\bibfnamefont {Y.}~\bibnamefont {Dubitsky}}, \bibinfo
  {author} {\bibfnamefont {A.}~\bibnamefont {Zaopo}}, \ and\ \bibinfo {author}
  {\bibfnamefont {M.~J.}\ \bibnamefont {Rosseinsky}},\ }\bibfield  {title}
  {\enquote {\bibinfo {title} {{The Mott-Hubbard insulating state and orbital
  degeneracy in the superconducting C$_{60}^{3-}$ fulleride family}},}\
  }\href@noop {} {\bibfield  {journal} {\bibinfo  {journal} {Nat. Mater.}\
  }\textbf {\bibinfo {volume} {2}},\ \bibinfo {pages} {605} (\bibinfo {year}
  {2003})}\BibitemShut {NoStop}%
\bibitem [{\citenamefont {Chibotaru}(2005)}]{Chibotaru2005}%
  \BibitemOpen
  \bibfield  {author} {\bibinfo {author} {\bibfnamefont {L.~F.}\ \bibnamefont
  {Chibotaru}},\ }\bibfield  {title} {\enquote {\bibinfo {title}
  {{Spin-Vibronic Superexchange in Mott-Hubbard Fullerides}},}\ }\href
  {\doibase 10.1103/PhysRevLett.94.186405} {\bibfield  {journal} {\bibinfo
  {journal} {Phys. Rev. Lett.}\ }\textbf {\bibinfo {volume} {94}},\ \bibinfo
  {pages} {186405} (\bibinfo {year} {2005})}\BibitemShut {NoStop}%
\bibitem [{\citenamefont {Allemand}\ \emph {et~al.}(1991)\citenamefont
  {Allemand}, \citenamefont {Khemani}, \citenamefont {Koch}, \citenamefont
  {Wudl}, \citenamefont {Holczer}, \citenamefont {Donovan}, \citenamefont
  {Gr\"{u}ner},\ and\ \citenamefont {Thompson}}]{Allemand1991}%
  \BibitemOpen
  \bibfield  {author} {\bibinfo {author} {\bibfnamefont {P.~M.}\ \bibnamefont
  {Allemand}}, \bibinfo {author} {\bibfnamefont {K.~C.}\ \bibnamefont
  {Khemani}}, \bibinfo {author} {\bibfnamefont {A.}~\bibnamefont {Koch}},
  \bibinfo {author} {\bibfnamefont {F.}~\bibnamefont {Wudl}}, \bibinfo {author}
  {\bibfnamefont {K.}~\bibnamefont {Holczer}}, \bibinfo {author} {\bibfnamefont
  {S.}~\bibnamefont {Donovan}}, \bibinfo {author} {\bibfnamefont
  {G.}~\bibnamefont {Gr\"{u}ner}}, \ and\ \bibinfo {author} {\bibfnamefont
  {J.~D.}\ \bibnamefont {Thompson}},\ }\bibfield  {title} {\enquote {\bibinfo
  {title} {{Organic molecular soft ferromagnetism in a fullerene C$_{60}$}},}\
  }\href@noop {} {\bibfield  {journal} {\bibinfo  {journal} {Science}\ }\textbf
  {\bibinfo {volume} {253}},\ \bibinfo {pages} {301} (\bibinfo {year}
  {1991})}\BibitemShut {NoStop}%
\bibitem [{\citenamefont {Kawamoto}(1997)}]{Kawamoto1997}%
  \BibitemOpen
  \bibfield  {author} {\bibinfo {author} {\bibfnamefont {T.}~\bibnamefont
  {Kawamoto}},\ }\bibfield  {title} {\enquote {\bibinfo {title} {{A theoretical
  model for ferromagnetism of TDAE-C$_{60}$}},}\ }\href {\doibase
  http://dx.doi.org/10.1016/S0038-1098(96)00584-4} {\bibfield  {journal}
  {\bibinfo  {journal} {Solid State Commun.}\ }\textbf {\bibinfo {volume}
  {101}},\ \bibinfo {pages} {231} (\bibinfo {year} {1997})}\BibitemShut
  {NoStop}%
\bibitem [{\citenamefont {Sato}\ \emph {et~al.}(1997)\citenamefont {Sato},
  \citenamefont {Yamabe},\ and\ \citenamefont {Tanaka}}]{Sato1997}%
  \BibitemOpen
  \bibfield  {author} {\bibinfo {author} {\bibfnamefont {T.}~\bibnamefont
  {Sato}}, \bibinfo {author} {\bibfnamefont {T.}~\bibnamefont {Yamabe}}, \ and\
  \bibinfo {author} {\bibfnamefont {K.}~\bibnamefont {Tanaka}},\ }\bibfield
  {title} {\enquote {\bibinfo {title} {Magnetic ordering in fullerene
  charge-transfer complexes},}\ }\href {\doibase 10.1103/PhysRevB.56.307}
  {\bibfield  {journal} {\bibinfo  {journal} {Phys. Rev. B}\ }\textbf {\bibinfo
  {volume} {56}},\ \bibinfo {pages} {307} (\bibinfo {year} {1997})}\BibitemShut
  {NoStop}%
\bibitem [{\citenamefont {Kambe}\ \emph {et~al.}(2007)\citenamefont {Kambe},
  \citenamefont {Kajiyoshi}, \citenamefont {Fujiwara},\ and\ \citenamefont
  {Oshima}}]{Kambe2007}%
  \BibitemOpen
  \bibfield  {author} {\bibinfo {author} {\bibfnamefont {T.}~\bibnamefont
  {Kambe}}, \bibinfo {author} {\bibfnamefont {K.}~\bibnamefont {Kajiyoshi}},
  \bibinfo {author} {\bibfnamefont {M.}~\bibnamefont {Fujiwara}}, \ and\
  \bibinfo {author} {\bibfnamefont {K.}~\bibnamefont {Oshima}},\ }\bibfield
  {title} {\enquote {\bibinfo {title} {{Antiferromagnetic Ordering Driven by
  the Molecular Orbital Order of ${\mathrm{C}}_{60}$ in
  ${\ensuremath{\alpha}}^{\ensuremath{'}}$-Tetra-Kis-(Dimethylamino)-Ethylene-${\mathrm{C}}_{60}$}},}\
  }\href {\doibase 10.1103/PhysRevLett.99.177205} {\bibfield  {journal}
  {\bibinfo  {journal} {Phys. Rev. Lett.}\ }\textbf {\bibinfo {volume} {99}},\
  \bibinfo {pages} {177205} (\bibinfo {year} {2007})}\BibitemShut {NoStop}%
\bibitem [{\citenamefont {Amsharov}\ \emph {et~al.}(2011)\citenamefont
  {Amsharov}, \citenamefont {Kr\"{a}mer},\ and\ \citenamefont
  {Jansen}}]{Amsharov2011}%
  \BibitemOpen
  \bibfield  {author} {\bibinfo {author} {\bibfnamefont {K.~Yu.}\ \bibnamefont
  {Amsharov}}, \bibinfo {author} {\bibfnamefont {Y.}~\bibnamefont
  {Kr\"{a}mer}}, \ and\ \bibinfo {author} {\bibfnamefont {M.}~\bibnamefont
  {Jansen}},\ }\bibfield  {title} {\enquote {\bibinfo {title} {{Direct
  Observation of the Transition from Static to Dynamic Jahn-Teller Effects in
  the [Cs(THF)$_4$]C$_{60}$ Fulleride}},}\ }\href {\doibase
  10.1002/anie.201105360} {\bibfield  {journal} {\bibinfo  {journal} {Angew.
  Chem. Int. Ed.}\ }\textbf {\bibinfo {volume} {50}},\ \bibinfo {pages} {11640}
  (\bibinfo {year} {2011})}\BibitemShut {NoStop}%
\bibitem [{\citenamefont {Francis}\ \emph {et~al.}(2012)\citenamefont
  {Francis}, \citenamefont {Scharinger}, \citenamefont {N\'emeth},
  \citenamefont {Kamar\'as},\ and\ \citenamefont {Kuntscher}}]{Francis2012}%
  \BibitemOpen
  \bibfield  {author} {\bibinfo {author} {\bibfnamefont {E.~A.}\ \bibnamefont
  {Francis}}, \bibinfo {author} {\bibfnamefont {S.}~\bibnamefont {Scharinger}},
  \bibinfo {author} {\bibfnamefont {K.}~\bibnamefont {N\'emeth}}, \bibinfo
  {author} {\bibfnamefont {K.}~\bibnamefont {Kamar\'as}}, \ and\ \bibinfo
  {author} {\bibfnamefont {C.~A.}\ \bibnamefont {Kuntscher}},\ }\bibfield
  {title} {\enquote {\bibinfo {title} {{Pressure-induced transition from the
  dynamic to static Jahn-Teller effect in (Ph${}_{4}$P)${}_{2}$IC${}_{60}$}},}\
  }\href {\doibase 10.1103/PhysRevB.85.195428} {\bibfield  {journal} {\bibinfo
  {journal} {Phys. Rev. B}\ }\textbf {\bibinfo {volume} {85}},\ \bibinfo
  {pages} {195428} (\bibinfo {year} {2012})}\BibitemShut {NoStop}%
\bibitem [{\citenamefont {Konarev}\ \emph {et~al.}(2013)\citenamefont
  {Konarev}, \citenamefont {Kuzmin}, \citenamefont {Simonov}, \citenamefont
  {Yudanova}, \citenamefont {Khasanov}, \citenamefont {Saito},\ and\
  \citenamefont {Lyubovskaya}}]{Konarev2013}%
  \BibitemOpen
  \bibfield  {author} {\bibinfo {author} {\bibfnamefont {D.~V.}\ \bibnamefont
  {Konarev}}, \bibinfo {author} {\bibfnamefont {A.~V.}\ \bibnamefont {Kuzmin}},
  \bibinfo {author} {\bibfnamefont {S.~V.}\ \bibnamefont {Simonov}}, \bibinfo
  {author} {\bibfnamefont {E.~I.}\ \bibnamefont {Yudanova}}, \bibinfo {author}
  {\bibfnamefont {S.~S.}\ \bibnamefont {Khasanov}}, \bibinfo {author}
  {\bibfnamefont {G.}~\bibnamefont {Saito}}, \ and\ \bibinfo {author}
  {\bibfnamefont {R.~N.}\ \bibnamefont {Lyubovskaya}},\ }\bibfield  {title}
  {\enquote {\bibinfo {title} {{Experimental observation of C$_{60}$ LUMO
  splitting in the C$_{60}^{2-}$ dianions due to the Jahn-Teller effect.
  Comparison with the C$_{60}^{\cdot -}$ radical anions}},}\ }\href {\doibase
  10.1039/C3CP44359K} {\bibfield  {journal} {\bibinfo  {journal} {Phys. Chem.
  Chem. Phys.}\ }\textbf {\bibinfo {volume} {15}},\ \bibinfo {pages} {9136}
  (\bibinfo {year} {2013})}\BibitemShut {NoStop}%
\bibitem [{\citenamefont {Auerbach}\ \emph {et~al.}(1994)\citenamefont
  {Auerbach}, \citenamefont {Manini},\ and\ \citenamefont
  {Tosatti}}]{Auerbach1994}%
  \BibitemOpen
  \bibfield  {author} {\bibinfo {author} {\bibfnamefont {A.}~\bibnamefont
  {Auerbach}}, \bibinfo {author} {\bibfnamefont {N.}~\bibnamefont {Manini}}, \
  and\ \bibinfo {author} {\bibfnamefont {E.}~\bibnamefont {Tosatti}},\
  }\bibfield  {title} {\enquote {\bibinfo {title} {{Electron-vibron
  interactions in charged fullerenes. I. Berry phases}},}\ }\href {\doibase
  10.1103/PhysRevB.49.12998} {\bibfield  {journal} {\bibinfo  {journal} {Phys.
  Rev. B}\ }\textbf {\bibinfo {volume} {49}},\ \bibinfo {pages} {12998}
  (\bibinfo {year} {1994})}\BibitemShut {NoStop}%
\bibitem [{\citenamefont {Manini}\ \emph {et~al.}(1994)\citenamefont {Manini},
  \citenamefont {Tosatti},\ and\ \citenamefont {Auerbach}}]{Manini1994}%
  \BibitemOpen
  \bibfield  {author} {\bibinfo {author} {\bibfnamefont {N.}~\bibnamefont
  {Manini}}, \bibinfo {author} {\bibfnamefont {E.}~\bibnamefont {Tosatti}}, \
  and\ \bibinfo {author} {\bibfnamefont {A.}~\bibnamefont {Auerbach}},\
  }\bibfield  {title} {\enquote {\bibinfo {title} {{Electron-vibron
  interactions in charged fullerenes. II. Pair energies and spectra}},}\ }\href
  {\doibase 10.1103/PhysRevB.49.13008} {\bibfield  {journal} {\bibinfo
  {journal} {Phys. Rev. B}\ }\textbf {\bibinfo {volume} {49}},\ \bibinfo
  {pages} {13008} (\bibinfo {year} {1994})}\BibitemShut {NoStop}%
\bibitem [{\citenamefont {O'Brien}(1996)}]{OBrien1996}%
  \BibitemOpen
  \bibfield  {author} {\bibinfo {author} {\bibfnamefont {M.~C.~M.}\
  \bibnamefont {O'Brien}},\ }\bibfield  {title} {\enquote {\bibinfo {title}
  {{Vibronic energies in
  ${\mathrm{C}}_{60}$${\mathrm{}}^{\mathit{n}\mathrm{-}}$ and the Jahn-Teller
  effect}},}\ }\href {\doibase 10.1103/PhysRevB.53.3775} {\bibfield  {journal}
  {\bibinfo  {journal} {Phys. Rev. B}\ }\textbf {\bibinfo {volume} {53}},\
  \bibinfo {pages} {3775} (\bibinfo {year} {1996})}\BibitemShut {NoStop}%
\bibitem [{\citenamefont {Chancey}\ and\ \citenamefont
  {O'Brien}(1997)}]{Chancey1997}%
  \BibitemOpen
  \bibfield  {author} {\bibinfo {author} {\bibfnamefont {C.~C.}\ \bibnamefont
  {Chancey}}\ and\ \bibinfo {author} {\bibfnamefont {M.~C.~M.}\ \bibnamefont
  {O'Brien}},\ }\href@noop {} {\emph {\bibinfo {title} {The Jahn--Teller Effect
  in C$_{60}$ and Other Icosahedral Complexes}}}\ (\bibinfo  {publisher}
  {Princeton University Press},\ \bibinfo {address} {Princeton},\ \bibinfo
  {year} {1997})\BibitemShut {NoStop}%
\bibitem [{\citenamefont {Gunnarsson}\ \emph {et~al.}(1995)\citenamefont
  {Gunnarsson}, \citenamefont {Handschuh}, \citenamefont {Bechthold},
  \citenamefont {Kessler}, \citenamefont {Gantef\"{o}r},\ and\ \citenamefont
  {Eberhardt}}]{Gunnarsson1995}%
  \BibitemOpen
  \bibfield  {author} {\bibinfo {author} {\bibfnamefont {O.}~\bibnamefont
  {Gunnarsson}}, \bibinfo {author} {\bibfnamefont {H.}~\bibnamefont
  {Handschuh}}, \bibinfo {author} {\bibfnamefont {P.~S.}\ \bibnamefont
  {Bechthold}}, \bibinfo {author} {\bibfnamefont {B.}~\bibnamefont {Kessler}},
  \bibinfo {author} {\bibfnamefont {G.}~\bibnamefont {Gantef\"{o}r}}, \ and\
  \bibinfo {author} {\bibfnamefont {W.}~\bibnamefont {Eberhardt}},\ }\bibfield
  {title} {\enquote {\bibinfo {title} {{Photoemission Spectra of C$_{60}^-$:
  Electron-Phonon Coupling, Jahn-Teller Effect, and Superconductivity in the
  Fullerides}},}\ }\href {\doibase 10.1103/PhysRevLett.74.1875} {\bibfield
  {journal} {\bibinfo  {journal} {Phys. Rev. Lett.}\ }\textbf {\bibinfo
  {volume} {74}},\ \bibinfo {pages} {1875} (\bibinfo {year}
  {1995})}\BibitemShut {NoStop}%
\bibitem [{\citenamefont {Winter}\ and\ \citenamefont
  {Kuzmany}(1996)}]{Winter1996}%
  \BibitemOpen
  \bibfield  {author} {\bibinfo {author} {\bibfnamefont {J.}~\bibnamefont
  {Winter}}\ and\ \bibinfo {author} {\bibfnamefont {H.}~\bibnamefont
  {Kuzmany}},\ }\bibfield  {title} {\enquote {\bibinfo {title} {Landau damping
  and lifting of vibrational degeneracy in metallic potassium fulleride},}\
  }\href {\doibase 10.1103/PhysRevB.53.655} {\bibfield  {journal} {\bibinfo
  {journal} {Phys. Rev. B}\ }\textbf {\bibinfo {volume} {53}},\ \bibinfo
  {pages} {655} (\bibinfo {year} {1996})}\BibitemShut {NoStop}%
\bibitem [{\citenamefont {Hands}\ \emph {et~al.}(2008)\citenamefont {Hands},
  \citenamefont {Dunn}, \citenamefont {Bates}, \citenamefont {Hope},
  \citenamefont {Meech},\ and\ \citenamefont {Andrews}}]{Hands2008}%
  \BibitemOpen
  \bibfield  {author} {\bibinfo {author} {\bibfnamefont {I.~D.}\ \bibnamefont
  {Hands}}, \bibinfo {author} {\bibfnamefont {J.~L.}\ \bibnamefont {Dunn}},
  \bibinfo {author} {\bibfnamefont {C.~A.}\ \bibnamefont {Bates}}, \bibinfo
  {author} {\bibfnamefont {M.~J.}\ \bibnamefont {Hope}}, \bibinfo {author}
  {\bibfnamefont {S.~R.}\ \bibnamefont {Meech}}, \ and\ \bibinfo {author}
  {\bibfnamefont {D.~L.}\ \bibnamefont {Andrews}},\ }\bibfield  {title}
  {\enquote {\bibinfo {title} {{Vibronic interactions in the visible and
  near-infrared spectra of ${\mathrm{C}}_{60}^{\ensuremath{-}}$ anions}},}\
  }\href {\doibase 10.1103/PhysRevB.77.115445} {\bibfield  {journal} {\bibinfo
  {journal} {Phys. Rev. B}\ }\textbf {\bibinfo {volume} {77}},\ \bibinfo
  {pages} {115445} (\bibinfo {year} {2008})}\BibitemShut {NoStop}%
\bibitem [{\citenamefont {Varma}\ \emph {et~al.}(1991)\citenamefont {Varma},
  \citenamefont {Zaanen},\ and\ \citenamefont {Raghavachari}}]{Varma1991}%
  \BibitemOpen
  \bibfield  {author} {\bibinfo {author} {\bibfnamefont {C.~M.}\ \bibnamefont
  {Varma}}, \bibinfo {author} {\bibfnamefont {J.}~\bibnamefont {Zaanen}}, \
  and\ \bibinfo {author} {\bibfnamefont {K.}~\bibnamefont {Raghavachari}},\
  }\bibfield  {title} {\enquote {\bibinfo {title} {{Superconductivity in the
  Fullerenes}},}\ }\href@noop {} {\bibfield  {journal} {\bibinfo  {journal}
  {Science}\ }\textbf {\bibinfo {volume} {254}},\ \bibinfo {pages} {989}
  (\bibinfo {year} {1991})}\BibitemShut {NoStop}%
\bibitem [{\citenamefont {Schluter}\ \emph {et~al.}(1992)\citenamefont
  {Schluter}, \citenamefont {Lannoo}, \citenamefont {Needels}, \citenamefont
  {Baraff},\ and\ \citenamefont {Tom\'anek}}]{Schluter1992}%
  \BibitemOpen
  \bibfield  {author} {\bibinfo {author} {\bibfnamefont {M.}~\bibnamefont
  {Schluter}}, \bibinfo {author} {\bibfnamefont {M.}~\bibnamefont {Lannoo}},
  \bibinfo {author} {\bibfnamefont {M.}~\bibnamefont {Needels}}, \bibinfo
  {author} {\bibfnamefont {G.~A.}\ \bibnamefont {Baraff}}, \ and\ \bibinfo
  {author} {\bibfnamefont {D.}~\bibnamefont {Tom\'anek}},\ }\bibfield  {title}
  {\enquote {\bibinfo {title} {{Electron-phonon coupling and superconductivity
  in alkali-intercalated ${\mathrm{C}}_{60}$ solid}},}\ }\href {\doibase
  10.1103/PhysRevLett.68.526} {\bibfield  {journal} {\bibinfo  {journal} {Phys.
  Rev. Lett.}\ }\textbf {\bibinfo {volume} {68}},\ \bibinfo {pages} {526}
  (\bibinfo {year} {1992})}\BibitemShut {NoStop}%
\bibitem [{\citenamefont {Faulhaber}\ \emph {et~al.}(1993)\citenamefont
  {Faulhaber}, \citenamefont {Ko},\ and\ \citenamefont
  {Briddon}}]{Faulhaber1993}%
  \BibitemOpen
  \bibfield  {author} {\bibinfo {author} {\bibfnamefont {J.~C.~R.}\
  \bibnamefont {Faulhaber}}, \bibinfo {author} {\bibfnamefont {D.~Y.~K.}\
  \bibnamefont {Ko}}, \ and\ \bibinfo {author} {\bibfnamefont {P.~R.}\
  \bibnamefont {Briddon}},\ }\bibfield  {title} {\enquote {\bibinfo {title}
  {{Vibronic coupling in ${\mathrm{C}}_{60}$ and
  ${\mathrm{C}}_{60}^{3\mathrm{\ensuremath{-}}}$}},}\ }\href {\doibase
  10.1103/PhysRevB.48.661} {\bibfield  {journal} {\bibinfo  {journal} {Phys.
  Rev. B}\ }\textbf {\bibinfo {volume} {48}},\ \bibinfo {pages} {661} (\bibinfo
  {year} {1993})}\BibitemShut {NoStop}%
\bibitem [{\citenamefont {Antropov}\ \emph {et~al.}(1993)\citenamefont
  {Antropov}, \citenamefont {Gunnarsson},\ and\ \citenamefont
  {Liechtenstein}}]{Antropov1993}%
  \BibitemOpen
  \bibfield  {author} {\bibinfo {author} {\bibfnamefont {V.~P.}\ \bibnamefont
  {Antropov}}, \bibinfo {author} {\bibfnamefont {O.}~\bibnamefont
  {Gunnarsson}}, \ and\ \bibinfo {author} {\bibfnamefont {A.~I.}\ \bibnamefont
  {Liechtenstein}},\ }\bibfield  {title} {\enquote {\bibinfo {title} {{Phonons,
  electron-phonon, and electron-plasmon coupling in ${\mathrm{C}}_{60}$
  compounds}},}\ }\href {\doibase 10.1103/PhysRevB.48.7651} {\bibfield
  {journal} {\bibinfo  {journal} {Phys. Rev. B}\ }\textbf {\bibinfo {volume}
  {48}},\ \bibinfo {pages} {7651} (\bibinfo {year} {1993})}\BibitemShut
  {NoStop}%
\bibitem [{\citenamefont {Breda}\ \emph {et~al.}(1998)\citenamefont {Breda},
  \citenamefont {Broglia}, \citenamefont {Col\`{o}}, \citenamefont {Roman},
  \citenamefont {Alasia}, \citenamefont {Onida}, \citenamefont {Ponomarev},\
  and\ \citenamefont {Vigezzi}}]{Breda1998}%
  \BibitemOpen
  \bibfield  {author} {\bibinfo {author} {\bibfnamefont {N.}~\bibnamefont
  {Breda}}, \bibinfo {author} {\bibfnamefont {R.~A.}\ \bibnamefont {Broglia}},
  \bibinfo {author} {\bibfnamefont {G.}~\bibnamefont {Col\`{o}}}, \bibinfo
  {author} {\bibfnamefont {H.~E.}\ \bibnamefont {Roman}}, \bibinfo {author}
  {\bibfnamefont {F.}~\bibnamefont {Alasia}}, \bibinfo {author} {\bibfnamefont
  {G.}~\bibnamefont {Onida}}, \bibinfo {author} {\bibfnamefont
  {V.}~\bibnamefont {Ponomarev}}, \ and\ \bibinfo {author} {\bibfnamefont
  {E.}~\bibnamefont {Vigezzi}},\ }\bibfield  {title} {\enquote {\bibinfo
  {title} {Electron–phonon coupling in charged buckminsterfullerene},}\
  }\href {\doibase https://doi.org/10.1016/S0009-2614(98)00131-6} {\bibfield
  {journal} {\bibinfo  {journal} {Chem. Phys. Lett.}\ }\textbf {\bibinfo
  {volume} {286}},\ \bibinfo {pages} {350} (\bibinfo {year}
  {1998})}\BibitemShut {NoStop}%
\bibitem [{\citenamefont {Manini}\ \emph {et~al.}(2001)\citenamefont {Manini},
  \citenamefont {Corso}, \citenamefont {Fabrizio},\ and\ \citenamefont
  {Tosatti}}]{Manini2001}%
  \BibitemOpen
  \bibfield  {author} {\bibinfo {author} {\bibfnamefont {N.}~\bibnamefont
  {Manini}}, \bibinfo {author} {\bibfnamefont {A.~Dal}\ \bibnamefont {Corso}},
  \bibinfo {author} {\bibfnamefont {M.}~\bibnamefont {Fabrizio}}, \ and\
  \bibinfo {author} {\bibfnamefont {E.}~\bibnamefont {Tosatti}},\ }\bibfield
  {title} {\enquote {\bibinfo {title} {Electron-vibration coupling constants in
  positively charged fullerene},}\ }\href {\doibase 10.1080/13642810110062663}
  {\bibfield  {journal} {\bibinfo  {journal} {Phil. Mag. B}\ }\textbf {\bibinfo
  {volume} {81}},\ \bibinfo {pages} {793} (\bibinfo {year} {2001})}\BibitemShut
  {NoStop}%
\bibitem [{\citenamefont {Saito}(2002)}]{Saito2002}%
  \BibitemOpen
  \bibfield  {author} {\bibinfo {author} {\bibfnamefont {M.}~\bibnamefont
  {Saito}},\ }\bibfield  {title} {\enquote {\bibinfo {title} {{Electron-phonon
  coupling of electron- or hole-injected ${\mathrm{C}}_{60}$}},}\ }\href
  {\doibase 10.1103/PhysRevB.65.220508} {\bibfield  {journal} {\bibinfo
  {journal} {Phys. Rev. B}\ }\textbf {\bibinfo {volume} {65}},\ \bibinfo
  {pages} {220508} (\bibinfo {year} {2002})}\BibitemShut {NoStop}%
\bibitem [{\citenamefont {Frederiksen}\ \emph {et~al.}(2008)\citenamefont
  {Frederiksen}, \citenamefont {Franke}, \citenamefont {Arnau}, \citenamefont
  {Schulze}, \citenamefont {Pascual},\ and\ \citenamefont
  {Lorente}}]{Frederiksen2008}%
  \BibitemOpen
  \bibfield  {author} {\bibinfo {author} {\bibfnamefont {T.}~\bibnamefont
  {Frederiksen}}, \bibinfo {author} {\bibfnamefont {K.~J.}\ \bibnamefont
  {Franke}}, \bibinfo {author} {\bibfnamefont {A.}~\bibnamefont {Arnau}},
  \bibinfo {author} {\bibfnamefont {G.}~\bibnamefont {Schulze}}, \bibinfo
  {author} {\bibfnamefont {J.~I.}\ \bibnamefont {Pascual}}, \ and\ \bibinfo
  {author} {\bibfnamefont {N.}~\bibnamefont {Lorente}},\ }\bibfield  {title}
  {\enquote {\bibinfo {title} {Dynamic jahn-teller effect in electronic
  transport through single ${\text{c}}_{60}$ molecules},}\ }\href {\doibase
  10.1103/PhysRevB.78.233401} {\bibfield  {journal} {\bibinfo  {journal} {Phys.
  Rev. B}\ }\textbf {\bibinfo {volume} {78}},\ \bibinfo {pages} {233401}
  (\bibinfo {year} {2008})}\BibitemShut {NoStop}%
\bibitem [{\citenamefont {Laflamme~Janssen}\ \emph {et~al.}(2010)\citenamefont
  {Laflamme~Janssen}, \citenamefont {C\^ot\'e}, \citenamefont {Louie},\ and\
  \citenamefont {Cohen}}]{LaflammeJanssen2010}%
  \BibitemOpen
  \bibfield  {author} {\bibinfo {author} {\bibfnamefont {J.}~\bibnamefont
  {Laflamme~Janssen}}, \bibinfo {author} {\bibfnamefont {M.}~\bibnamefont
  {C\^ot\'e}}, \bibinfo {author} {\bibfnamefont {S.~G.}\ \bibnamefont {Louie}},
  \ and\ \bibinfo {author} {\bibfnamefont {M.~L.}\ \bibnamefont {Cohen}},\
  }\bibfield  {title} {\enquote {\bibinfo {title} {{Electron-phonon coupling in
  ${\text{C}}_{60}$ using hybrid functionals}},}\ }\href {\doibase
  10.1103/PhysRevB.81.073106} {\bibfield  {journal} {\bibinfo  {journal} {Phys.
  Rev. B}\ }\textbf {\bibinfo {volume} {81}},\ \bibinfo {pages} {073106}
  (\bibinfo {year} {2010})}\BibitemShut {NoStop}%
\bibitem [{\citenamefont {Wang}\ \emph {et~al.}(2005)\citenamefont {Wang},
  \citenamefont {Woo},\ and\ \citenamefont {Wang}}]{Wang2005}%
  \BibitemOpen
  \bibfield  {author} {\bibinfo {author} {\bibfnamefont {X.-B.}\ \bibnamefont
  {Wang}}, \bibinfo {author} {\bibfnamefont {H.-K.}\ \bibnamefont {Woo}}, \
  and\ \bibinfo {author} {\bibfnamefont {L.-S.}\ \bibnamefont {Wang}},\
  }\bibfield  {title} {\enquote {\bibinfo {title} {{Vibrational cooling in a
  cold ion trap: Vibrationally resolved photoelectron spectroscopy of cold
  C$_{60}^-$ anions}},}\ }\href
  {http://scitation.aip.org/content/aip/journal/jcp/123/5/10.1063/1.1998787}
  {\bibfield  {journal} {\bibinfo  {journal} {J. Chem. Phys.}\ }\textbf
  {\bibinfo {volume} {123}},\ \bibinfo {eid} {051106} (\bibinfo {year}
  {2005})}\BibitemShut {NoStop}%
\bibitem [{\citenamefont {Iwahara}\ \emph {et~al.}(2010)\citenamefont
  {Iwahara}, \citenamefont {Sato}, \citenamefont {Tanaka},\ and\ \citenamefont
  {Chibotaru}}]{Iwahara2010}%
  \BibitemOpen
  \bibfield  {author} {\bibinfo {author} {\bibfnamefont {N.}~\bibnamefont
  {Iwahara}}, \bibinfo {author} {\bibfnamefont {T.}~\bibnamefont {Sato}},
  \bibinfo {author} {\bibfnamefont {K.}~\bibnamefont {Tanaka}}, \ and\ \bibinfo
  {author} {\bibfnamefont {L.~F.}\ \bibnamefont {Chibotaru}},\ }\bibfield
  {title} {\enquote {\bibinfo {title} {{Vibronic coupling in
  ${\text{C}}_{60}^{\ensuremath{-}}$ anion revisited: Derivations from
  photoelectron spectra and DFT calculations}},}\ }\href {\doibase
  10.1103/PhysRevB.82.245409} {\bibfield  {journal} {\bibinfo  {journal} {Phys.
  Rev. B}\ }\textbf {\bibinfo {volume} {82}},\ \bibinfo {pages} {245409}
  (\bibinfo {year} {2010})}\BibitemShut {NoStop}%
\bibitem [{\citenamefont {Faber}\ \emph {et~al.}(2011)\citenamefont {Faber},
  \citenamefont {Janssen}, \citenamefont {C{\^o}t{\'e}}, \citenamefont
  {Runge},\ and\ \citenamefont {Blase}}]{Faber2011}%
  \BibitemOpen
  \bibfield  {author} {\bibinfo {author} {\bibfnamefont {C.}~\bibnamefont
  {Faber}}, \bibinfo {author} {\bibfnamefont {J.~L.}\ \bibnamefont {Janssen}},
  \bibinfo {author} {\bibfnamefont {M.}~\bibnamefont {C{\^o}t{\'e}}}, \bibinfo
  {author} {\bibfnamefont {E.}~\bibnamefont {Runge}}, \ and\ \bibinfo {author}
  {\bibfnamefont {X.}~\bibnamefont {Blase}},\ }\bibfield  {title} {\enquote
  {\bibinfo {title} {{Electron-phonon coupling in the C$_{60}$ fullerene within
  the many-body $GW$ approach}},}\ }\href {\doibase 10.1103/PhysRevB.84.155104}
  {\bibfield  {journal} {\bibinfo  {journal} {Phys. Rev. B}\ }\textbf {\bibinfo
  {volume} {84}},\ \bibinfo {pages} {155104} (\bibinfo {year}
  {2011})}\BibitemShut {NoStop}%
\bibitem [{\citenamefont {Jahn}\ and\ \citenamefont {Teller}(1937)}]{Jahn1937}%
  \BibitemOpen
  \bibfield  {author} {\bibinfo {author} {\bibfnamefont {H.~A.}\ \bibnamefont
  {Jahn}}\ and\ \bibinfo {author} {\bibfnamefont {E.}~\bibnamefont {Teller}},\
  }\bibfield  {title} {\enquote {\bibinfo {title} {{Stability of Polyatomic
  Molecules in Degenerate Electronic States. I. Orbital Degeneracy}},}\
  }\href@noop {} {\bibfield  {journal} {\bibinfo  {journal} {Proc. R. Soc.
  Lond. A}\ }\textbf {\bibinfo {volume} {161}},\ \bibinfo {pages} {220}
  (\bibinfo {year} {1937})}\BibitemShut {NoStop}%
\bibitem [{\citenamefont {Bersuker}\ and\ \citenamefont
  {Polinger}(1989)}]{Bersuker1989}%
  \BibitemOpen
  \bibfield  {author} {\bibinfo {author} {\bibfnamefont {I.~B.}\ \bibnamefont
  {Bersuker}}\ and\ \bibinfo {author} {\bibfnamefont {V.~Z.}\ \bibnamefont
  {Polinger}},\ }\href@noop {} {\emph {\bibinfo {title} {Vibronic Interactions
  in Molecules and Crystals}}}\ (\bibinfo  {publisher} {Springer--Verlag},\
  \bibinfo {address} {Berlin},\ \bibinfo {year} {1989})\BibitemShut {NoStop}%
\bibitem [{\citenamefont {Dunn}\ and\ \citenamefont {Bates}(1995)}]{Dunn1995}%
  \BibitemOpen
  \bibfield  {author} {\bibinfo {author} {\bibfnamefont {J.~L.}\ \bibnamefont
  {Dunn}}\ and\ \bibinfo {author} {\bibfnamefont {C.~A.}\ \bibnamefont
  {Bates}},\ }\bibfield  {title} {\enquote {\bibinfo {title} {{Analysis of the
  $T_{1u}\otimes h_{g}$ Jahn-Teller system as a model for ${\mathrm{C}}_{60}$
  molecules}},}\ }\href {\doibase 10.1103/PhysRevB.52.5996} {\bibfield
  {journal} {\bibinfo  {journal} {Phys. Rev. B}\ }\textbf {\bibinfo {volume}
  {52}},\ \bibinfo {pages} {5996} (\bibinfo {year} {1995})}\BibitemShut
  {NoStop}%
\bibitem [{\citenamefont {Alqannas}\ \emph {et~al.}(2013)\citenamefont
  {Alqannas}, \citenamefont {Lakin}, \citenamefont {Farrow},\ and\
  \citenamefont {Dunn}}]{Alqannas2013}%
  \BibitemOpen
  \bibfield  {author} {\bibinfo {author} {\bibfnamefont {H.~S.}\ \bibnamefont
  {Alqannas}}, \bibinfo {author} {\bibfnamefont {A.~J.}\ \bibnamefont {Lakin}},
  \bibinfo {author} {\bibfnamefont {J.~A.}\ \bibnamefont {Farrow}}, \ and\
  \bibinfo {author} {\bibfnamefont {J.~L.}\ \bibnamefont {Dunn}},\ }\bibfield
  {title} {\enquote {\bibinfo {title} {{Interplay between Coulomb and
  Jahn-Teller effects in icosahedral systems with triplet electronic states
  coupled to $h$-type vibrations}},}\ }\href {\doibase
  10.1103/PhysRevB.88.165430} {\bibfield  {journal} {\bibinfo  {journal} {Phys.
  Rev. B}\ }\textbf {\bibinfo {volume} {88}},\ \bibinfo {pages} {165430}
  (\bibinfo {year} {2013})}\BibitemShut {NoStop}%
\bibitem [{\citenamefont {O'Brien}(1971)}]{OBrien1971}%
  \BibitemOpen
  \bibfield  {author} {\bibinfo {author} {\bibfnamefont {M.~C.~M.}\
  \bibnamefont {O'Brien}},\ }\bibfield  {title} {\enquote {\bibinfo {title}
  {{The Jahn-Teller effect in a $p$ state equally coupled to $E_g$ and $T_{2g}$
  vibrations}},}\ }\href@noop {} {\bibfield  {journal} {\bibinfo  {journal} {J.
  Phys. C: Solid State Phys.}\ }\textbf {\bibinfo {volume} {4}},\ \bibinfo
  {pages} {2524} (\bibinfo {year} {1971})}\BibitemShut {NoStop}%
\bibitem [{\citenamefont {O'Brien}(1969)}]{OBrien1969}%
  \BibitemOpen
  \bibfield  {author} {\bibinfo {author} {\bibfnamefont {M.~C.~M.}\
  \bibnamefont {O'Brien}},\ }\bibfield  {title} {\enquote {\bibinfo {title}
  {{Dynamic Jahn-Teller Effect in an Orbital Triplet State Coupled to Both
  ${E}_{g}$ and ${T}_{2g}$ Vibrations}},}\ }\href {\doibase
  10.1103/PhysRev.187.407} {\bibfield  {journal} {\bibinfo  {journal} {Phys.
  Rev.}\ }\textbf {\bibinfo {volume} {187}},\ \bibinfo {pages} {407} (\bibinfo
  {year} {1969})}\BibitemShut {NoStop}%
\bibitem [{\citenamefont {Altmann}\ and\ \citenamefont
  {Herzig}(1994)}]{Altmann1994}%
  \BibitemOpen
  \bibfield  {author} {\bibinfo {author} {\bibfnamefont {S.~L.}\ \bibnamefont
  {Altmann}}\ and\ \bibinfo {author} {\bibfnamefont {P.}~\bibnamefont
  {Herzig}},\ }\href@noop {} {\emph {\bibinfo {title} {{Point-Group Theory
  Tables}}}}\ (\bibinfo  {publisher} {Claredon Press},\ \bibinfo {address}
  {Oxford},\ \bibinfo {year} {1994})\BibitemShut {NoStop}%
\bibitem [{\citenamefont {Condon}\ and\ \citenamefont
  {Shortley}(1951)}]{Condon1953}%
  \BibitemOpen
  \bibfield  {author} {\bibinfo {author} {\bibfnamefont {E.~U.}\ \bibnamefont
  {Condon}}\ and\ \bibinfo {author} {\bibfnamefont {G.~H.}\ \bibnamefont
  {Shortley}},\ }\href@noop {} {\emph {\bibinfo {title} {The Theory of Atomic
  Spectra}}}\ (\bibinfo  {publisher} {Cambridge University Press},\ \bibinfo
  {address} {Cambridge},\ \bibinfo {year} {1951})\BibitemShut {NoStop}%
\bibitem [{\citenamefont {Romestain}\ and\ \citenamefont
  {Merle~d'Aubign\'e}(1971)}]{Romestain1971}%
  \BibitemOpen
  \bibfield  {author} {\bibinfo {author} {\bibfnamefont {R.}~\bibnamefont
  {Romestain}}\ and\ \bibinfo {author} {\bibfnamefont {Y.}~\bibnamefont
  {Merle~d'Aubign\'e}},\ }\bibfield  {title} {\enquote {\bibinfo {title}
  {{Jahn-Teller Effect of an Orbital Triplet Coupled to Both ${E}_{g}$ and
  ${T}_{2g}$ Modes of Vibrations: Symmetry of the Vibronic States}},}\ }\href
  {\doibase 10.1103/PhysRevB.4.4611} {\bibfield  {journal} {\bibinfo  {journal}
  {Phys. Rev. B}\ }\textbf {\bibinfo {volume} {4}},\ \bibinfo {pages} {4611}
  (\bibinfo {year} {1971})}\BibitemShut {NoStop}%
\bibitem [{Note1()}]{Note1}%
  \BibitemOpen
  \bibinfo {note} {Note that the nuclear part $|\chi \rangle $ is not
  normalized, and thus the weights of LS terms in the vibronic state are not
  equal (see also Eq. (\ref {Eq:chi}).}\BibitemShut {Stop}%
\bibitem [{\citenamefont {Ham}(1968)}]{Ham1968}%
  \BibitemOpen
  \bibfield  {author} {\bibinfo {author} {\bibfnamefont {F.~S.}\ \bibnamefont
  {Ham}},\ }\bibfield  {title} {\enquote {\bibinfo {title} {{Effect of Linear
  Jahn-Teller Coupling on Paramagnetic Resonance in a $^{2}E$ State}},}\ }\href
  {\doibase 10.1103/PhysRev.166.307} {\bibfield  {journal} {\bibinfo  {journal}
  {Phys. Rev.}\ }\textbf {\bibinfo {volume} {166}},\ \bibinfo {pages} {307}
  (\bibinfo {year} {1968})}\BibitemShut {NoStop}%
\bibitem [{\citenamefont {Iwahara}(2018)}]{Iwahara2018}%
  \BibitemOpen
  \bibfield  {author} {\bibinfo {author} {\bibfnamefont {N.}~\bibnamefont
  {Iwahara}},\ }\bibfield  {title} {\enquote {\bibinfo {title} {Berry phase of
  adiabatic electronic configurations in fullerene anions},}\ }\href {\doibase
  10.1103/PhysRevB.97.075413} {\bibfield  {journal} {\bibinfo  {journal} {Phys.
  Rev. B}\ }\textbf {\bibinfo {volume} {97}},\ \bibinfo {pages} {075413}
  (\bibinfo {year} {2018})}\BibitemShut {NoStop}%
\bibitem [{\citenamefont {Racah}(1942)}]{Racah1942}%
  \BibitemOpen
  \bibfield  {author} {\bibinfo {author} {\bibfnamefont {G.}~\bibnamefont
  {Racah}},\ }\bibfield  {title} {\enquote {\bibinfo {title} {{Theory of
  Complex Spectra. II}},}\ }\href {\doibase 10.1103/PhysRev.62.438} {\bibfield
  {journal} {\bibinfo  {journal} {Phys. Rev.}\ }\textbf {\bibinfo {volume}
  {62}},\ \bibinfo {pages} {438} (\bibinfo {year} {1942})}\BibitemShut
  {NoStop}%
\bibitem [{\citenamefont {Racah}(1943)}]{Racah1943}%
  \BibitemOpen
  \bibfield  {author} {\bibinfo {author} {\bibfnamefont {G.}~\bibnamefont
  {Racah}},\ }\bibfield  {title} {\enquote {\bibinfo {title} {{Theory of
  Complex Spectra. III}},}\ }\href {\doibase 10.1103/PhysRev.63.367} {\bibfield
   {journal} {\bibinfo  {journal} {Phys. Rev.}\ }\textbf {\bibinfo {volume}
  {63}},\ \bibinfo {pages} {367} (\bibinfo {year} {1943})}\BibitemShut
  {NoStop}%
\bibitem [{\citenamefont {Bethune}\ \emph {et~al.}(1991)\citenamefont
  {Bethune}, \citenamefont {Meijer}, \citenamefont {Tang}, \citenamefont
  {Rosen}, \citenamefont {Golden}, \citenamefont {Seki}, \citenamefont
  {Brown},\ and\ \citenamefont {de~Vries}}]{Bethune1991}%
  \BibitemOpen
  \bibfield  {author} {\bibinfo {author} {\bibfnamefont {D.~S.}\ \bibnamefont
  {Bethune}}, \bibinfo {author} {\bibfnamefont {G.}~\bibnamefont {Meijer}},
  \bibinfo {author} {\bibfnamefont {W.~C.}\ \bibnamefont {Tang}}, \bibinfo
  {author} {\bibfnamefont {H.~J.}\ \bibnamefont {Rosen}}, \bibinfo {author}
  {\bibfnamefont {W.~G.}\ \bibnamefont {Golden}}, \bibinfo {author}
  {\bibfnamefont {H.}~\bibnamefont {Seki}}, \bibinfo {author} {\bibfnamefont
  {C.~A.}\ \bibnamefont {Brown}}, \ and\ \bibinfo {author} {\bibfnamefont
  {M.~S.}\ \bibnamefont {de~Vries}},\ }\bibfield  {title} {\enquote {\bibinfo
  {title} {{Vibrational Raman and infrared spectra of chromatographically
  separated C$_{60}$ and C$_{70}$ fullerene clusters}},}\ }\href {\doibase
  http://dx.doi.org/10.1016/0009-2614(91)90312-W} {\bibfield  {journal}
  {\bibinfo  {journal} {Chem. Phys. Lett.}\ }\textbf {\bibinfo {volume}
  {179}},\ \bibinfo {pages} {181} (\bibinfo {year} {1991})}\BibitemShut
  {NoStop}%
\bibitem [{\citenamefont {Sookhun}\ \emph {et~al.}(2003)\citenamefont
  {Sookhun}, \citenamefont {Dunn},\ and\ \citenamefont {Bates}}]{Sookhun2003}%
  \BibitemOpen
  \bibfield  {author} {\bibinfo {author} {\bibfnamefont {S.}~\bibnamefont
  {Sookhun}}, \bibinfo {author} {\bibfnamefont {J.~L.}\ \bibnamefont {Dunn}}, \
  and\ \bibinfo {author} {\bibfnamefont {C.~A.}\ \bibnamefont {Bates}},\
  }\bibfield  {title} {\enquote {\bibinfo {title} {{Jahn-Teller effects in the
  fullerene anion ${\mathrm{C}}_{60}^{2\ensuremath{-}}$}},}\ }\href {\doibase
  10.1103/PhysRevB.68.235403} {\bibfield  {journal} {\bibinfo  {journal} {Phys.
  Rev. B}\ }\textbf {\bibinfo {volume} {68}},\ \bibinfo {pages} {235403}
  (\bibinfo {year} {2003})}\BibitemShut {NoStop}%
\bibitem [{\citenamefont {Dunn}\ and\ \citenamefont {Li}(2005)}]{Dunn2005}%
  \BibitemOpen
  \bibfield  {author} {\bibinfo {author} {\bibfnamefont {J.~L.}\ \bibnamefont
  {Dunn}}\ and\ \bibinfo {author} {\bibfnamefont {H.}~\bibnamefont {Li}},\
  }\bibfield  {title} {\enquote {\bibinfo {title} {{Jahn-Teller effects in the
  fullerene anion ${\mathrm{C}}_{60}^{3\ensuremath{-}}$}},}\ }\href {\doibase
  10.1103/PhysRevB.71.115411} {\bibfield  {journal} {\bibinfo  {journal} {Phys.
  Rev. B}\ }\textbf {\bibinfo {volume} {71}},\ \bibinfo {pages} {115411}
  (\bibinfo {year} {2005})}\BibitemShut {NoStop}%
\bibitem [{Note2()}]{Note2}%
  \BibitemOpen
  \bibinfo {note} {Note that due to bielectronic interaction the static JT
  energy for $n = 2, 4$ and 3 is slightly smaller than the expected respective
  values $4E_\protect \text {JT}^{(1)}$ and $3E_\protect \text {JT}^{(1)}$,
  where $E_\protect \text {JT}^{(1)}$ is the static JT energy for $n = 1$
  (Table \ref {Table:E}).}\BibitemShut {Stop}%
\bibitem [{SM()}]{SM}%
  \BibitemOpen
  \href@noop {} {}\bibinfo {note} {See for the numerical values of vibronic
  levels and the range of the temperature Supplemental Materials at
  [URL].}\BibitemShut {Stop}%
\bibitem [{\citenamefont {Tomita}\ \emph {et~al.}(2005)\citenamefont {Tomita},
  \citenamefont {Andersen}, \citenamefont {Bonderup}, \citenamefont
  {Hvelplund}, \citenamefont {Liu}, \citenamefont {Nielsen}, \citenamefont
  {Pedersen}, \citenamefont {Rangama}, \citenamefont {Hansen},\ and\
  \citenamefont {Echt}}]{Tomita2005}%
  \BibitemOpen
  \bibfield  {author} {\bibinfo {author} {\bibfnamefont {S.}~\bibnamefont
  {Tomita}}, \bibinfo {author} {\bibfnamefont {J.~U.}\ \bibnamefont
  {Andersen}}, \bibinfo {author} {\bibfnamefont {E.}~\bibnamefont {Bonderup}},
  \bibinfo {author} {\bibfnamefont {P.}~\bibnamefont {Hvelplund}}, \bibinfo
  {author} {\bibfnamefont {B.}~\bibnamefont {Liu}}, \bibinfo {author}
  {\bibfnamefont {S.~B.}\ \bibnamefont {Nielsen}}, \bibinfo {author}
  {\bibfnamefont {U.~V.}\ \bibnamefont {Pedersen}}, \bibinfo {author}
  {\bibfnamefont {J.}~\bibnamefont {Rangama}}, \bibinfo {author} {\bibfnamefont
  {K.}~\bibnamefont {Hansen}}, \ and\ \bibinfo {author} {\bibfnamefont
  {O.}~\bibnamefont {Echt}},\ }\bibfield  {title} {\enquote {\bibinfo {title}
  {{Dynamic Jahn-Teller Effects in Isolated ${\mathrm{C}}_{60}^{-}$ Studied by
  Near-Infrared Spectroscopy in a Storage Ring}},}\ }\href {\doibase
  10.1103/PhysRevLett.94.053002} {\bibfield  {journal} {\bibinfo  {journal}
  {Phys. Rev. Lett.}\ }\textbf {\bibinfo {volume} {94}},\ \bibinfo {pages}
  {053002} (\bibinfo {year} {2005})}\BibitemShut {NoStop}%
\bibitem [{Note3()}]{Note3}%
  \BibitemOpen
  \bibinfo {note} {The vibronic level with $J=3$ splits into $T_{2u}$ and $G_u$
  levels \cite {Altmann1994} due to weak higher order vibronic coupling.
  Although the side band is attributed to the ground $T_{1u}$ to the $T_{2u}$
  excitations, all the quasi degenerate levels $(J = 3,2,1)$ including the
  $T_{2u}$ vibronic level are populated and contribute to the side
  band.}\BibitemShut {Stop}%
\bibitem [{Note4()}]{Note4}%
  \BibitemOpen
  \bibinfo {note} {The bielectronic energy for the ground adiabatic state is
  smaller than for vibronic ground state (Table \ref {Table:E}) because the JT
  dynamics contribute to a stronger mixing of the electronic terms of a given
  spin multiplicity.}\BibitemShut {Stop}%
\bibitem [{\citenamefont {Tomita}\ \emph {et~al.}(2006)\citenamefont {Tomita},
  \citenamefont {Andersen}, \citenamefont {Cederquist}, \citenamefont
  {Concina}, \citenamefont {Echt}, \citenamefont {Forster}, \citenamefont
  {Hansen}, \citenamefont {Huber}, \citenamefont {Hvelplund}, \citenamefont
  {Jensen}, \citenamefont {Liu}, \citenamefont {Manil}, \citenamefont
  {Maunoury}, \citenamefont {Nielsen}, \citenamefont {Rangama}, \citenamefont
  {Schmidt},\ and\ \citenamefont {Zettergren}}]{Tomita2006}%
  \BibitemOpen
  \bibfield  {author} {\bibinfo {author} {\bibfnamefont {S.}~\bibnamefont
  {Tomita}}, \bibinfo {author} {\bibfnamefont {J.~U.}\ \bibnamefont
  {Andersen}}, \bibinfo {author} {\bibfnamefont {H.}~\bibnamefont
  {Cederquist}}, \bibinfo {author} {\bibfnamefont {B.}~\bibnamefont {Concina}},
  \bibinfo {author} {\bibfnamefont {O.}~\bibnamefont {Echt}}, \bibinfo {author}
  {\bibfnamefont {J.~S.}\ \bibnamefont {Forster}}, \bibinfo {author}
  {\bibfnamefont {K.}~\bibnamefont {Hansen}}, \bibinfo {author} {\bibfnamefont
  {B.~A.}\ \bibnamefont {Huber}}, \bibinfo {author} {\bibfnamefont
  {P.}~\bibnamefont {Hvelplund}}, \bibinfo {author} {\bibfnamefont
  {J.}~\bibnamefont {Jensen}}, \bibinfo {author} {\bibfnamefont
  {B.}~\bibnamefont {Liu}}, \bibinfo {author} {\bibfnamefont {B.}~\bibnamefont
  {Manil}}, \bibinfo {author} {\bibfnamefont {L.}~\bibnamefont {Maunoury}},
  \bibinfo {author} {\bibfnamefont {S.~Br\o{}ndsted}\ \bibnamefont {Nielsen}},
  \bibinfo {author} {\bibfnamefont {J.}~\bibnamefont {Rangama}}, \bibinfo
  {author} {\bibfnamefont {H.~T.}\ \bibnamefont {Schmidt}}, \ and\ \bibinfo
  {author} {\bibfnamefont {H.}~\bibnamefont {Zettergren}},\ }\bibfield  {title}
  {\enquote {\bibinfo {title} {{Lifetimes of C$_{60}^{2-}$ and C$_{70}^{2-}$
  dianions in a storage ring}},}\ }\href {\doibase 10.1063/1.2155435}
  {\bibfield  {journal} {\bibinfo  {journal} {J. Chem. Phys.}\ }\textbf
  {\bibinfo {volume} {124}},\ \bibinfo {pages} {024310} (\bibinfo {year}
  {2006})}\BibitemShut {NoStop}%
\bibitem [{\citenamefont {Rao}\ and\ \citenamefont {Jena}(1985)}]{Rao1985}%
  \BibitemOpen
  \bibfield  {author} {\bibinfo {author} {\bibfnamefont {B.~K.}\ \bibnamefont
  {Rao}}\ and\ \bibinfo {author} {\bibfnamefont {P.}~\bibnamefont {Jena}},\
  }\bibfield  {title} {\enquote {\bibinfo {title} {{Physics of small metal
  clusters: Topology, magnetism, and electronic structure}},}\ }\href {\doibase
  10.1103/PhysRevB.32.2058} {\bibfield  {journal} {\bibinfo  {journal} {Phys.
  Rev. B}\ }\textbf {\bibinfo {volume} {32}},\ \bibinfo {pages} {2058}
  (\bibinfo {year} {1985})}\BibitemShut {NoStop}%
\bibitem [{\citenamefont {Ham}\ and\ \citenamefont {Leung}(1993)}]{Ham1993}%
  \BibitemOpen
  \bibfield  {author} {\bibinfo {author} {\bibfnamefont {F.~S.}\ \bibnamefont
  {Ham}}\ and\ \bibinfo {author} {\bibfnamefont {C.-H.}\ \bibnamefont
  {Leung}},\ }\bibfield  {title} {\enquote {\bibinfo {title} {{Dynamic
  Jahn-Teller effect for a double acceptor or acceptor-bound exciton in
  semiconductors: Mechanism for an inverted level ordering}},}\ }\href
  {\doibase 10.1103/PhysRevLett.71.3186} {\bibfield  {journal} {\bibinfo
  {journal} {Phys. Rev. Lett.}\ }\textbf {\bibinfo {volume} {71}},\ \bibinfo
  {pages} {3186} (\bibinfo {year} {1993})}\BibitemShut {NoStop}%
\bibitem [{\citenamefont {Abragam}\ and\ \citenamefont
  {Bleaney}(1970)}]{Abragam1970}%
  \BibitemOpen
  \bibfield  {author} {\bibinfo {author} {\bibfnamefont {A.}~\bibnamefont
  {Abragam}}\ and\ \bibinfo {author} {\bibfnamefont {B.}~\bibnamefont
  {Bleaney}},\ }\href@noop {} {\emph {\bibinfo {title} {Electron Paramagnetic
  Resonance of Transition Ions}}}\ (\bibinfo  {publisher} {Claredon Press},\
  \bibinfo {address} {Oxford},\ \bibinfo {year} {1970})\BibitemShut {NoStop}%
\bibitem [{\citenamefont {Varshalovich}\ \emph {et~al.}(1988)\citenamefont
  {Varshalovich}, \citenamefont {Moskalev},\ and\ \citenamefont
  {Khersonskii}}]{Varshalovich1988}%
  \BibitemOpen
  \bibfield  {author} {\bibinfo {author} {\bibfnamefont {D.~A.}\ \bibnamefont
  {Varshalovich}}, \bibinfo {author} {\bibfnamefont {A.~N.}\ \bibnamefont
  {Moskalev}}, \ and\ \bibinfo {author} {\bibfnamefont {V.~K.}\ \bibnamefont
  {Khersonskii}},\ }\href@noop {} {\emph {\bibinfo {title} {Quantum Theory of
  Angular Momentum}}}\ (\bibinfo  {publisher} {World Scientific},\ \bibinfo
  {address} {Singapore},\ \bibinfo {year} {1988})\BibitemShut {NoStop}%
\bibitem [{\citenamefont {Child}\ and\ \citenamefont
  {Longuet-Higgins}(1961)}]{Child1961}%
  \BibitemOpen
  \bibfield  {author} {\bibinfo {author} {\bibfnamefont {M.~S.}\ \bibnamefont
  {Child}}\ and\ \bibinfo {author} {\bibfnamefont {H.~C.}\ \bibnamefont
  {Longuet-Higgins}},\ }\bibfield  {title} {\enquote {\bibinfo {title}
  {{Studies of the Jahn-Teller effect III. The rotational and vibrational
  spectra of symmetric-top molecules in electronically degenerate states}},}\
  }\href {\doibase 10.1098/rsta.1961.0017} {\bibfield  {journal} {\bibinfo
  {journal} {Phil. Trans. R. Soc. A}\ }\textbf {\bibinfo {volume} {254}},\
  \bibinfo {pages} {259} (\bibinfo {year} {1961})}\BibitemShut {NoStop}%
\bibitem [{\citenamefont {Ponzellini}(2014)}]{Ponzellini}%
  \BibitemOpen
  \bibfield  {author} {\bibinfo {author} {\bibfnamefont {P.}~\bibnamefont
  {Ponzellini}},\ }\emph {\bibinfo {title} {Computation of the paramagnetic
  g-factor for the fullerene monocation and monoanion}},\ \href@noop {}
  {Master's thesis},\ \bibinfo  {school} {Milan University} (\bibinfo {year}
  {2014})\BibitemShut {NoStop}%
\bibitem [{Note5()}]{Note5}%
  \BibitemOpen
  \bibinfo {note} {The vibronic coupling parameters derived from the
  photoelectron spectra \cite {Iwahara2010} could be slightly overestimated
  because the dependence of intensities on the absorbed photon energy ($\hslash
  \omega _\protect \text {ph}$) was neglected since $\omega _\mu /\omega
  _\protect \text {ph} \ll 1$. Within the second order perturbation theory, the
  intensity is proportional to the product of $g_\mu ^2$ and $\omega _\protect
  \text {ph}$. Using this relation, the vibronic coupling parameters for high
  frequency modes are estimated to be reduced by about 3-4 \%.}\BibitemShut
  {Stop}%
\bibitem [{\citenamefont {Trulove}\ \emph {et~al.}(1995)\citenamefont
  {Trulove}, \citenamefont {Carlin}, \citenamefont {Eaton},\ and\ \citenamefont
  {Eaton}}]{Trulove1995}%
  \BibitemOpen
  \bibfield  {author} {\bibinfo {author} {\bibfnamefont {P.~C.}\ \bibnamefont
  {Trulove}}, \bibinfo {author} {\bibfnamefont {R.~T.}\ \bibnamefont {Carlin}},
  \bibinfo {author} {\bibfnamefont {G.~R.}\ \bibnamefont {Eaton}}, \ and\
  \bibinfo {author} {\bibfnamefont {S.~S.}\ \bibnamefont {Eaton}},\ }\bibfield
  {title} {\enquote {\bibinfo {title} {{Determination of the singlet-triplet
  energy separation for C$_{60}^{2-}$ in DMSO by electron paramagnetic
  resonance}},}\ }\href {\doibase 10.1021/ja00128a014} {\bibfield  {journal}
  {\bibinfo  {journal} {J. Am. Chem. Soc.}\ }\textbf {\bibinfo {volume}
  {117}},\ \bibinfo {pages} {6265} (\bibinfo {year} {1995})}\BibitemShut
  {NoStop}%
\bibitem [{\citenamefont {Brouet}\ \emph
  {et~al.}(2002{\natexlab{a}})\citenamefont {Brouet}, \citenamefont {Alloul},
  \citenamefont {Garaj},\ and\ \citenamefont {Forr\'o}}]{Brouet2002c}%
  \BibitemOpen
  \bibfield  {author} {\bibinfo {author} {\bibfnamefont {V.}~\bibnamefont
  {Brouet}}, \bibinfo {author} {\bibfnamefont {H.}~\bibnamefont {Alloul}},
  \bibinfo {author} {\bibfnamefont {S.}~\bibnamefont {Garaj}}, \ and\ \bibinfo
  {author} {\bibfnamefont {L.}~\bibnamefont {Forr\'o}},\ }\bibfield  {title}
  {\enquote {\bibinfo {title} {Persistence of molecular excitations in metallic
  fullerides and their role in a possible metal to insulator transition at high
  temperatures},}\ }\href {\doibase 10.1103/PhysRevB.66.155124} {\bibfield
  {journal} {\bibinfo  {journal} {Phys. Rev. B}\ }\textbf {\bibinfo {volume}
  {66}},\ \bibinfo {pages} {155124} (\bibinfo {year}
  {2002}{\natexlab{a}})}\BibitemShut {NoStop}%
\bibitem [{\citenamefont {Brouet}\ \emph
  {et~al.}(2002{\natexlab{b}})\citenamefont {Brouet}, \citenamefont {Alloul},
  \citenamefont {Garaj},\ and\ \citenamefont {Forr\'o}}]{Brouet2002a}%
  \BibitemOpen
  \bibfield  {author} {\bibinfo {author} {\bibfnamefont {V.}~\bibnamefont
  {Brouet}}, \bibinfo {author} {\bibfnamefont {H.}~\bibnamefont {Alloul}},
  \bibinfo {author} {\bibfnamefont {S.}~\bibnamefont {Garaj}}, \ and\ \bibinfo
  {author} {\bibfnamefont {L.}~\bibnamefont {Forr\'o}},\ }\bibfield  {title}
  {\enquote {\bibinfo {title} {{Gaps and excitations in fullerides with
  partially filled bands: NMR study of ${\mathrm{Na}}_{2}{\mathrm{C}}_{60}$ and
  ${\mathrm{K}}_{4}{\mathrm{C}}_{60}$}},}\ }\href {\doibase
  10.1103/PhysRevB.66.155122} {\bibfield  {journal} {\bibinfo  {journal} {Phys.
  Rev. B}\ }\textbf {\bibinfo {volume} {66}},\ \bibinfo {pages} {155122}
  (\bibinfo {year} {2002}{\natexlab{b}})}\BibitemShut {NoStop}%
\bibitem [{\citenamefont {Zimmer}\ \emph {et~al.}(1994)\citenamefont {Zimmer},
  \citenamefont {Helmle}, \citenamefont {Mehring},\ and\ \citenamefont
  {Rachdi}}]{Zimmer1994}%
  \BibitemOpen
  \bibfield  {author} {\bibinfo {author} {\bibfnamefont {G.}~\bibnamefont
  {Zimmer}}, \bibinfo {author} {\bibfnamefont {M.}~\bibnamefont {Helmle}},
  \bibinfo {author} {\bibfnamefont {M.}~\bibnamefont {Mehring}}, \ and\
  \bibinfo {author} {\bibfnamefont {F.}~\bibnamefont {Rachdi}},\ }\bibfield
  {title} {\enquote {\bibinfo {title} {{Lattice Dynamics and $^{13}$C
  Paramagnetic Shift in K$_4$C$_{60}$}},}\ }\href
  {http://stacks.iop.org/0295-5075/27/i=7/a=009} {\bibfield  {journal}
  {\bibinfo  {journal} {Europhys. Lett.}\ }\textbf {\bibinfo {volume} {27}},\
  \bibinfo {pages} {543} (\bibinfo {year} {1994})}\BibitemShut {NoStop}%
\bibitem [{\citenamefont {Zimmer}\ \emph {et~al.}(1995)\citenamefont {Zimmer},
  \citenamefont {Mehring}, \citenamefont {Goze},\ and\ \citenamefont
  {Rachdi}}]{Zimmer1995}%
  \BibitemOpen
  \bibfield  {author} {\bibinfo {author} {\bibfnamefont {G.}~\bibnamefont
  {Zimmer}}, \bibinfo {author} {\bibfnamefont {M.}~\bibnamefont {Mehring}},
  \bibinfo {author} {\bibfnamefont {C.}~\bibnamefont {Goze}}, \ and\ \bibinfo
  {author} {\bibfnamefont {F.}~\bibnamefont {Rachdi}},\ }\bibfield  {title}
  {\enquote {\bibinfo {title} {{Rotational dynamics of
  ${\mathrm{C}}_{60}^{4\mathrm{\ensuremath{-}}}$ and electronic excitation in
  ${\mathrm{Rb}}_{4}$${\mathrm{C}}_{60}$}},}\ }\href {\doibase
  10.1103/PhysRevB.52.13300} {\bibfield  {journal} {\bibinfo  {journal} {Phys.
  Rev. B}\ }\textbf {\bibinfo {volume} {52}},\ \bibinfo {pages} {13300}
  (\bibinfo {year} {1995})}\BibitemShut {NoStop}%
\bibitem [{\citenamefont {Lukyanchuk}\ \emph {et~al.}(1995)\citenamefont
  {Lukyanchuk}, \citenamefont {Kirova}, \citenamefont {Rachdi}, \citenamefont
  {Goze}, \citenamefont {Molinie},\ and\ \citenamefont
  {Mehring}}]{Lukyanchuk1995}%
  \BibitemOpen
  \bibfield  {author} {\bibinfo {author} {\bibfnamefont {I.}~\bibnamefont
  {Lukyanchuk}}, \bibinfo {author} {\bibfnamefont {N.}~\bibnamefont {Kirova}},
  \bibinfo {author} {\bibfnamefont {F.}~\bibnamefont {Rachdi}}, \bibinfo
  {author} {\bibfnamefont {C.}~\bibnamefont {Goze}}, \bibinfo {author}
  {\bibfnamefont {P.}~\bibnamefont {Molinie}}, \ and\ \bibinfo {author}
  {\bibfnamefont {M.}~\bibnamefont {Mehring}},\ }\bibfield  {title} {\enquote
  {\bibinfo {title} {{Electronic localization in
  ${\mathrm{Rb}}_{4}$${\mathrm{C}}_{60}$ from bulk magnetic measurements}},}\
  }\href {\doibase 10.1103/PhysRevB.51.3978} {\bibfield  {journal} {\bibinfo
  {journal} {Phys. Rev. B}\ }\textbf {\bibinfo {volume} {51}},\ \bibinfo
  {pages} {3978} (\bibinfo {year} {1995})}\BibitemShut {NoStop}%
\bibitem [{\citenamefont {Ricc\`o}\ \emph {et~al.}(2003)\citenamefont
  {Ricc\`o}, \citenamefont {Fumera}, \citenamefont {Shiroka}, \citenamefont
  {Ligabue}, \citenamefont {Bucci},\ and\ \citenamefont {Bolzoni}}]{Ricco2003}%
  \BibitemOpen
  \bibfield  {author} {\bibinfo {author} {\bibfnamefont {M.}~\bibnamefont
  {Ricc\`o}}, \bibinfo {author} {\bibfnamefont {G.}~\bibnamefont {Fumera}},
  \bibinfo {author} {\bibfnamefont {T.}~\bibnamefont {Shiroka}}, \bibinfo
  {author} {\bibfnamefont {O.}~\bibnamefont {Ligabue}}, \bibinfo {author}
  {\bibfnamefont {C.}~\bibnamefont {Bucci}}, \ and\ \bibinfo {author}
  {\bibfnamefont {F.}~\bibnamefont {Bolzoni}},\ }\bibfield  {title} {\enquote
  {\bibinfo {title} {{Metal-to-insulator evolution in
  $({\mathrm{NH}}_{3})_{x}{\mathrm{NaK}}_{2}{\mathrm{C}}_{60}:$ An NMR
  study}},}\ }\href {\doibase 10.1103/PhysRevB.68.035102} {\bibfield  {journal}
  {\bibinfo  {journal} {Phys. Rev. B}\ }\textbf {\bibinfo {volume} {68}},\
  \bibinfo {pages} {035102} (\bibinfo {year} {2003})}\BibitemShut {NoStop}%
\bibitem [{\citenamefont {Konarev}\ \emph {et~al.}(2003)\citenamefont
  {Konarev}, \citenamefont {Khasanov}, \citenamefont {Saito}, \citenamefont
  {Vorontsov}, \citenamefont {Otsuka}, \citenamefont {Lyubovskaya},\ and\
  \citenamefont {Antipin}}]{Konarev2003}%
  \BibitemOpen
  \bibfield  {author} {\bibinfo {author} {\bibfnamefont {D.~V.}\ \bibnamefont
  {Konarev}}, \bibinfo {author} {\bibfnamefont {S.~S.}\ \bibnamefont
  {Khasanov}}, \bibinfo {author} {\bibfnamefont {G.}~\bibnamefont {Saito}},
  \bibinfo {author} {\bibfnamefont {I.~I.}\ \bibnamefont {Vorontsov}}, \bibinfo
  {author} {\bibfnamefont {A.}~\bibnamefont {Otsuka}}, \bibinfo {author}
  {\bibfnamefont {R.~N.}\ \bibnamefont {Lyubovskaya}}, \ and\ \bibinfo {author}
  {\bibfnamefont {Y.~M.}\ \bibnamefont {Antipin}},\ }\bibfield  {title}
  {\enquote {\bibinfo {title} {{Crystal Structure and Magnetic Properties of an
  Ionic C$_{60}$ Complex with Decamethylcobaltocene:
  (Cp*$_2$Co)$_2$C$_{60}$(C$_6$H$_4$Cl$_2$, C$_6$H$_5$CN)$_2$.
  Singlet−Triplet Transitions in the C$_{60}^{2-}$ Anion}},}\ }\href
  {\doibase 10.1021/ic0340074} {\bibfield  {journal} {\bibinfo  {journal}
  {Inorg. Chem.}\ }\textbf {\bibinfo {volume} {42}},\ \bibinfo {pages} {3706}
  (\bibinfo {year} {2003})}\BibitemShut {NoStop}%
\bibitem [{\citenamefont {Konarev}\ \emph {et~al.}(2017)\citenamefont
  {Konarev}, \citenamefont {Troyanov}, \citenamefont {Otsuka}, \citenamefont
  {Yamochi}, \citenamefont {Saito},\ and\ \citenamefont
  {Lyubovskaya}}]{Konarev2017}%
  \BibitemOpen
  \bibfield  {author} {\bibinfo {author} {\bibfnamefont {D.~V.}\ \bibnamefont
  {Konarev}}, \bibinfo {author} {\bibfnamefont {S.~I.}\ \bibnamefont
  {Troyanov}}, \bibinfo {author} {\bibfnamefont {A.}~\bibnamefont {Otsuka}},
  \bibinfo {author} {\bibfnamefont {H.}~\bibnamefont {Yamochi}}, \bibinfo
  {author} {\bibfnamefont {G.}~\bibnamefont {Saito}}, \ and\ \bibinfo {author}
  {\bibfnamefont {R.~N.}\ \bibnamefont {Lyubovskaya}},\ }\bibfield  {title}
  {\enquote {\bibinfo {title} {{Fullerene C$_{60}$ dianion salt,
  (Me$_4$N$^+$)$_2$(C$_{60}^{2-}$){\textperiodcentered}(TPC)$_2${\textperiodcentered}2C$_6$H$_4$Cl$_2$,
  where TPC is triptycene, obtained by a multicomponent approach}},}\
  }\href@noop {} {\bibfield  {journal} {\bibinfo  {journal} {New J. Chem.}\
  }\textbf {\bibinfo {volume} {41}},\ \bibinfo {pages} {4779} (\bibinfo {year}
  {2017})}\BibitemShut {NoStop}%
\bibitem [{\citenamefont {Jegli\ifmmode~\check{c}\else \v{c}\fi{}}\ \emph
  {et~al.}(2009)\citenamefont {Jegli\ifmmode~\check{c}\else \v{c}\fi{}},
  \citenamefont {Ar\ifmmode~\check{c}\else \v{c}\fi{}on}, \citenamefont
  {Poto\ifmmode~\check{c}\else \v{c}\fi{}nik}, \citenamefont {Ganin},
  \citenamefont {Takabayashi}, \citenamefont {Rosseinsky},\ and\ \citenamefont
  {Prassides}}]{Jeglic2009}%
  \BibitemOpen
  \bibfield  {author} {\bibinfo {author} {\bibfnamefont {P.}~\bibnamefont
  {Jegli\ifmmode~\check{c}\else \v{c}\fi{}}}, \bibinfo {author} {\bibfnamefont
  {D.}~\bibnamefont {Ar\ifmmode~\check{c}\else \v{c}\fi{}on}}, \bibinfo
  {author} {\bibfnamefont {A.}~\bibnamefont {Poto\ifmmode~\check{c}\else
  \v{c}\fi{}nik}}, \bibinfo {author} {\bibfnamefont {A.~Y.}\ \bibnamefont
  {Ganin}}, \bibinfo {author} {\bibfnamefont {Y.}~\bibnamefont {Takabayashi}},
  \bibinfo {author} {\bibfnamefont {M.~J.}\ \bibnamefont {Rosseinsky}}, \ and\
  \bibinfo {author} {\bibfnamefont {K.}~\bibnamefont {Prassides}},\ }\bibfield
  {title} {\enquote {\bibinfo {title} {Low-moment antiferromagnetic ordering in
  triply charged cubic fullerides close to the metal-insulator transition},}\
  }\href {\doibase 10.1103/PhysRevB.80.195424} {\bibfield  {journal} {\bibinfo
  {journal} {Phys. Rev. B}\ }\textbf {\bibinfo {volume} {80}},\ \bibinfo
  {pages} {195424} (\bibinfo {year} {2009})}\BibitemShut {NoStop}%
\bibitem [{Note6()}]{Note6}%
  \BibitemOpen
  \bibinfo {note} {The activation energy of C$_{60}^{2-}$ in gas phase has been
  estimated to be 120 $\pm $ 20 meV by analyzing the decay rate from
  C$_{60}^{2-}$ to C$_{60}^-+e^-$, where $e^-$ is an electron \cite
  {Tomita2006}. However, the singlet-triplet excitation is relatively small
  value in their analysis and many approximations are employed for the
  treatment of the complicated process, and hence, the error bar of the gap
  would be large.}\BibitemShut {Stop}%
\bibitem [{\citenamefont {Yoshizawa}\ \emph {et~al.}(1993)\citenamefont
  {Yoshizawa}, \citenamefont {Sato}, \citenamefont {Tanaka}, \citenamefont
  {Yamabe},\ and\ \citenamefont {Okahara}}]{Yoshizawa1993}%
  \BibitemOpen
  \bibfield  {author} {\bibinfo {author} {\bibfnamefont {K.}~\bibnamefont
  {Yoshizawa}}, \bibinfo {author} {\bibfnamefont {T.}~\bibnamefont {Sato}},
  \bibinfo {author} {\bibfnamefont {K.}~\bibnamefont {Tanaka}}, \bibinfo
  {author} {\bibfnamefont {T.}~\bibnamefont {Yamabe}}, \ and\ \bibinfo {author}
  {\bibfnamefont {K.}~\bibnamefont {Okahara}},\ }\bibfield  {title} {\enquote
  {\bibinfo {title} {{ESR study of TDAE-C$_{60}$ and TDAE-C$_{70}$ in
  solution}},}\ }\href {\doibase
  http://dx.doi.org/10.1016/0009-2614(93)89148-B} {\bibfield  {journal}
  {\bibinfo  {journal} {Chem. Phys. Lett.}\ }\textbf {\bibinfo {volume}
  {213}},\ \bibinfo {pages} {498} (\bibinfo {year} {1993})}\BibitemShut
  {NoStop}%
\bibitem [{\citenamefont {Ihara}\ \emph {et~al.}(2011)\citenamefont {Ihara},
  \citenamefont {Alloul}, \citenamefont {Wzietek}, \citenamefont {Pontiroli},
  \citenamefont {Mazzani},\ and\ \citenamefont {Ricc\`o}}]{Ihara2011}%
  \BibitemOpen
  \bibfield  {author} {\bibinfo {author} {\bibfnamefont {Y.}~\bibnamefont
  {Ihara}}, \bibinfo {author} {\bibfnamefont {H.}~\bibnamefont {Alloul}},
  \bibinfo {author} {\bibfnamefont {P.}~\bibnamefont {Wzietek}}, \bibinfo
  {author} {\bibfnamefont {D.}~\bibnamefont {Pontiroli}}, \bibinfo {author}
  {\bibfnamefont {M.}~\bibnamefont {Mazzani}}, \ and\ \bibinfo {author}
  {\bibfnamefont {M.}~\bibnamefont {Ricc\`o}},\ }\bibfield  {title} {\enquote
  {\bibinfo {title} {{Spin dynamics at the Mott transition and in the metallic
  state of the Cs$_3$C$_{60}$ superconducting phases}},}\ }\href@noop {}
  {\bibfield  {journal} {\bibinfo  {journal} {Europhys. Lett.}\ }\textbf
  {\bibinfo {volume} {94}},\ \bibinfo {pages} {37007} (\bibinfo {year}
  {2011})}\BibitemShut {NoStop}%
\bibitem [{\citenamefont {Boeddinghaus}\ \emph {et~al.}(2014)\citenamefont
  {Boeddinghaus}, \citenamefont {Klein}, \citenamefont {Wahl}, \citenamefont
  {Jakes}, \citenamefont {Eichel},\ and\ \citenamefont
  {F\"{a}ssler}}]{Boeddinghaus2014}%
  \BibitemOpen
  \bibfield  {author} {\bibinfo {author} {\bibfnamefont {M.~Bele}\ \bibnamefont
  {Boeddinghaus}}, \bibinfo {author} {\bibfnamefont {W.}~\bibnamefont {Klein}},
  \bibinfo {author} {\bibfnamefont {B.}~\bibnamefont {Wahl}}, \bibinfo {author}
  {\bibfnamefont {P.}~\bibnamefont {Jakes}}, \bibinfo {author} {\bibfnamefont
  {R.-A.}\ \bibnamefont {Eichel}}, \ and\ \bibinfo {author} {\bibfnamefont
  {T.~F.}\ \bibnamefont {F\"{a}ssler}},\ }\bibfield  {title} {\enquote
  {\bibinfo {title} {{C$_{60}^{3-}$ versus C$_{60}^{4-}$/C$_{60}^{2-}$ -
  Synthesis and Characterization of Five Salts Containing Discrete Fullerene
  Anions}},}\ }\href {\doibase 10.1002/zaac.201300607} {\bibfield  {journal}
  {\bibinfo  {journal} {Z. Anorg. Allg. Chem.}\ }\textbf {\bibinfo {volume}
  {640}},\ \bibinfo {pages} {701} (\bibinfo {year} {2014})}\BibitemShut
  {NoStop}%
\bibitem [{\citenamefont {G\"{u}tlich}\ \emph {et~al.}(1994)\citenamefont
  {G\"{u}tlich}, \citenamefont {Hauser},\ and\ \citenamefont
  {Spiering}}]{Gutlich1994}%
  \BibitemOpen
  \bibfield  {author} {\bibinfo {author} {\bibfnamefont {P.}~\bibnamefont
  {G\"{u}tlich}}, \bibinfo {author} {\bibfnamefont {A.}~\bibnamefont {Hauser}},
  \ and\ \bibinfo {author} {\bibfnamefont {H.}~\bibnamefont {Spiering}},\
  }\bibfield  {title} {\enquote {\bibinfo {title} {{Thermal and Optical
  Switching of Iron(II) Complexes}},}\ }\href {\doibase 10.1002/anie.199420241}
  {\bibfield  {journal} {\bibinfo  {journal} {Angew. Chem. Int. Ed.}\ }\textbf
  {\bibinfo {volume} {33}},\ \bibinfo {pages} {2024} (\bibinfo {year}
  {1994})}\BibitemShut {NoStop}%
\bibitem [{\citenamefont {Martin}\ and\ \citenamefont
  {Ritchie}(1993)}]{Martin1993}%
  \BibitemOpen
  \bibfield  {author} {\bibinfo {author} {\bibfnamefont {R.~L.}\ \bibnamefont
  {Martin}}\ and\ \bibinfo {author} {\bibfnamefont {J.~P.}\ \bibnamefont
  {Ritchie}},\ }\bibfield  {title} {\enquote {\bibinfo {title} {{Coulomb and
  exchange interactions in
  ${\mathrm{C}}_{60}^{\mathit{n}\mathrm{\ensuremath{-}}}$}},}\ }\href {\doibase
  10.1103/PhysRevB.48.4845} {\bibfield  {journal} {\bibinfo  {journal} {Phys.
  Rev. B}\ }\textbf {\bibinfo {volume} {48}},\ \bibinfo {pages} {4845}
  (\bibinfo {year} {1993})}\BibitemShut {NoStop}%
\bibitem [{\citenamefont {L\"{u}ders}\ \emph {et~al.}(2002)\citenamefont
  {L\"{u}ders}, \citenamefont {Bordoni}, \citenamefont {Manini}, \citenamefont
  {Corso}, \citenamefont {Fabrizio},\ and\ \citenamefont
  {Tosatti}}]{Luder2002}%
  \BibitemOpen
  \bibfield  {author} {\bibinfo {author} {\bibfnamefont {M.}~\bibnamefont
  {L\"{u}ders}}, \bibinfo {author} {\bibfnamefont {A.}~\bibnamefont {Bordoni}},
  \bibinfo {author} {\bibfnamefont {N.}~\bibnamefont {Manini}}, \bibinfo
  {author} {\bibfnamefont {A.~Dal}\ \bibnamefont {Corso}}, \bibinfo {author}
  {\bibfnamefont {M.}~\bibnamefont {Fabrizio}}, \ and\ \bibinfo {author}
  {\bibfnamefont {E.}~\bibnamefont {Tosatti}},\ }\bibfield  {title} {\enquote
  {\bibinfo {title} {Coulomb couplings in positively charged fullerene},}\
  }\href {\doibase 10.1080/13642810208220729} {\bibfield  {journal} {\bibinfo
  {journal} {Philos. Mag. B}\ }\textbf {\bibinfo {volume} {82}},\ \bibinfo
  {pages} {1611} (\bibinfo {year} {2002})}\BibitemShut {NoStop}%
\bibitem [{\citenamefont {Nomura}\ \emph {et~al.}(2012)\citenamefont {Nomura},
  \citenamefont {Nakamura},\ and\ \citenamefont {Arita}}]{Nomura2012}%
  \BibitemOpen
  \bibfield  {author} {\bibinfo {author} {\bibfnamefont {Y.}~\bibnamefont
  {Nomura}}, \bibinfo {author} {\bibfnamefont {K.}~\bibnamefont {Nakamura}}, \
  and\ \bibinfo {author} {\bibfnamefont {R.}~\bibnamefont {Arita}},\ }\bibfield
   {title} {\enquote {\bibinfo {title} {{{\it Ab initio} derivation of
  electronic low-energy models for C$_{60}$ and aromatic compounds}},}\ }\href
  {\doibase 10.1103/PhysRevB.85.155452} {\bibfield  {journal} {\bibinfo
  {journal} {Phys. Rev. B}\ }\textbf {\bibinfo {volume} {85}},\ \bibinfo
  {pages} {155452} (\bibinfo {year} {2012})}\BibitemShut {NoStop}%
\bibitem [{\citenamefont {Nikolaev}\ and\ \citenamefont
  {Michel}(2002)}]{Nikolaev2002}%
  \BibitemOpen
  \bibfield  {author} {\bibinfo {author} {\bibfnamefont {A.~V.}\ \bibnamefont
  {Nikolaev}}\ and\ \bibinfo {author} {\bibfnamefont {K.~H.}\ \bibnamefont
  {Michel}},\ }\bibfield  {title} {\enquote {\bibinfo {title} {{Molecular
  terms, magnetic moments, and optical transitions of molecular ions
  C$_{60}^{m\pm}$}},}\ }\href {\doibase 10.1063/1.1497644} {\bibfield
  {journal} {\bibinfo  {journal} {J. Chem. Phys.}\ }\textbf {\bibinfo {volume}
  {117}},\ \bibinfo {pages} {4761} (\bibinfo {year} {2002})}\BibitemShut
  {NoStop}%
\bibitem [{\citenamefont {Judd}(1967)}]{Judd1967}%
  \BibitemOpen
  \bibfield  {author} {\bibinfo {author} {\bibfnamefont {B.~R.}\ \bibnamefont
  {Judd}},\ }\href@noop {} {\emph {\bibinfo {title} {Second Quantization and
  Atomic Spectroscopy}}}\ (\bibinfo  {publisher} {The Johns Hopkins Press},\
  \bibinfo {address} {Baltimore},\ \bibinfo {year} {1967})\BibitemShut
  {NoStop}%
\bibitem [{\citenamefont {Polinger}\ and\ \citenamefont
  {Bersuker}(1979)}]{Polinger1979}%
  \BibitemOpen
  \bibfield  {author} {\bibinfo {author} {\bibfnamefont {V.~Z.}\ \bibnamefont
  {Polinger}}\ and\ \bibinfo {author} {\bibfnamefont {G.~I.}\ \bibnamefont
  {Bersuker}},\ }\bibfield  {title} {\enquote {\bibinfo {title} {Multimode
  jahn-teller effect for an e term with strong vibronic coupling i. local and
  resonant states},}\ }\href {\doibase 10.1002/pssb.2220950210} {\bibfield
  {journal} {\bibinfo  {journal} {phys. status solidi (b)}\ }\textbf {\bibinfo
  {volume} {95}},\ \bibinfo {pages} {403} (\bibinfo {year} {1979})}\BibitemShut
  {NoStop}%
\bibitem [{\citenamefont {Manini}\ and\ \citenamefont
  {Tosatti}(1998)}]{Manini1998}%
  \BibitemOpen
  \bibfield  {author} {\bibinfo {author} {\bibfnamefont {N.}~\bibnamefont
  {Manini}}\ and\ \bibinfo {author} {\bibfnamefont {E.}~\bibnamefont
  {Tosatti}},\ }\bibfield  {title} {\enquote {\bibinfo {title} {Exact
  zero-point energy shift in the $e\ensuremath{\bigotimes}(\mathrm{nE})$,
  $t\ensuremath{\bigotimes}(\mathrm{nH})$ many-modes dynamic jahn-teller
  systems at strong coupling},}\ }\href {\doibase 10.1103/PhysRevB.58.782}
  {\bibfield  {journal} {\bibinfo  {journal} {Phys. Rev. B}\ }\textbf {\bibinfo
  {volume} {58}},\ \bibinfo {pages} {782} (\bibinfo {year} {1998})}\BibitemShut
  {NoStop}%
\end{thebibliography}

%

\clearpage
\begin{center}
\textbf{
Supplemental Materials\\
for
\\
``Dynamical Jahn-Teller effect of fullerene anions''
}
\end{center}
\setcounter{section}{0}
\setcounter{equation}{0}
\setcounter{figure}{0}

\renewcommand\thetable{S\arabic{table}}
\renewcommand\thefigure{S\arabic{figure}}
\renewcommand\thepage{S\arabic{page}}
\renewcommand\theequation{S\arabic{equation}}

\section*{Vibronic states}
\begin{figure*}[tb]
\begin{center}
\begin{tabular}{ll}
(a) & (b) \\
\includegraphics[width=6.5cm]{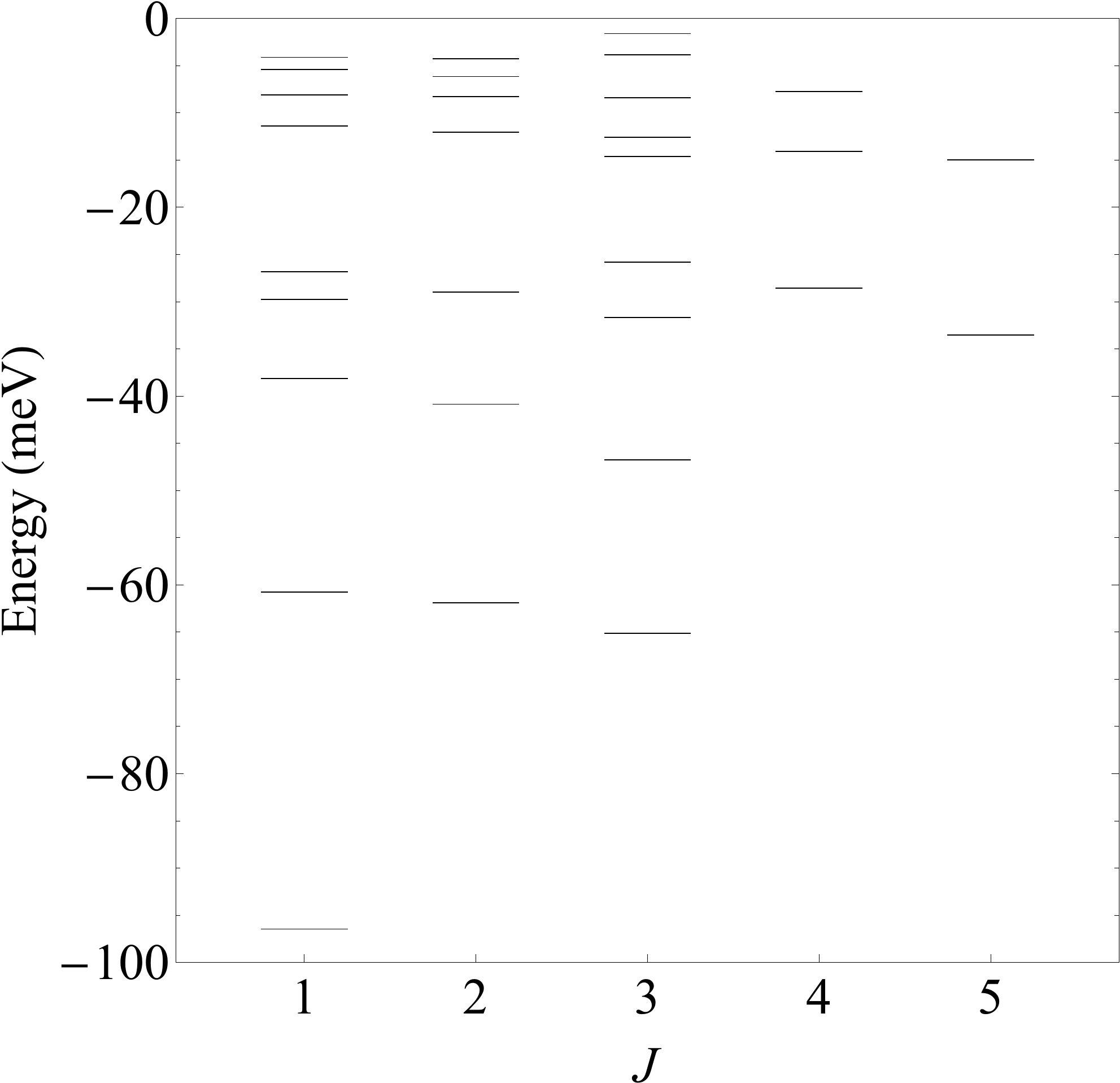}
&
\includegraphics[width=6.5cm]{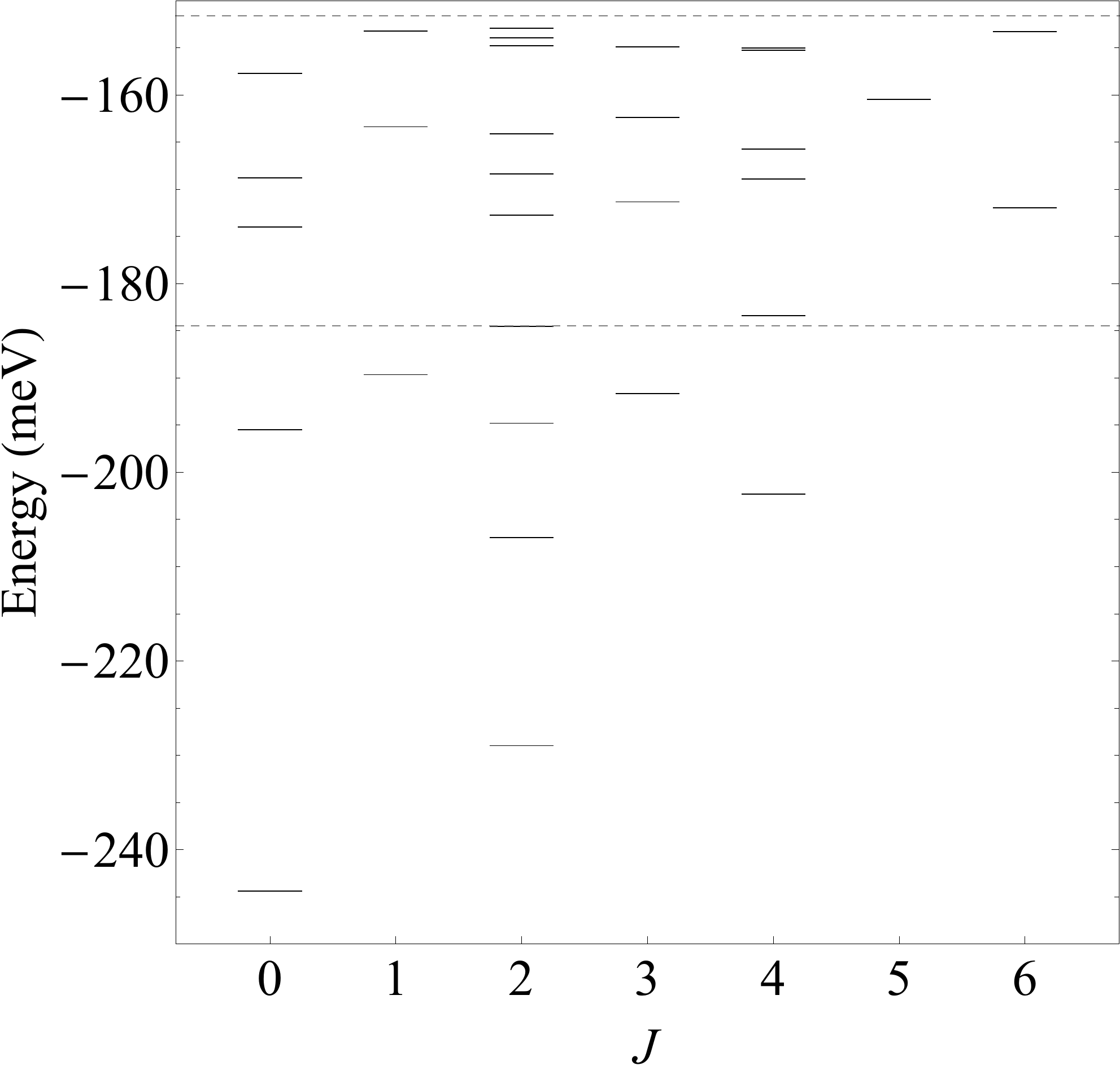}
\\
(c) \\
\multicolumn{2}{c}{
\includegraphics[width=13cm]{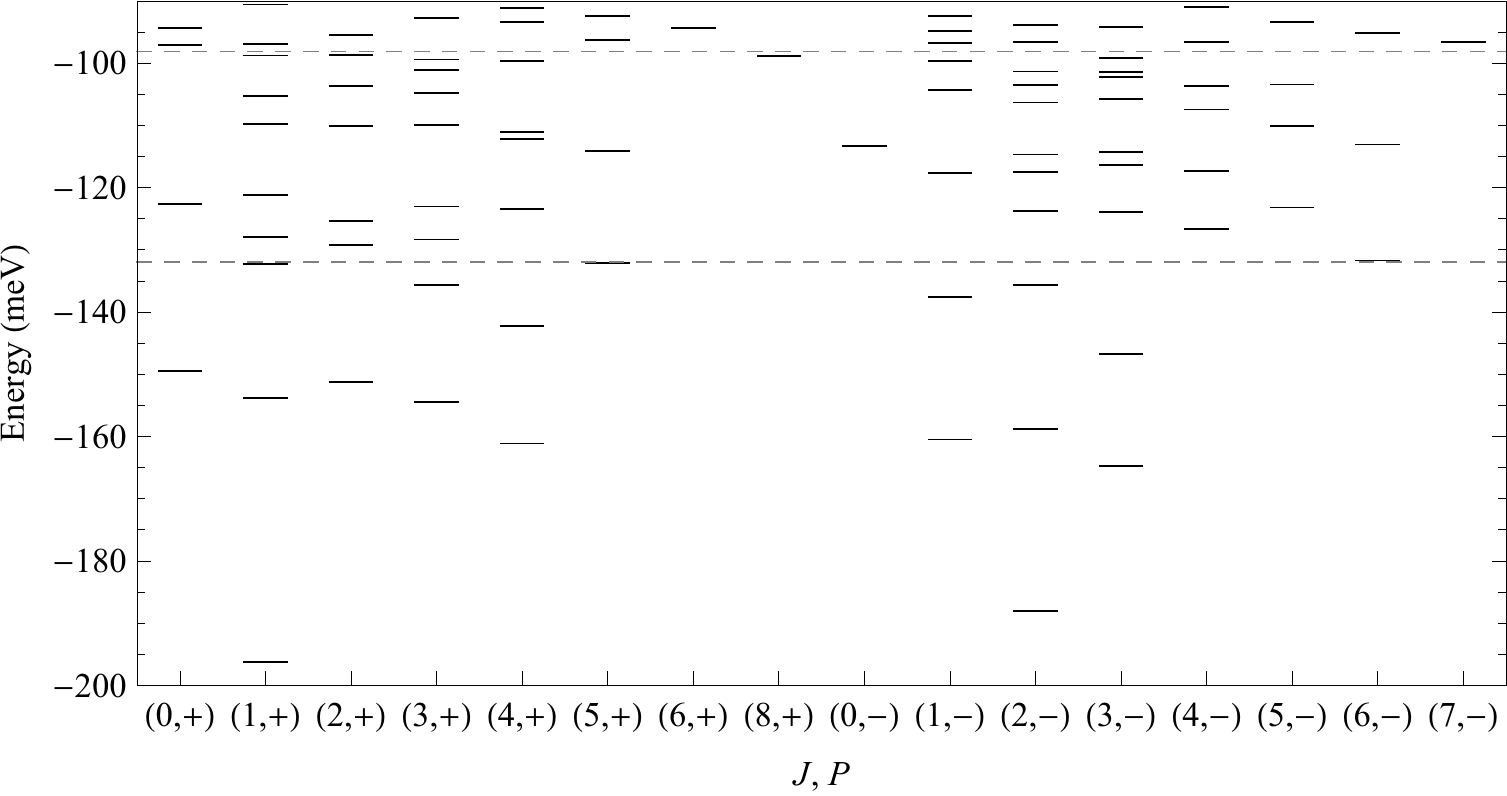}
}
\end{tabular}
\end{center}
\caption{Low-lying energy levels of (a) C$^{-}_{60}$, (b) C$_{60}^{2-}$, and (c) C$_{60}^{3-}$ (meV). 
(c) The dashed lines represent the ground ($= -132$ meV) and the first excited ($= -98$ meV) high-spin levels, and short solid lines indicate the vibronic levels. The vibronic levels are shown for each $(J,P)$.  $\pm$ in the parenthesis mean $\pm 1$.
}
\label{Fig:ESM}
\end{figure*}

The vibronic levels calculated using numerical diagonalization of the model Jahn-Teller Hamiltonian are shown in 
Table \ref{Table:E1} (C$_{60}^-$), Table \ref{Table:E2} (C$_{60}^{2-}$), and Table \ref{Table:E3} (C$_{60}^{3-}$). 
The energy levels are also shown in Figure \ref{Fig:ESM}.

\begin{table}[b]
\begin{ruledtabular}
\caption{
Vibronic energy levels of C$_{60}^{-}$ (meV). The numbers in the parentheses correspond to $J$.
}
\label{Table:E1}
\begin{tabular}{cccccccc}
  & (1)       & 9 &$-4.123$ &   &  (3)     & 9 &$-1.607$  \\
1 & $-96.469$ &   & (2)     & 1 &$-65.135$ &   &(4)         \\
2 & $-60.753$ & 1 &$-61.918$& 2 &$-46.786$ & 1 &$-28.549$  \\
3 & $-38.126$ & 2 &$-40.873$& 3 &$-31.703$ & 2 & $-14.108$ \\
4 & $-29.757$ & 3 &$-29.004$& 4 &$-25.806$ & 3 & $-7.724$ \\
5 & $-26.841$ & 4 &$-12.071$& 5 &$-14.620$ &   &(5)         \\ 
6 & $-11.395$ & 5 &$-8.2642$& 6 &$-12.575$ & 1 & $-33.554$  \\
7 & $-8.099$  & 6 &$-6.155$ & 7 &$-8.417$  & 2 &$-14.985$   \\
8 & $-5.411$  & 7 &$-4.265$ & 8 &$-3.843$                   \\
\end{tabular}                                  
\end{ruledtabular}                    
\end{table}

\begin{table}[tb]
\begin{ruledtabular}
\caption{
Vibronic energy levels of C$_{60}^{2-}$ (meV). The numbers in the parentheses correspond to $J$.
}
\label{Table:E2}
\begin{tabular}{cccccccc}
  & (0)        &   & (2)         & 10 & $-152.955$ & 4 & $-165.749$ \\
1 & $-244.384$ & 1 & $-228.975$  &    &  (3)       & 5 & $-155.291$ \\
2 & $-195.495$ & 2 & $-206.904$  & 1  & $-191.666$ & 6 & $-155.025$ \\ 
3 & $-174.029$ & 3 & $-194.800$  & 2  & $-171.345$ &   & (5)        \\
4 & $-168.821$ & 4 & $-184.548$  & 3  & $-162.395$ & 1 & $-160.500$ \\
5 & $-157.717$ & 5 & $-172.749$  & 4  & $-154.904$ &   & (6)        \\
  & (1)        & 6 & $-168.370$  &    & (4)        & 1 & $-171.972$ \\  
1 & $-189.665$ & 7 & $-164.121$  & 1  & $-202.335$ & 2 & $-153.287$ \\
2 & $-163.386$ & 8 & $-154.793$  & 2  & $-183.401$ \\
3 & $-153.257$ & 9 & $-153.938$  & 3  & $-168.910$ \\
\end{tabular}       
\end{ruledtabular}    
\end{table}

\begin{table}[tb]
\begin{ruledtabular}
\caption{
Vibronic energy levels of C$_{60}^{3-}$ (meV). The numbers in the parentheses correspond to $(J,P)$.
}
\label{Table:E3}
\begin{tabular}{cccccccc}
  & (0, +1)    &   & (3, +1)   &   & (0, $-1$)  &   & (3, $-1$)  \\
1 & $-149.514$ & 3 &$-128.35$  & 1 & $-113.346$ & 3 &$-123.937$  \\ 
2 & $-122.589$ & 4 &$-123.062$ &   & (1, $-1$)  & 4 &$-116.344$  \\
3 & $ -97.096$ & 5 &$-109.974$ & 1 &$-160.476$  & 5 &$-114.233$  \\
4 & $ -94.341$ & 6 &$-104.776$ & 2 &$-137.641$  & 6 &$-105.732$  \\
  & (1, +1)    & 7 &$-101.145$ & 3 &$-117.72$   & 7 &$-102.237$  \\
1 & $-196.208$ & 8 &$ -99.425$ & 4 &$-104.31$   & 8 &$-101.396$  \\
2 & $-153.873$ & 9 &$ -92.712$ & 5 &$ -99.715$  & 9 &$ -99.173$  \\
3 & $-132.303$ &   & (4, +1)   & 6 &$ -96.8$    &10&$-94.158$    \\
4 & $-127.979$ & 1 &$-161.124$ & 7 &$ -94.821$  &   & (4, $-1$)  \\
5 & $-121.190$ & 2 &$-142.233$ & 8 &$ -92.428$  & 1 &$-126.641$  \\
6 & $-109.736$ & 3 &$-123.431$ & 9 &$ -90.096$  & 2 &$-117.299$  \\
7 & $-105.268$ & 4 &$-112.199$ &   & (2, $-1$)  & 3 &$-107.463$  \\
8 & $ -98.765$ & 5 &$-111.134$ & 1 &$-188.066$  & 4 &$-103.696$  \\         
9 & $ -96.990$ & 6 &$ -99.635$ & 2 &$-158.762$  & 5 &$ -96.592$  \\ 
10& $ -90.602$ & 7 &$ -93.371$ & 3 &$-135.616$  & 6 &$ -91.030$  \\
  & (2, +1)    & 8 &$ -91.121$ & 4 &$-123.771$  &   & (5, $-1$)  \\
1 &$-151.29$   &   & (5, +1)   & 5 &$-117.508$  & 1 &$-123.197$  \\
2 &$-129.185$  & 1 &$-132.168$ & 6 &$-114.68$   & 2 &$-110.125$  \\
3 &$-125.337$  & 2 &$-114.092$ & 7 &$-106.323$  & 3 &$-103.455$  \\
4 &$-110.146$  & 3 &$ -96.314$ & 8 &$-103.52$   & 4 &$ -93.411$  \\
5 &$-103.722$  & 4 &$ -92.431$ & 9 &$-101.341$  &   &(6, $-1$)   \\
6 &$ -98.728$  &   & (6, +1)   & 10&$ -96.625$  & 1 &$-131.736$  \\
7 &$ -95.528$  & 1 &$ -94.366 $& 11&$ -93.915$  & 2 &$-113.08$   \\
  & (3, +1)    &   & (8, +1)   &   &(3, $-1$)   & 3 &$ -95.162$  \\
1 &$-154.422$  & 1 &$ -98.803$ & 1 &$-164.704$  &   & (7, $-1$)  \\
2 &$-135.694$  &   &  &          2 &$-146.775$  & 1 &$ -96.666$  \\
\end{tabular}
\end{ruledtabular}
\end{table}

\section*{Maximal temperature}
\begin{figure*}
\begin{center}
\begin{tabular}{lll}
(a) & (b) & (c) \\
\includegraphics[width=5.5cm]{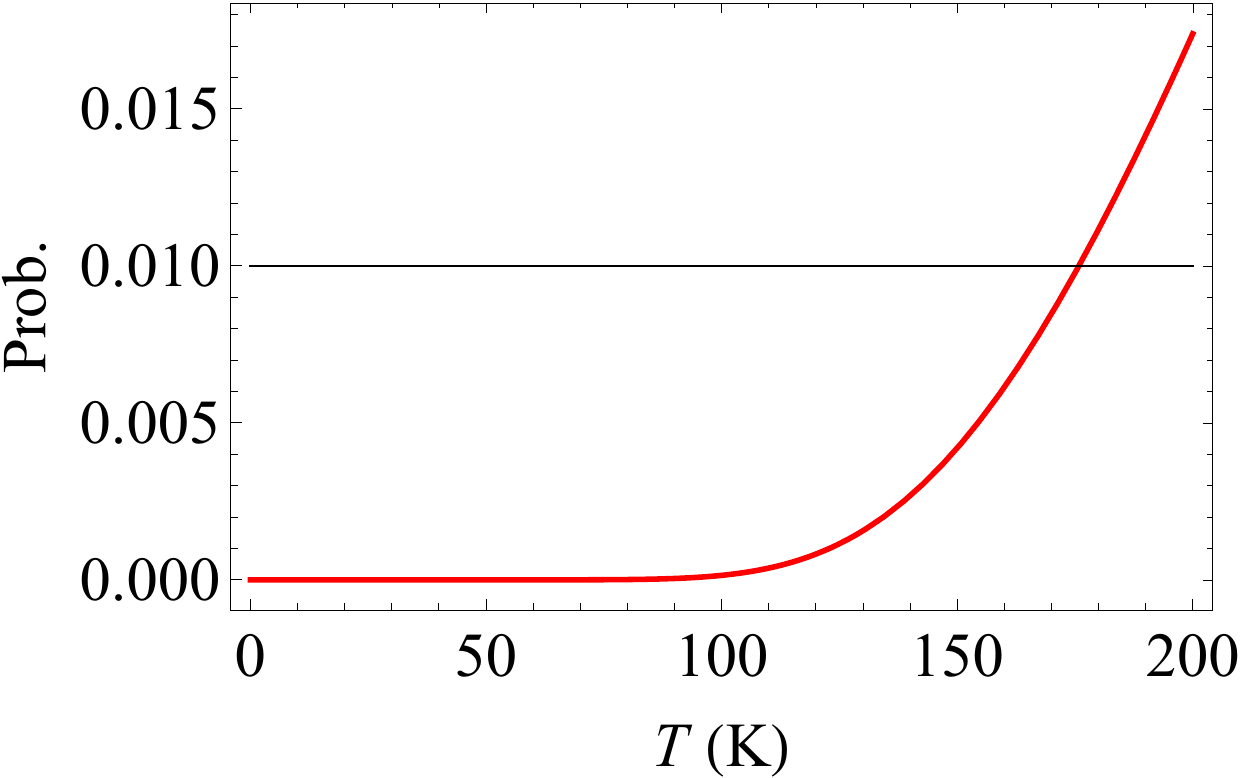}
& 
\includegraphics[width=5.5cm]{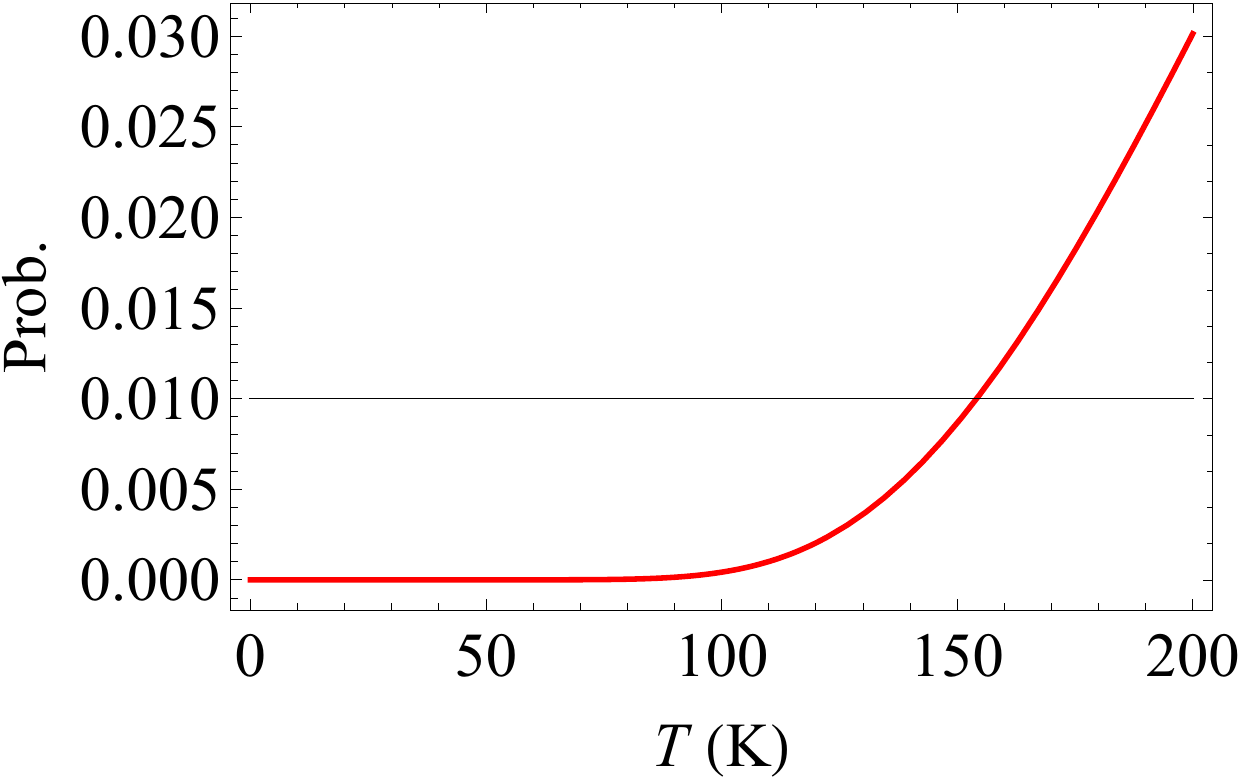}
&
\includegraphics[width=5.5cm]{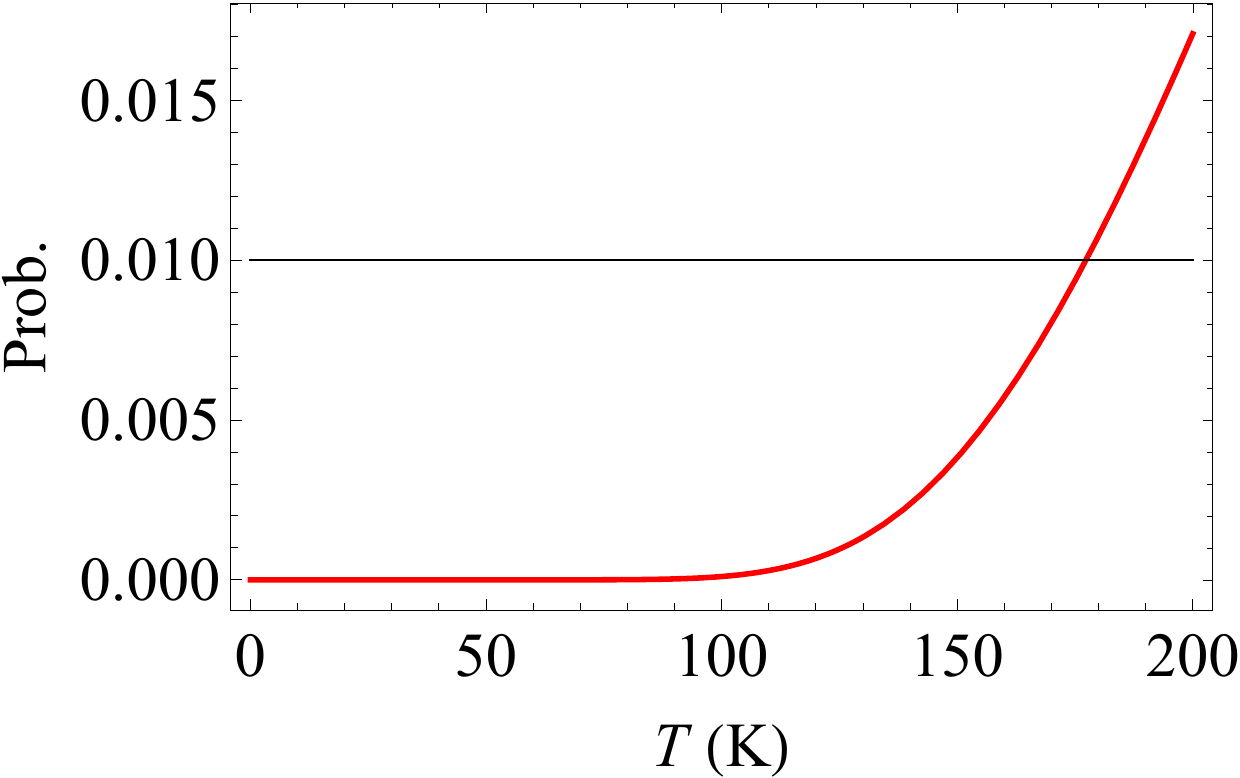}
\end{tabular}
\end{center}
\caption{
Probability that the levels at high-energy levels are occupied.
The levels are (a) 0 meV for C$_{60}^-$, (b) $-150$ meV for C$_{60}^{2-}$, and (c) $-90$ meV for C$_{60}^{3-}$.
}
\label{Fig:prob}
\end{figure*}

The maximal temperature for the simulation of spin gap is determined based on the condition that the sum of the probabilities of the high-energy levels are not thermally occupied: 
\begin{eqnarray}
 p = \sum_{J (P)} (2J+1) e^{-E_0/k_\text{B}T}/Z.
\end{eqnarray}
Here, $E_0$ is the high energy obtained energy levels.
The probabilities with respect to temperature are shown in Fig. \ref{Fig:prob}. 
In all cases, the probability $p$ is small (ca 1-2 \%) up to $T \approx 180$ K. 
Therefore, we calculated entropy in the main text up to the temperature. 

\section*{Vibronic levels of effective single mode Jahn-Teller model}
\begin{figure}[tb]
\includegraphics[width=8cm]{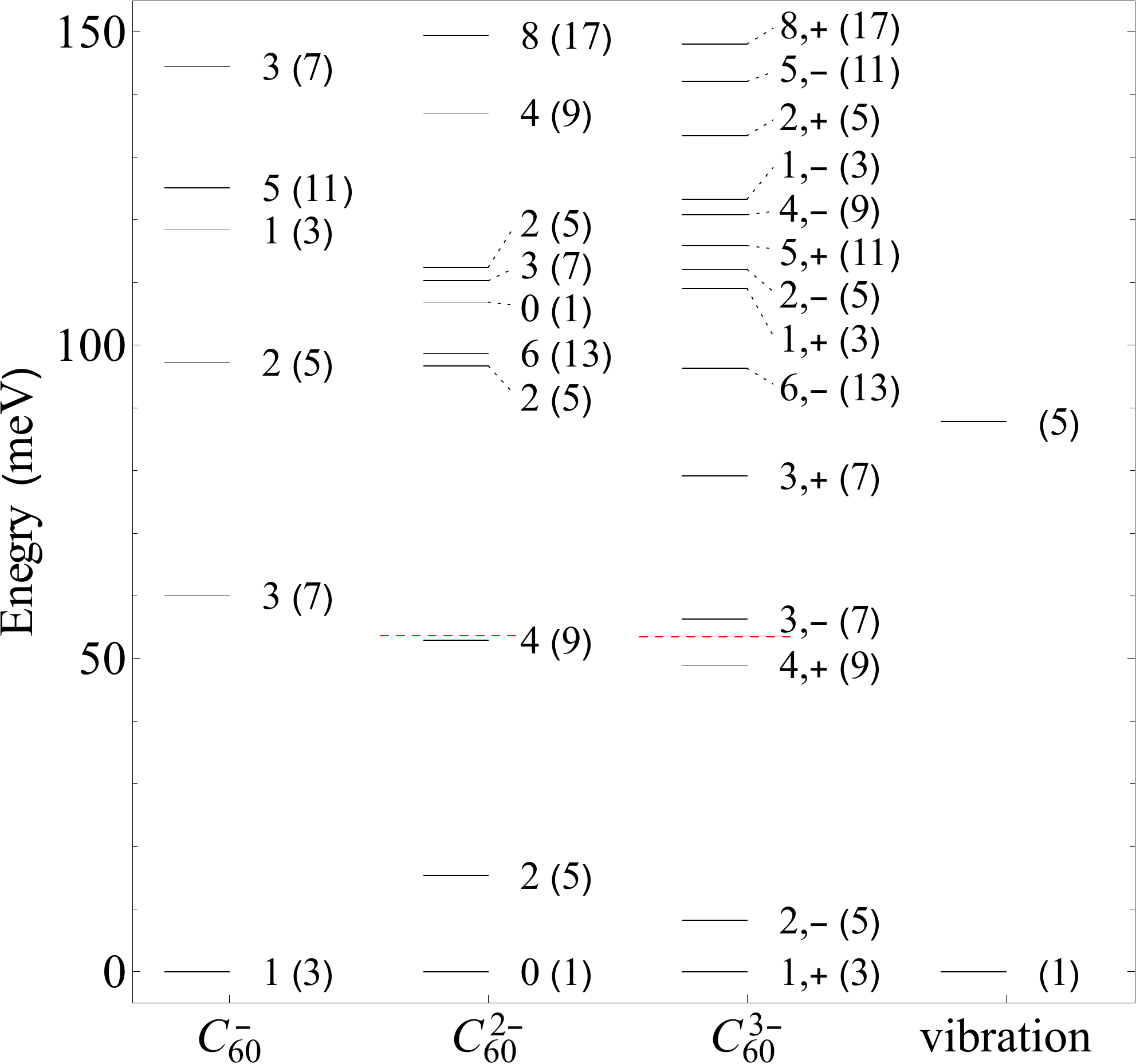}
\caption{
Vibornic energy levels of the effective $p^n \otimes d$ Jahn-Teller model}.
\label{SMFig:Eeff}
\end{figure}

In Fig. \ref{SMFig:Eeff}, the vibronic levels of the effective single mode model are shown. 
Compared with Fig. 4 in the main text, up to higher energy levels are shown. 

\end{document}